\input epsf.tex
\documentclass[12pt]{article}
\usepackage{graphicx}
\usepackage{psfig,picinpar,floatflt,amssymb,epsf,array,bbm,mathrsfs}
\csname @addtoreset\endcsname{equation}{section}

\def\bseq{\begin{subequation}}  
\def\eseq{\end{subequation}}
\def\bsea{\begin{subeqnarray}}  
\def\esea{\end{subeqnarray}}

\def\Tilde#1{\widetilde{#1}}                    

\newcommand{\beq}{\begin{equation}}
\newcommand{\eeq}{\end{equation}}
\newcommand{\bea}{\begin{eqnarray}}
\newcommand{\eea}{\end{eqnarray}}
\newcommand{\ena}{\end{eqnarray}}

\renewcommand{\a}{\alpha}
\renewcommand{\b}{\beta}

\renewcommand{\d}{\delta}
\newcommand{\pa}{\partial}
\newcommand{\g}{\gamma}
\newcommand{\G}{\Gamma}

\newcommand{\D}{\Delta}
\newcommand{\e}{\epsilon}

\renewcommand{\l}{\lambda}
\renewcommand{\L}{\Lambda}

\newcommand{\p}{\pi}

\newcommand{\s}{\sigma}

\newcommand{\Db}{\overline{D}}

\newcommand{\Phib}{\overline{\Phi}}
\newcommand{\adot}{\dot{\alpha}}
\newcommand{\bdot}{\dot{\beta}}
\newcommand{\gdot}{\dot{\gamma}}

\newcommand{\thb}{\overline{\theta}}

\renewcommand{\thb}{\overline{\theta}}

\def\Mb{\kern 2pt\mathchoice
        {
         \vbox{\hrule width10pt height 0.4pt depth 0pt
         \kern 1.2pt\hbox{\kern -2pt$\displaystyle M$}}}
        {
         \vbox{\hrule width10pt height 0.4pt depth 0pt
         \kern 1.2pt\hbox{\kern -2pt$\textstyle M$}}}
        {
\vbox{\hrule width6pt height 0.4pt depth 0pt
         \kern 1.0pt\hbox{\kern -2pt$\scriptstyle M$}}}
        {
         \vbox{\hrule width5pt height 0.4pt depth 0pt
         \kern 0.8pt\hbox{\kern -2pt$\scriptscriptstyle M$}}}}

\def\Sb{\kern 2pt\mathchoice
        {
         \vbox{\hrule width6pt height 0.4pt depth 0pt
         \kern 1.2pt\hbox{\kern -2pt$\displaystyle S$}}}
        {
         \vbox{\hrule width6pt height 0.4pt depth 0pt
         \kern 1.2pt\hbox{\kern -2pt$\textstyle S$}}}
        {
         \vbox{\hrule width3.5pt height 0.4pt depth 0pt
         \kern 1.0pt\hbox{\kern -2pt$\scriptstyle S$}}}
        {
         \vbox{\hrule width3pt height 0.4pt depth 0pt
         \kern 0.8pt\hbox{\kern -2pt$\scriptscriptstyle S$}}}}

\def\Rb{\kern 2pt\mathchoice
        {
         \vbox{\hrule width5.5pt height 0.4pt depth 0pt
         \kern 1.2pt\hbox{\kern -2.5pt$\displaystyle R$}}}
        {
         \vbox{\hrule width5.5pt height 0.4pt depth 0pt
         \kern 1.2pt\hbox{\kern -2.5pt$\textstyle R$}}}
        {
         \vbox{\hrule width3.5pt height 0.4pt depth 0pt
         \kern 1.0pt\hbox{\kern -2.2pt$\scriptstyle R$}}}
        {
         \vbox{\hrule width3pt height 0.4pt depth 0pt
         \kern 0.8pt\hbox{\kern -2.2pt$\scriptscriptstyle R$}}}}

  \def\pp{{\mathchoice
      {
          \kern 1pt%
          \raise 1pt
          \vbox{\hrule width5pt height0.4pt depth0pt
            \kern -2pt
            \hbox{\kern 2.3pt
              \vrule width0.4pt height6pt depth0pt
              }
            \kern -2pt
            \hrule width5pt height0.4pt depth0pt}%
            \kern 1pt
       }
        {
          \kern 1pt%
          \raise 1pt
          \vbox{\hrule width4.3pt height0.4pt depth0pt
            \kern -1.8pt
            \hbox{\kern 1.95pt
              \vrule width0.4pt height5.4pt depth0pt
              }
            \kern -1.8pt
            \hrule width4.3pt height0.4pt depth0pt}%
            \kern 1pt
        }
        {
          \kern 0.5pt%
          \raise 1pt
          \vbox{\hrule width4.0pt height0.3pt depth0pt
            \kern -1.9pt  
            \hbox{\kern 1.85pt
              \vrule width0.3pt height5.7pt depth0pt
              }
            \kern -1.9pt
            \hrule width4.0pt height0.3pt depth0pt}%
            \kern 0.5pt
        }
        {
          \kern 0.5pt%
          \raise 1pt
          \vbox{\hrule width3.6pt height0.3pt depth0pt
            \kern -1.5pt
            \hbox{\kern 1.65pt
              \vrule width0.3pt height4.5pt depth0pt
              }
            \kern -1.5pt
            \hrule width3.6pt height0.3pt depth0pt}%
            \kern 0.5pt
        }
    }}

  \def\mm{{\mathchoice
               {
                 \kern 1pt
           \raise 1pt    \vbox{\hrule width5pt height0.4pt depth0pt
                  \kern 2pt
                  \hrule width5pt height0.4pt depth0pt}
                 \kern 1pt}
               {
                \kern 1pt
           \raise 1pt \vbox{\hrule width4.3pt height0.4pt depth0pt
                  \kern 1.8pt
                  \hrule width4.3pt height0.4pt depth0pt}
                 \kern 1pt}
               {
                \kern 0.5pt
           \raise 1pt
                \vbox{\hrule width4.0pt height0.3pt depth0pt
                  \kern 1.9pt
                  \hrule width4.0pt height0.3pt depth0pt}
                \kern 1pt}
               {
               \kern 0.5pt
         \raise 1pt  \vbox{\hrule width3.6pt height0.3pt depth0pt
                  \kern 1.5pt
                  \hrule width3.6pt height0.3pt depth0pt}
               \kern 0.5pt}
               }}

\def\pd{{\kern0.5pt
           + \kern-5.05pt \raise5.8pt\hbox{$\textstyle.$}\kern
0.5pt}}

\def\pmd{{\kern0.5pt
          \pm \kern-5.05pt
\raise6.3pt\hbox{$\textstyle.$}\kern1.5pt}}

\def\md{{\mathchoice
   {
      {{\kern 1pt - \kern-6.2pt \raise5pt\hbox{$\textstyle.$}\kern
1pt}}}
    {
      {{\kern 1pt - \kern-6.2pt \raise5pt\hbox{$\textstyle.$}\kern
1pt}}}
    {
      {\kern0.5pt - \kern-5.05pt
\raise3.4pt\hbox{$\textstyle.$}\kern0.5pt}}
    {
      {\kern0.5pt - \kern-5.05pt
\raise3.4pt\hbox{$\textstyle.$}\kern0.5pt}}}}

\newcommand{\ad}{{\dot{\alpha}}}
\newcommand{\bd}{{\dot{\beta}}}
\newcommand{\pab}{{\overline{\pa}}}
\newcommand{\Del}{\nabla}

\newcommand{\boldnabla}{  \nabla \hspace{-0.12in}{\nabla}}

\renewcommand{\th}{\theta}
\renewcommand{\adot}{\dot{\alpha}}
\renewcommand{\bdot}{\dot{\beta}}
\renewcommand{\gdot}{\dot{\gamma}}

\newcommand{\boldbox}{  \Box \hspace{-0.121in}{\Box} \hspace{-0.119in}{\Box}}

\begin{document}

\begin{titlepage}
{\hbox to\hsize{October  2005 \hfill
{Bicocca--FT--05--24}}}

\begin{center}
\vglue .06in
{\Large\bf NON(ANTI)COMMUTATIVE SYM THEORY: RENORMALIZATION IN SUPERSPACE}
\\[.45in]
Marcus T. Grisaru\footnote{marcus.grisaru@mcgill.ca}\\
{\it Physics Department, McGill University \\
Montreal, QC Canada H3A 2T8 }
\\
[.2in]
Silvia Penati\footnote{silvia.penati@mib.infn.it} ~and~
Alberto Romagnoni\footnote{alberto.romagnoni@mib.infn.it}\\
{\it Dipartimento di Fisica dell'Universit\`a degli studi di
Milano-Bicocca,\\
and INFN, Sezione di Milano, piazza della Scienza 3, I-20126 Milano,
Italy}\\[.8in]

{\bf ABSTRACT}\\[.0015in]
\end{center}
We present a systematic investigation of one--loop renormalizability for
nonanticommutative $N=1/2$, $U({\cal N})$ SYM theory in superspace. We
first discuss classical gauge invariance of the pure gauge theory and
show that in contradistinction to the ordinary
anticommutative case, different representations of supercovariant
derivatives and field strengths do not lead to equivalent descriptions of the theory.
Subsequently
we develop background field methods which allow us to compute a manifestly
covariant gauge effective action. One--loop evaluation of divergent
contributions reveals that the theory simply obtained from the ordinary one
by trading products for star products is not renormalizable. In the case of SYM
with no matter we present a $N=1/2$ improved action which we show to be
one--loop renormalizable and which is perfectly compatible with the algebraic
structure of the star product. For this action we compute the beta functions.
A brief discussion on the inclusion of chiral matter is also presented.

${~~~}$ \newline
\\[.01in]

\end{titlepage}


\section{Introduction and Conclusions}

In the recent past the study of field theories defined on noncommutative spaces
has received new impetus from the realization that, in a string theory context,
the low-energy dynamics of D-branes is affected by the presence of world-volume
fluxes in a manner which can be described by such field theories
\cite{OV}-\cite{lerda}.
The effect of
noncommutativity manifests itself in the appearance of nontrivial product
properties: the multiplication of fields is no longer commutative but described
by a so-called $\ast$-product. Similarly, when the basic setup is
supersymmetric, the description in superspace of the corresponding superfield
theories is by means of modified anticommutation relations for the spinor
coordinates. In the simplest case, that of a constant graviphoton flux ${\cal
F}_{\a \b}$, the superspace geometry is modified through a nontrivial
anticommutator $\{ \theta^\a , \theta^\b \}= {\cal F}^{\a \b}$ instead of the
usual vanishing one. (Since one keeps the anticommutator of the conjugate
variables at zero, $\{\overline{\theta}^\ad , \overline{\theta}^\bd \}=0$, it is
necessary to work in Euclidean space where dotted and undotted spinors are
unrelated.) Again, the product of superfields is now a $\ast$-product. The
nontrivial anticommutativity (NAC) rule usually leads to a deformation of SUSY  which
corresponds to partial breaking of supersymmetry. This manifests itself by the
appearance, in the component actions, of additional couplings with reduced
invariance properties. For example, $N=1$ global SUSY is broken down to $N=1/2$
with supersymmetry broken in the antichiral sector.

The study of the NAC supersymmetric geometry was initiated in a number of
papers \cite {ferrara, KPT, ferrara2} and the corresponding description of the
resulting field theories first given by Seiberg \cite{seiberg} and subsequently by a
number of other authors. A partial list consists of, in components, refs. 
\cite{AIO}-\cite{Ryttov} 
and in superspace, refs. \cite{Britto0}-\cite{ABBP2} (it should be emphasized that the starting
point is always the superfield description with a nontrivial $\ast$-product and the
subsequent decomposition into components). Of interest has been the question of their
renormalizability. In the absence of deformations the high-energy behavior of
SUSY theories is very much softened by the standard boson-fermion
cancellations; to what extent does this behavior survive the NAC deformation?

The renormalizability  issue has been studied in components, and, by the
present authors, in superspace. We have considered both  deformed WZ and SYM
models. For the former case it was possible to give a complete answer: already
at the one--loop level renormalizability is lost but by adding a single new
coupling in the classical theory, dependent on the deformation parameter, it
can be restored not only at one loop but to all orders of
perturbation theory \cite{BF, R}. Divergences are still only logarithmic so
that the induced SUSY breaking is soft. This follows also from the fact that
the NAC deformation can be mimicked by means of a spurion field.
The same conclusions were reached in components \cite{LR}.

The situation for NAC SYM is more complicated. It is straightforward to start
with a superspace description $S \sim \int {\rm Tr}(W^\a * W_\a)$ where the
$W$'s themselves contain implicit star products and go to components (although
a fermion field shift is necessary in order to maintain the usual component
transformation rules), but such an approach requires choosing a WZ gauge {\em
ab initio} and leaves open the question whether in the presence of the star
product something has been lost. Nonetheless, proceeding with the component
approach, a number of results have been obtained. In particular Jack, Jones and
Worthy \cite{JJW} have provided the most complete results of the one--loop
quantum properties of the deformed component theory and shown that by
adding new, deformation-parameter-dependent terms to the classical theory the
one--loop effective action can be renormalized. In the meantime we had examined
the superspace situation. At the one--loop level, in a background field approach
(which maintains control on the gauge invariance of the theory) it was shown
\cite{pr}  that new divergences were indeed present which could not be removed
by renormalization. The important point however was the check that indeed, even
in the presence of NAC, the effective action is gauge invariant and therefore
the safety of going to WZ gauge is ensured.

We have continued the superspace work and in this paper we show how, by again
adding new terms to the classical action, we can make the superspace theory
one-loop renormalizable, generally confirming the work of \cite{JJW}. We prove that
subtraction of one--loop divergences does not require renormalization of the
NAC parameter. Therefore, renormalization does not deform the star product. We have
also been able to show that the supersymmetry breaking due to the NAC geometry is soft. 
In the rest of this Section we give a detailed summary of our procedure and results
and, to save the reader from leafing to the end of this rather technical
report, we present also our conclusions.

\vspace{ 1cm}

{\em The classical action}

SYM is best described in a covariant approach by means of covariant
derivatives and field strengths satisfying a number of constraints. For
quantization purposes these constraints have to be solved in terms of
unconstrained superfields, and this role is played by the gauge prepotential $V
= V^AT_A$ where $T_A$ are the generators of the gauge group $U({\cal N})$. (In
the ordinary theory it would be sufficient to study $SU({\cal N})$ and $U(1)$
separately but, as we shall see, in the NAC case the two subgroups  are
intimately linked and must be considered together). The field strength $W_\a$ (and
its conjugate) is expressed in terms of exponentials of the prepotential and
their spinor derivatives. Of course everything involves the star product.

Whereas in Minkowski space $V$ is usually considered real, it turns out that in
order to maintain certain conjugation properties of the (spinor) covariant
derivatives it is necessary to choose $V$ imaginary. In both the ordinary and
NAC case the description in terms of $V$ introduces a certain asymmetry between
some of the geometric quantities and their conjugates, and one has the choice of  {\em
gauge chiral representation} or {\em gauge antichiral representation}. In the
first case the prepotential covariantizes only the chiral spinor derivative
${D}_\a $ while the antichiral derivative $\overline{D}_\ad$ needs not be
covariantized. In the second case the opposite is true. The corresponding field
strengths also differ slightly in their prepotential dependence. We denote them
as $W_\a~ , \Tilde{W}_\ad$ and $\Tilde{W}_\a~ , \overline{W}_\ad$. The choice of
representation makes no difference in the usual, AC, case. It turns out that in
the NAC case, and especially in the background field method, this choice is not
totally arbitrary.

In the AC case, and in the absence of instantons, the choice among four
possible actions $ S \sim \int W^\a W_\a$, its complex conjugate, or
the corresponding antichiral gauge quantities has no consequences; they all
lead to the same component action. This is no longer true in the NAC case.
Whereas in the usual case the equivalence is established by making use of
cyclicity of the trace in quantities such as, for example, $\int
d^2\overline{\theta} {\rm Tr}( e^{-V} \overline{W}^\ad \overline{W}_\ad e^V )$, the
 cyclicity is lost unless
integration over $d^2 \theta$ is present. Even gauge invariance can be lost.
Our preferred choice, before quantization can proceed, is for the action
\beq
S = \frac{1}{2g^2}\int d^4x d^2 \overline{\theta}~ 
{\rm Tr} (\overline{W}^\ad \overline{W_\ad})
\label{actionchoice}
\eeq
{\em in antichiral representation}.

\vspace{1cm}

{ \em The U(1) problem}

In the NAC case  $U(1)$  is interlinked with $SU({\cal N})$, especially in the
transformation properties of the corresponding gauge fields. In particular, the $U(1)$ field
transforms  nontrivially under the $SU({\cal N})$ group. Consequently, although
 the action (\ref{actionchoice}) which contains both gauge fields is invariant under
 the full group, a separate  action
 \beq
 S_0 =\frac{1}{2g_0^2}\int d^4x d^2 \overline{\theta}~ {\rm Tr} (\overline{W}^\ad)
 {\rm Tr}(\overline{W_\ad})
 \label{actionU1}
\eeq
which describes just the $U(1)$ field (and which is also generated by quantum corrections)
is not. Only by completing this action with an additional piece  involving the cubic term
${\cal F}^{\rho \g}{\rm Tr} \left(
\pa_{\rho \dot{\rho}} \overline{\Gamma}^{\adot} \right) {\rm Tr}\left(
\overline{W}_{\adot} \overline{\Gamma}_{\g}^{~\dot{\rho}} \right)$ we can maintain gauge
invariance. Furthermore, as we shall see, the generation of this additional piece and
the quadratic part thereof complicates the choice of a suitable gauge-fixing term;
it forces us to a propagator choice away from Feynman gauge, thus leading to more cumbersome supergraph
evaluation.

\vspace{1cm}

{\em The background field method}

The computation of the effective action is best done using the background field method
since gauge invariance of the resulting effective action is maintained (although, in
principle problems could arise in the NAC case). Interestingly, unlike the AC case,
as a consequence of $V$ being pure imaginary, it is possible to perform a background-quantum
 splitting, as we did, where {\em all} quantities, quantum, background and total,
  can be chosen in
 antichiral representation, in contradistinction to the usual, Minkowski,
  case where the background has to be in vector representation.

The splitting can be performed in a manner analogous to the usual case, with
$ \nabla_\a = {\boldnabla}_\a =D_\a $, $\nabla_{\adot} = e_{\ast}^V
\ast \boldnabla_{\adot}\ast e_{\ast}^{-V} $ for the derivatives and
$\overline{\Phi} =
 \bold{\overline{\Phi}} $, $
   \Phi = e^V_{\ast} \ast \bold{\Phi}\ast e^{-V}_\ast $
   for chiral superfields, where plain and boldface letters indicate
 quantum and background quantities, respectively; and one-loop Feynman rules can be derived as
in the AC case \cite{superspace}.

A slight complication manifests itself in computing
matter or ghost contributions to the effective action. Using the doubling
 trick explained in that reference one separates contributions from only chiral
 superfields and from antichiral superfields, but using the  fact that these superfields
 are complex conjugates of each other it is easy to see that these contributions are equal.
In the NAC case chiral and antichiral fields are not related by complex conjugation.
Nonetheless, by explicit examination of the
manner in which one--loop contributions are obtained, it is possible to show that they
still contribute equally  and only one of them need be computed explicitly.
 After the splitting,
the evaluation of contributions from the gauge
fields themselves presents no new problem. Their quantization proceeds in straightforward
manner (except for the complications with the $U(1)$ fields noted above).

\vspace{1 cm}

{\em Structure of the divergent terms}

Before computing the one-loop divergences it is useful to fix the general
structure of local terms allowed by gauge invariance and other symmetries. Knowing
this  is a useful guide to the subsequent calculation and allows us to
eliminate from the beginning diagrams which would not lead to the appropriate structures.

Gauge invariance of the  background field action implies that allowed terms
can only depend on background field strengths and gauge connections but not on the background
 prepotential. They will also depend on the
deformation parameter. Furthermore, if written as integrals over full superspace, they may
 also depend
explicitly on $\overline{\theta}^\ad$ as a consequence of the explicit breaking of $N=1$ SUSY.
As in other recent studies \cite{LR, Beren, R} it is also helpful to take advantage
of additional global (pseudo)symmetries of the classical action which lead to further restrictions
(note that possible anomalies would not affect these symmetries of the local terms). In the
 present case we do have one global R--symmetry and we can assign specific R--weights to
 the various quantities that can appear in the effective action. The resulting divergent
 structures (we employ a cutoff $\Lambda$ for convenience and  also use dimensional
 arguments), are quite limited in number and can be classified prior to any calculations.

 \vspace{1cm}

{\em One-loop divergences and lack of renormalizability}

We have found convenient to perform perturbative calculations in
momentum superspace by Fourier transforming all the superspace 
variables and in particular introducing spinorial momenta $\pi_\a$
conjugate to the derivatives $\pa_\a$. In this setup the presence of the
star product is signaled by the appearance at the vertices of phase factors  
of the form  ${\rm exp}[{\pi}_\a {\cal F}^{\a \b} {\pi}_\b]$.   
D--algebra rules now require that in the integration 
over spinorial loop momenta exactly one $\pi^2$ and one $\overline{\pi}^2$ be left
for each loop. It is important to note that, while in the planar diagrams the resulting
phase factor does not depend on the loop momenta and powers of
$\pi$ and $\overline{\pi}$ only come from propagators and vertices, in the nonplanar ones
powers of $\pi$ come also from the expansion of the nontrivial phase factor
and new divergent contributions proportional to the NAC parameter arise. This is the way
NAC geometry affects the UV properties of our theory.

We have computed the divergences of the gauge effective action stemming from vector,
as well as the (chiral) matter and ghost loops. We find contributions from planar
diagrams (these are of course the standard, renormalizable ones),
as well as ones proportional to the deformation parameter and its square coming from
nonplanar diagrams. In the background
 field method none of these arise from the vector loops themselves. They are proportional to
\bea
\G^{(1)}_{gauge} &\rightarrow&  \frac{1}{2}(-3 + N_f)  \times \Big\{
\int d^4x~d^4 \theta~ \Big[ {\cal N}{\rm Tr}\left(
\overline{\Gamma}^{\adot} ~ \overline{W}_{\adot}\right) - {\rm
Tr}\left( \overline{\Gamma}^{\adot}  \right){\rm Tr}\left(
\overline{W}_{\adot}\right) \Big] \nonumber \\
&& \qquad \qquad \qquad \qquad - 4 i {\cal F}^{\rho \g} \int d^4x~d^4 \theta~~
\overline{\theta}^2~{\rm
Tr} \left( \pa_{\rho \dot{\rho}}  \overline{\Gamma}^{\adot} \right)
{\rm Tr}\left( \overline{W}_{\adot}
\overline{\Gamma}_{\g}^{~\dot{\rho}} \right) \nonumber \\
&& \qquad \qquad \qquad \qquad + ~{\cal F}^2 \int d^4x~d^4 \theta~~ \overline{\theta}^2~{\rm
Tr} \left(\overline{\Gamma}^{\adot}\overline{W}_{\adot}\right){\rm
Tr}\left(\overline{W}^{\bdot}\overline{W}_{\bdot}\right) \Big\}
\eea
As shown in \cite{pr}, whereas the first and last terms are separately gauge-invariant,
 gauge invariance of the second and third terms is true only when they appear in this
 particular combination.

The resulting divergences cannot be renormalized away. This can be seen at the
superspace level,
but it can also be verified by going to components, in agreement with \cite{JJW}.

\vspace{1cm}

{\em Renormalizable deformations and beta-functions}

As was the case in the Wess-Zumino model, and is also the case in the component
 discussions of
NAC SYM, it is possible to deform the classical action in such a way as to
produce a one--loop renormalizable theory. The manner in which we have proceeded is to
start {\em ab initio} with a deformed action containing all possible terms allowed by
gauge invariance, R--symmetry, and dimensional considerations. As mentioned earlier,
one of the complications that arise in performing the calculations is the
appearance of a separate $U(1)$ term with, a priori, a different weight from its gauge
invariant combination partner; the simplest way to proceed, as we did, is to
accept the fact that Feynman gauge for this field is  ultimately not a simple
choice and we must use a more complicated propagator. Aside from this, the calculations
proceed apace.

We find  that in the presence of new ``classical'' terms, vector loops themselves
contribute now to the one-loop divergences. Once they are all calculated we end up
with an obviously one-loop renormalizable situation depending on a number of
arbitrary coupling constants. We then compute the $\b$-functions and we find that
at this one-loop level they allow for specific restrictions on these constants.
In particular, there are two choices which lead to minimal deformed actions which are
one-loop renormalizable:
\bea
&& S_{min} =
~\frac{1}{2~g^2}\int d^4x~d^4\theta~ {\rm Tr}\left(
\overline{\Gamma}^{\adot}~\overline{W}_{\adot}\right) \nonumber \\
&& \qquad
+ \frac{1}{2~g_0^2~{\cal N}} \int d^4x~d^4 \theta~ \Bigg[ ~~ {\rm
Tr}\left( \overline{\Gamma}^{\adot} \right){\rm Tr}\left(
\overline{W}_{\adot}\right) \nonumber \\
&& \qquad \qquad  \qquad \qquad \qquad\qquad + 4 i {\cal
F}^{\rho \g} \overline{\theta}^2~{\rm Tr} \left(
\pa_{\rho \dot{\rho}} \overline{\Gamma}^{\adot} \right) {\rm Tr}\left(
\overline{W}_{\adot} \overline{\Gamma}_{\g}^{~\dot{\rho}} \right)
 \nonumber \\
&& \qquad \qquad \qquad \qquad \qquad \qquad - {\cal F}^2
\overline{\theta}^2~{\rm Tr}
\left(\overline{\Gamma}^{\adot}\overline{W}_{\adot}\right){\rm
Tr}\left(\overline{W}^{\bdot}\overline{W}_{\bdot}\right) ~~~\Bigg] \nonumber \\
&& \qquad + \frac{1}{l^2} ~{\cal F}^2 \int d^4 x ~d^4
\theta~\overline{\theta}^2 ~{\rm Tr} \left( \overline{\Gamma}^{\adot}~
\overline{W}_{\adot}~\overline{W}^{\bdot}~\overline{W}_{\bdot}\right)
\label{min1}
\end{eqnarray}
or
\bea
&& S'_{min} =
~\frac{1}{2~g^2} \int d^4 x ~ d^4 \theta~ \Bigg[ ~~{\rm Tr}\left(
\overline{\Gamma}^{\adot}~\overline{W}_{\adot}\right) \nonumber \\
&& \qquad  \qquad \qquad \qquad  \qquad \qquad+ {\cal F}^2
~ \overline{\theta}^2~{\rm Tr}
\left(\overline{\Gamma}^{\adot}\overline{W}_{\adot}\right){\rm
Tr}\left(\overline{W}^{\bdot}\overline{W}_{\bdot}\right) \Bigg] \nonumber \\
&& \qquad
+ \frac{1}{2~g_0^2~{\cal N}} \int d^4x~d^4 \theta~ \Bigg[ ~~{\rm
Tr}\left( \overline{\Gamma}^{\adot} \right){\rm Tr}\left(
\overline{W}_{\adot}\right) \nonumber \\
&& \qquad  \qquad \qquad \qquad  \qquad \qquad+ 4 i {\cal
F}^{\rho \g} ~ \overline{\theta}^2~{\rm Tr} \left(
\pa_{\rho \dot{\rho}} \overline{\Gamma}^{\adot} \right) {\rm Tr}\left(
\overline{W}_{\adot} \overline{\Gamma}_{\g}^{~\dot{\rho}} \right) \Bigg]
 \nonumber \\
&& \qquad + \frac{1}{l^2} ~{\cal F}^2 \int d^4 x ~d^4
\theta~\overline{\theta}^2 ~{\rm Tr} \left( \overline{\Gamma}^{\adot}~
\overline{W}_{\adot}~\overline{W}^{\bdot}~\overline{W}_{\bdot}\right)
\label{min2}
\end{eqnarray}
depending on just three arbitrary coupling constants, of which two, the $SU({\cal N})$
and the $U(1)$  $g$ and $g_0$, reflect the corresponding situation in ordinary AC SYM,
whereas the third one is a new feature of NAC.

\vspace{1cm}

{\em Final comments for the exhausted reader}

Let us summarize this summary:

We have examined in superspace the quantum properties of NAC SYM,
concluding that, as in components, suitable deformations of the classical
actions are necessary in order to achieve renormalizability. This is only
at the one--loop level; the higher--loop situation remains to be studied but the hope
is that, just as in the Wess-Zumino model \cite{BF, R},
these deformations are sufficient to achieve the same results at all loops.
We note that the deformed actions can also be written
purely in terms of star products and that the deformation parameter needs no renormalization.

The reader may feel that the corresponding analysis in components is simpler, but
it does require going to WZ gauge with implications  that were not {\em a priori}
clear to us. We also think that some of the  other issues that arose in the course of
our investigation are of interest, if only to superspace {\em aficionados}.

Our work complements, and should be compared to that of \cite{JJW}. In fact, by taking our
superspace results to WZ component gauge it is possible to make a direct
comparison. In general we are in
agreement with the results there although some of the details may differ and are also obscured
by the choice of WZ gauge and elimination of auxiliary fields.
Working in superspace, or with the full complement of auxiliary fields,
automatically obviates the need for a nonlinear, field--dependent wave-function
renormalization (see also \cite{JJW, JJW3}) and makes the discussion cleaner.
Aside from this, the only significant difference seems to be the following:
the new term in the modified component action of that reference that is required
for renormalization
(besides the splitting between $g$ and $g_0$ terms) would come from a superspace
expression of the form $\int \overline{\th}^2 {\rm Tr}(\overline{\G}){\rm Tr}(\overline{W})
{\rm Tr}(\overline{W} \overline{W})$. Instead, our corresponding new term  has the form
$\int \overline{\th}^2 {\rm Tr}(\overline{\G} \overline{W} \overline{W} \overline{W})$.
The two terms differ only in the color structure. It would be nice to understand
the reason for this mismatch.

We remarked earlier that starting with the original deformed action the complete
contribution to the one-loop divergences comes only from chiral
matter or ghost fields. It follows immediately that at least the gauge sector
is completely finite in a theory with the field content of $N=4$ theory consisting of one
gauge and three matter chiral fields (see \ref{eff-action1}).
The situation for the final modified actions is more complicated.
In such a theory one would expect that  corresponding modifications in the chiral
matter sector would lead to additional one--loop contributions to the gauge effective
action. We conjecture that under those circumstances the modified actions may very well
maintain the finite properties of the undeformed $N=4$ theory.

In ref.\cite{lerda} the classical action for NAC SYM in components, in the WZ gauge
has been derived directly from string theory by computing the low energy limit of
string scattering amplitudes in the presence of a RR two--form flux. The resulting action
has the form (\ref{actionchoice}) and it corresponds to the natural NAC generalization of the
ordinary action where products have been promoted to star products. It would be interesting
to investigate how to extend that analysis to reproduce the renormalizable actions 
(\ref{min1}, \ref{min2}) with the correct number of coupling constants 
directly from string theory.


\section{SYM theories in $N=1/2$ superspace} \label{NACSYM}

Nonanticommutative $N=(\frac12 , 0)$ superspace can be defined as a truncation
of euclidean $N=(1,1)$ superspace endowed with nonstandard hermitian
conjugation rules for the spinorial variables \cite{seiberg, us, ILZ}. It is
described by the set of coordinates $(x^{\a\adot}, \theta^\a ,
{\overline{\theta}}^{\adot})$, $(\theta^\a)^\ast = i \theta_\a$,
$({\overline{\theta}}^{\adot})^\ast = i \overline{\theta}_{\adot}$,
satisfying
\beq \big\{ \theta^{\alpha}, \theta^{\beta} \big\} = 2 {\cal F}^{\alpha \beta}
\qquad \quad {\rm the~ rest} = 0 \label{nc} \eeq
where ${\cal F}^{\alpha
\beta}$ is a $2 \times 2$ symmetric, constant matrix. This algebra is
consistent only in euclidean signature where the chiral and antichiral sectors
are totally independent and not related by complex conjugation.

We use chiral representation \cite{superspace} for supercharges and
covariant derivatives
\bea
&& \overline{Q}_{\adot} = i ( \overline{\pa}_{\adot} - i \theta^{\a} \pa_{\a \adot} )
\quad , \quad Q_{\a} = i \pa_{\a}
\nonumber \\
&& \overline{D}_{\adot} =  \overline{\pa}_{\adot} \qquad \qquad \qquad , \quad
D_{\a} = \pa_{\a} + i {\overline{\theta}}^{\adot} \pa_{\a \adot}
\label{DQ}
\eea
While the algebra of covariant derivatives is not modified,
the algebra of supercharges receives an extra contribution from (\ref{nc})
and the supersymmetry in the antichiral sector is explicitly broken
\cite{seiberg}.

We note that in euclidean signature the complex conjugation relations are
\bea
&& Q_\a^\ast = i Q^\a ~\qquad , \qquad \overline{Q}_{\adot}^\ast = i \overline{Q}^{\adot}
\nonumber \\
&& D_\a^\ast = -i D^\a \qquad , \qquad \overline{D}_{\adot}^\ast = -i \overline{D}^{\adot}
\label{hc}
\eea

On the class of smooth superfunctions $\phi(x^{\a\adot}, \theta^\a ,
{\overline{\theta}}^{\adot})$, the NAC geometry (\ref{nc}) can be realized
by introducing the nonanticommutative but associative star product
\bea
\phi \ast \psi &\equiv& \phi e^{- \overleftarrow{\pa}_\a {\cal F}^{\a \b}
\overrightarrow{\pa}_\b} \psi
\nonumber \\
&=& \phi \psi - \phi \overleftarrow{\pa}_\a {\cal F}^{\a \b}
\overrightarrow{\pa}_\b \psi
+ \frac12 \phi \overleftarrow{\pa}_\a \overleftarrow{\pa}_\g {\cal F}^{\a \b}
{\cal F}^{\g \d}
\overrightarrow{\pa}_\d \overrightarrow{\pa}_\b \psi
\nonumber \\
&=& \phi \psi - \phi \overleftarrow{\pa}_\a {\cal F}^{\a \b}
\overrightarrow{\pa}_\b \psi - \frac12 {\cal F}^2 \pa^2\phi \, {\pa}^2 \psi
\label{star} \eea
where we have defined ${\cal F}^2 \equiv  {\cal F}^{\a \b}
{\cal F}_{\a \b}$. The covariant derivatives (\ref{DQ}) are still derivations
for this product. Therefore, the class of (anti)chiral superfields defined by
the constraints $\overline{D}_{\adot} \ast \Phi = D_\a \ast \overline{\Phi} =0$
is closed.

A property of the star product that we will often use is the following:
The trace of the $\ast$--product of a number of fields is not in general cyclic
unless it is integrated over $d^2\th$. Specifically, we have
\bea \label{first}
&& ~~~~~~~~{\rm Tr}(A \ast B) \neq {\rm Tr}(B \ast A)
\\
&& \int d^2\thb ~{\rm Tr}(A \ast B) \neq \int d^2\thb ~{\rm Tr}(B \ast A)
\eea
but
\beq
\int d^2\th ~{\rm Tr}(A \ast B) = \int d^2\th ~{\rm Tr}(B \ast A)
\label{second}
\eeq
However, even under $d^2\th$ integration the cyclicity property gets spoiled when the trace
appears multiplied by an extra function. In particular,
\beq
\int d^2\th ~{\rm Tr}(A \ast B) {\rm Tr } (C) \neq
\int d^2\th ~{\rm Tr}(B \ast A) {\rm Tr } (C)
\label{NACtrace}
\eeq
These properties can be easily proved by expanding the star product as in (\ref{star}).

\vskip 15pt

We now turn to the description of SYM theories in the non(anti)commutative case.
Supersymmetric Yang--Mills theories in $N=1/2$ superspace can be defined
as usual in terms of a scalar prepotential $V$ in the adjoint representation
of the gauge group ($V \equiv V_A T^A$, $T^A$ being the group generators)
\footnote{Since in the presence of non(anti)commutativity also the $U_\ast(1)$
gauge theory becomes nonabelian the relations we introduce hold nontrivially
for any gauge group, $U_\ast(1)$ included.}.
It is subject to the supergauge transformations
\beq
e_\ast^V \rightarrow  e_\ast^{V'} = e_\ast^{i \overline{\L}} \ast
e_\ast^V \ast e_\ast^{-i\L}
\label{gauge}
\eeq
where  $\L, \overline{\L}$ are chiral and antichiral superfields, respectively.

Supergauge covariant derivatives in superspace can be defined in the so--called
{\em gauge chiral representation} as
\beq
 \nabla_A \equiv (\nabla_\a , \nabla_{\adot}, \nabla_{\a \adot})
~=~ ( e_\ast^{-V} \ast D_\a \, e_\ast^V ~,~ \overline{D}_{\adot} ~,~
-i \{ \nabla_\a, \nabla_{\adot} \}_{\ast} )
\eeq
or, equivalently, in {\em gauge antichiral representation} as
\beq
 \overline{\nabla}_A \equiv (\overline{\nabla}_\a , \overline{\nabla}_{\adot},
\overline{\nabla}_{\a \adot})
~=~ (D_\a ~,~ e_\ast^{V} \ast \overline{D}_{\adot} \, e_\ast^{-V} ~,~
-i \{ \overline{\nabla}_\a, \overline{\nabla}_{\adot} \}_{\ast} )
\eeq

In contradistinction to the ordinary anticommutative case where the superfield
V is chosen to be real, in the NAC case, in order to preserve the c.c.
relations (\ref{hc}) also for the supergauge covariant derivatives, we choose
$V$ to be {\em pure imaginary}, $V^\dag = -V$.

The supergauge covariant derivatives can be expressed in terms of ordinary
superspace derivatives and a set of connections, as
$\nabla_A \equiv D_A - i \G_A$ or $\overline{\nabla}_A \equiv
\overline{D}_A - i \overline{\G}_A$.
In chiral representation they are
\beq
 \G_\a = ie_\ast^{-V} \ast D_\a \, e_\ast^V
\qquad , \qquad  \G_{\adot} = 0  \qquad , \qquad \G_{\a \adot} = -i \overline{D}_{\adot}
  \G_\a \label{connections}
\eeq
whereas in antichiral representation we have
\beq
\overline{\G}_{\a}=0     \qquad , \qquad  \overline{\G}_{\adot} =
 ie_\ast^{V} \ast \overline{D}_{\adot} \, e_\ast^{-V}
\qquad , \qquad \overline{\G}_{\a \adot} = -i D_\a  \overline{\G}_{\adot}
\eeq
The field strengths are defined as $\ast$--commutators of supergauge
covariant derivatives
\beq
{\rm chiral ~repr.} ~~~\quad W_\a = -\frac12 [ \nabla^{\adot}, \nabla_{\a \adot} ]_\ast
\qquad , \qquad \Tilde{W}_{\adot} = -\frac12 [ \nabla^{\a}, \nabla_{\a \adot} ]_\ast
\eeq
\beq
{\rm antich. ~repr.} \quad \Tilde{W}_\a =
-\frac12 [ \overline{\nabla}^{\adot}, \overline{\nabla}_{\a \adot} ]_\ast
\qquad , \qquad \overline{W}_{\adot} = -\frac12 [ \overline{\nabla}^{\a},
\overline{\nabla}_{\a \adot} ]_\ast
\eeq
and satisfy the Bianchi's identities $\nabla^\a \ast W_\a + \nabla^{\adot}
\ast \Tilde{W}_{\adot} =0$
and $\overline{\nabla}^\a \ast \Tilde{W}_\a + \overline{\nabla}^{\adot} \ast
\overline{W}_{\adot}=0$.
The superfield strengths in antichiral representation are related to the ones in
chiral representation as
\beq
\Tilde{W}_\a = e_\ast^V \ast W_\a \ast e_\ast^{-V} \qquad , \qquad
\overline{W}_{\adot} = e_\ast^{V} \ast \Tilde{W}_{\adot}   \ast e_\ast^{-V}
\label{Wrel}
\eeq
While $W_\a$ and $\overline{W}_{\adot}$ are ordinary chiral and antichiral superfields,
the tilde quantities are covariantly (anti)chiral.

Under supergauge transformations (\ref{gauge}) all the superfield strengths
transform covariantly. For infinitesimal transformations we have
\bea
\d W_\a = i [ \L , W_\a ]_\ast \qquad , \qquad
\d \Tilde{W}_{\adot} = i [ \L , \Tilde{W}_{\adot} ]_\ast
\nonumber \\
\d \Tilde{W}_\a = i [ \overline{\L} , \Tilde{W}_\a ]_\ast \qquad , \qquad
\d \overline{W}_{\adot} = i [ \overline{\L} , \overline{W}_{\adot} ]_\ast
\label{Wtransf}
\eea
If we expand $W_\a = W_\a^A T^A$ where $T^A$ are the group generators,
use the definitions
(\ref{id2bis}, \ref{id3}) and the identity (\ref{commut}) we can rewrite
\beq
\d W_\a^A = \frac{i}{2} d_{ABC} [ \L^B , W_\a^C ]_\ast - \frac12  f_{ABC}
\{ \L^B , W_\a^C \}_\ast
\label{Wtransf2}
\eeq
and similarly for the others. In the particular case of $U({\cal N})$, given the explicit
 expressions
(\ref{id3}) for $d_{ABC}$, the first term in $\d W_\a^A$ mixes $U(1)$ and $SU({\cal N})$
fields. In particular, the abelian,  $U(1)$ field strength $W_\a^0$ transforms
nontrivially under $SU({\cal N})$ and its
transform is given in terms of both $U(1)$ and $SU({\cal N})$ fields. In the
commutative limit this term goes to zero and we are back to the
ordinary theory where $SU({\cal N})$ fields only transform under $SU({\cal N})$
transformations while the abelian field is a singlet. As we shall see this is the
source of significant complications.

\vskip 15pt
In the ordinary anticommutative superspace, in the absence of instantonic effects,
any of the following actions
\bea
&& ~S = \int d^4x d^2 \th ~{\rm Tr} (W^\a W_\a) \quad ; \quad
\Tilde{S} = \int d^4x d^2 \thb ~{\rm Tr} (\Tilde{W}^{\adot} \Tilde{W}_{\adot})
\nonumber \\
&& \overline{S} = \int d^4x d^2 \thb ~{\rm Tr} (\overline{W}^{\adot} \overline{W}_{\adot})
\quad ; \quad \Tilde{\overline{S}} = \int d^4x d^2 \th ~{\rm Tr} (\Tilde{W}^\a \Tilde{W}_\a)
\label{Cactions}
\eea
can be used to describe pure gauge theory. In fact, any of these actions is gauge
invariant and, when reduced to components in the WZ gauge, describes the correct dynamics of
the physical degrees of freedom (gluons and gluinos) \cite{superspace}.
In particular, the actions $S$ and $\Tilde{\overline{S}}$, as well as $\overline{S}$ and
$\Tilde{S}$, are trivially identical as can be easily understood by using the
relations (\ref{Wrel}) and the cyclicity of the trace. Instead, $S$ and $\overline{S}$
differ by surface terms which are zero if we do not include instantons. The equivalence
of the actions (\ref{Cactions}) holds for any gauge group, $U(1)$ included.

A peculiarity of the NAC case is that in the presence of star products it is no
longer true  that the four actions (\ref{Cactions}) are all equivalent.
For example, let us consider $\Tilde{S}$ versus $\overline{S}$. By using the
relations (\ref{Wrel}) we have the following chain of relations
 \bea
\Tilde{S} = \int d^4x d^2 \thb ~{\rm Tr} (\Tilde{W}^{\adot}
\Tilde{W}_{\adot}) &=& \int d^4x d^2 \thb ~{\rm Tr} (e_\ast^{-V} \ast
\overline{W}^{\adot} \overline{W}_{\adot} \ast e_\ast^{V})
\nonumber \\
&\neq& \int d^4x d^2 \thb ~{\rm Tr} (\overline{W}^{\adot} \overline{W}_{\adot}) =
\overline{S}
\eea
since in this case the trace is not cyclic, as follows from
(\ref{first}). What is interesting from
the physical point of view is that the non--equivalence of the two actions has
important consequences for their gauge invariance. In fact, it is easy to show
that under transformations (\ref{Wtransf}) the action $\overline{S}$ {\em is} gauge
invariant whereas $\Tilde{S}$ {\em is not}. For what concerns $S$ and
$\Tilde{\overline{S}}$ instead, they are still equivalent and both gauge invariant
since they are defined as chiral integrals and the cyclicity of the trace can
be used in this case. Finally, as in the ordinary case, the two gauge invariant
actions $S$ and $\overline{S}$ are equivalent up to instantonic effects when reduced
to components in the WZ gauge \cite{seiberg}.

The situation is even worse if we consider only the $U(1)$ part of the actions
(\ref{Cactions}). We note that this part can be separated out in the form of
a product of single traces. Looking at the $U({\cal N})$ transformations of the abelian
superfield strengths as given in (\ref{Wtransf2}) one can prove that among the
abelian actions
 \bea
 && ~S_0 = \int d^4x d^2 \th ~{\rm Tr} (W^\a){\rm Tr}(
W_\a) \quad ; \quad \Tilde{S}_0 = \int d^4x d^2 \thb ~{\rm Tr}
(\Tilde{W}^{\adot} ){\rm Tr} (\Tilde{W}_{\adot})
\nonumber \\
&& \overline{S}_0 = \int d^4x d^2 \thb ~{\rm Tr} (\overline{W}^{\adot})
{\rm Tr}(\overline{W}_{\adot})
\quad ; \quad \Tilde{\overline{S}}_0 = \int d^4x d^2 \th ~{\rm Tr}
(\Tilde{W}^\a){\rm Tr}( \Tilde{W}_\a)
\label{abelian}
\eea
only $\Tilde{\overline{S}}_0$ is gauge invariant, whereas the others are {\em not} and
need to be completed by extra terms in order to restore gauge invariance. In particular,
we will be interested in the gauge invariant completion of $\overline{S}_0$ which reads
\cite{pr}
\beq
\int d^4x~d^2 \overline{\theta}~{\rm Tr}( \overline{W}^{\adot}){\rm Tr}(\overline{W}_{\adot})
+ 4 i {\cal F}^{\rho \g} \int d^4x~d^4
\theta~~ \overline{\theta}^2~{\rm Tr} \left( \pa_{\rho \dot{\rho}}
\overline{\Gamma}^{\adot} \right) {\rm Tr}\left(
\overline{W}_{\adot} \overline{\Gamma}_{\g}^{~\dot{\rho}} \right)
\label{ACgaugeinv}
\eeq
where ${\cal F}^{\rho \g}$ is the NAC parameter. We note that the lack of invariance
of the abelian actions in (\ref{abelian}) is due to the fact that the abelian gauge
field transforms nontrivially under $SU({\cal N})$ and its variation is proportional
to the $SU({\cal N})$ gauge fields (see eq. (\ref{Wtransf2})). We also note that
despite the nontrivial variation of the $U(1)$ part as described in $\overline{S}_0$,
the total action $\overline{S}$ which describes the propagation of $U(1)$ and $SU({\cal N})$
fields {\em is} gauge invariant since the gauge variation of the $U(1)$ fields gets
compensated by the gauge variation of the $SU({\cal N})$ fields. This is peculiar
to the NAC case and does not have direct correspondence in the ordinary anticommutative case.

Given the asymmetry between chiral and antichiral representations introduced by
the nonanticommutativity, it turns out that the choice of one representation
with respect to the other may be preferable from the point of view of technical
convenience. We find it preferable to work in antichiral representation for the
following reason: in the ordinary, anticommuting case we often switch between
full superspace integrals and chiral (or antichiral) integrals by using the
equivalence $\int d^4x d^4 \theta  ~{\rm Tr} [F(z)]    \equiv \int d^4x d^2
\theta
  ~{\rm Tr} [\overline{\nabla}^2 F(z)]  \equiv  \int d^4x
 d^2 \overline{\theta} ~{\rm Tr}[ \nabla^2 F(z)] $. However, in the NAC case
the second equality
fails if one is working in chiral representation, as can be seen by examining its
derivation
in the following sequence of equalities (star products understood in the NAC case):
\bea
&&\int d^4x d^4 \theta  ~{\rm Tr} [F(z)] = \int d^4x d^4 \theta  ~{\rm Tr}
\{e^{-V} [F(z)] e^V\} =
\int d^4x d^2 \overline{\theta}  ~{\rm Tr} \{D^2 e^{-V} [F(z)] e^V\}
\nonumber \\
&&= \int d^4x d^2 \overline{\theta}  ~{\rm Tr}  \{e^{-V} e^V D^2 e^{-V} [F(z)]
e^V\} =\int d^4x d^2 \overline{\theta}  ~{\rm Tr} \{ e^{-V} \nabla^2 [F(z)]
e^V\}
\eea
(Note that $\nabla^2 e^V=0$.) In the ordinary case one can use the
cyclicity of the trace to remove the exponentials after the last equality
and thus establish the
required equivalence. However, in the NAC case we know that the cyclicity of
the trace does not hold since a $ d^2 \theta$ integration is lacking. Thus the
first step above, which introduces the exponentials, is valid; however, after
the last step the exponentials cannot be removed and the usual equivalence
fails. By working in antichiral representation we generally manage to avoid
this problem since ${\nabla}_\a = D_\a$. We note that the same problem does not
occur for $\overline{\Del}_{\adot}$ since the surviving $d^2\th$ integration makes the
trace cyclic as in the ordinary case.

Therefore, from now on we choose to describe the gauge sector of the theory in
antichiral representation with the classical action
\beq
S_{inv} =
\frac{1}{2g^2} \int d^4x d^2 \thb ~{\rm Tr} (\overline{W}^{\adot}
\overline{W}_{\adot}) \label{invaction}
\eeq
or more generally with
\bea
&& S_{inv} =
\frac{1}{2g^2} \int d^4x d^2 \thb ~{\rm Tr} (\overline{W}^{\adot}
\overline{W}_{\adot})
\\
&&~~~~+ \frac{1}{2g_0^2}\int d^4x~d^4 \theta~\Big[ ~{\rm Tr}( \overline{\G}^{\adot})
{\rm Tr}(\overline{W}_{\adot})
+ 4 i {\cal F}^{\rho \g} \overline{\theta}^2~{\rm Tr} \left( \pa_{\rho \dot{\rho}}
\overline{\Gamma}^{\adot} \right) {\rm Tr}\left(
\overline{W}_{\adot} \overline{\Gamma}_{\g}^{~\dot{\rho}} \right) \Big]
\nonumber
\label{invaction2}
\eea
if we are interested in assigning different coupling constants to the
$SU({\cal N})$ and $U(1)$ gauge fields.
An equivalently convenient choice
would make use of $\Tilde{\overline{S}}$ in (\ref{Cactions}) and $\Tilde{\overline{S}}_0$
in (\ref{abelian}).


\section{Background field method in $N=1/2$ superspace} \label{BFmethod}

In the case of ordinary (anti)commutative superspace a suitable procedure
for performing perturbative quantum calculations for super Yang--Mills theories
is the background field method \cite{GSZ, superspace}.
This method can be applied to pure
gauge theories or to gauge theories in the presence of matter in arbitrary
representations of the gauge group (although at one loop, for matter in complex
 representations, the ``doubling trick'' \cite{superspace} cannot be used).
It consists of  a nonlinear
quantum--background splitting on the gauge superfields (reflecting the nonlinear nature of
the gauge transformations) which leads to separate
 background  and quantum gauge invariances. Gauge
fixing is then chosen which breaks the quantum  invariance while keeping
manifest invariance with respect to the background gauge transformations.
Therefore, at any given order in the loop expansion the contributions to the
effective action are expressed directly in terms of covariant derivatives and
field strengths ( without  explicit dependence on the prepotential $V$).

We generalize the background field method to the case of NAC super
Yang--Mills theories with chiral matter in a {\em real}
representation of the gauge group. We perform the splitting by
writing the covariant derivatives $\nabla_A$ in terms of { \em background}
covariant derivatives $\boldnabla_A$ and, in contradistinction to the usual Lorentzian case,
the pure imaginary {\em quantum} prepotential $V$. In
the ordinary case this splitting is usually done either in { \em chiral representation}
(the customary case) by writing
$
\nabla_\a =
e^{-V} {\boldnabla}_\a  e^V ~,~  \nabla_{\adot}
= \boldnabla_{\adot}
$
 or in {\em antichiral representation} where
$
\nabla_\a = {\boldnabla}_\a ,~\nabla_{\adot} =
 e^V  {\boldnabla}_{\adot}  e^{-V}$.
At the same time covariantly (anti)chiral superfields
(e.g. in the adjoint representation -- the case we consider here) are expressed in terms
of background covariantly (anti)chiral objects as $\Phi = \bold{\Phi} ~,~ \overline{\Phi} =
e^{-V} \bold{\overline{\Phi}} e^V$ or $\Phi = e^V \bold{\Phi}e^{-V} ~,~  \overline{\Phi} =
 \bold{\overline{\Phi}} $, respectively. The background covariant derivatives
  are assumed to be in {\em vector} representation.

In the NAC case, as discussed in the previous Section, it turns out to be more
convenient to work in quantum antichiral representation. It also turns out to
be possible and more convenient to choose the background derivatives in
antichiral representation. Thus, in the NAC case we use the splitting
\bea
\nabla_\a = {\boldnabla}_\a =D_\a \qquad &,& \qquad \nabla_{\adot} = e_{\ast}^V
\ast \boldnabla_{\adot}\ast e_{\ast}^{-V} = e_{\ast}^V \ast
e_{\ast}^U\ast\Db_{\adot} \ e_{\ast}^{-U} \ast e_{\ast}^{-V}
\nonumber\\
\overline{\Phi} =
 \bold{\overline{\Phi}} \qquad \qquad \qquad &,& \qquad
   \Phi = e^V_{\ast} \ast \bold{\Phi}\ast e^{-V}_\ast
   =e_\ast^V \ast e_\ast^U \ast \phi \ast e_\ast^{-U} \ast e^{-V}_\ast
\eea
The  splitting of the Euclidean prepotential $e^V _\ast
\rightarrow e^V_\ast \ast e^{U}_\ast$ where $U$ is the background prepotential,
is different from the Lorentzian case  where the reality of $V$ forces us
to choose a more complicated one \cite{superspace} and precludes the possibility
 of having all three types of derivatives in (anti)chiral representation.

The derivatives transform covariantly with respect to two types of
gauge transformations: quantum transformations
\bea
 \label{quantum}
&&
e_{\ast}^V \rightarrow e_{\ast}^{i \overline{\L}} \ast e_{\ast}^V \ast
e_{\ast}^{-i \L} \qquad , \qquad \qquad e_\ast ^U \rightarrow  e_\ast ^U
\nonumber \\
&& \nabla_A \rightarrow e_{\ast}^{i\overline{\L}} \ast \nabla_A \ast
e_{\ast}^{-i\overline{\L}}
\qquad , \qquad   {\boldnabla}_A \rightarrow  {\boldnabla}_A
\eea
with background covariantly (anti)chiral parameters, $\boldnabla_\a
\overline{\L} = \boldnabla_{\adot} \L = 0$, and background  transformations
\bea
  \label{background}
&&
e_{\ast}^V \rightarrow e_{\ast}^{i \overline{\l}} \ast e_{\ast}^V \ast
e_{\ast}^{-i \overline{\l}} \qquad , \qquad \qquad
e_{\ast}^U \rightarrow e_{\ast}^{i \overline{\l}} \ast e_{\ast}^U \ast
e_{\ast}^{-i \l}
\nonumber \\
&& \nabla_A \rightarrow e_{\ast}^{i\overline{\l}} \ast \nabla_A \ast
e_{\ast}^{-i\overline{\l}} \qquad , \qquad   {\boldnabla}_A
\rightarrow  e_{\ast}^{i\overline{\l}} \ast {\boldnabla}_A \ast
e_{\ast}^{-i\overline{\l}}
\nonumber \\
\eea
with ordinary chiral parameters $\Db_\ad \l = D_\a \overline{\l}=0$

Under quantum transformations background covariantly (anti)chiral
fields (in the adjoint representation) transform as $ \bold{\Phi}' =
e_\ast^{i \Lambda}\ast \bold{\Phi} \ast e_\ast ^{-i \Lambda}$,
$ \bold{\overline{\Phi}}' = e_\ast^{i \overline{\Lambda}}\ast
\bold{\overline{\Phi}} \ast e_\ast ^{-i \overline{\Lambda}}$; under background
transformations they both transform covariantly with parameter
$\overline{\lambda}$, $ \bold{\Phi}' = e_\ast^{i
  \overline{\lambda}}\ast \bold{\Phi} \ast e_\ast ^{-i
  \overline{\lambda}}$, $ \bold{\overline{\Phi}}' = e_\ast^{i
  \overline{\lambda}}\ast \bold{\overline{\Phi}} \ast e_\ast ^{-i
  \overline{\lambda}}$. Under both quantum and background
transformations the {\em   full} (anti)chiral fields transform
covariantly  with the parameters $\overline{\Lambda}$ and
$\overline{\lambda}$ respectively.

The classical action (\ref{invaction}) for a pure gauge theory, or more
generally the action
\bea
S &=& \frac{1}{2 g^2} \int d^4x ~d^2\overline{\theta} ~{\rm
Tr}(\overline{W}^{\adot} \overline{W}_{\adot}) +\int d^4x ~d^4\th~ {\rm
Tr}(e_\ast^{-V} \ast \overline{\Phi} \ast e^V_{\ast} \ast \Phi)
\nonumber \\
&& - \frac{1}{2} m \int d^4x ~d^2\th ~{\rm Tr}(\Phi^2) - \frac{1}{2}
\overline{m} \int d^4x~ d^2\overline{\th}~ {\rm Tr}(\overline{\Phi}^2)
\label{action1}
 \eea
 for gauge plus covariantly chiral matter is invariant
under the transformations (\ref{quantum}, \ref{background}). Background field
quantization consists in performing
 gauge--fixing which explicitly breaks the (\ref{quantum}) gauge invariance
while preserving manifest invariance of the effective action and correlation
functions under (\ref{background}). Choosing as in the ordinary
case the gauge--fixing functions as $f = \overline{\boldnabla}^2 \ast V$,
$\overline{f} = {\boldnabla}^2 \ast V$ the resulting gauge--fixed action has exactly
the same structure as in the ordinary case \cite{GSZ, superspace} with
products promoted to star products. Precisely,
$S_{tot} = S_{inv} + S_{GF} + S_{gh} $ where
$S_{gh}$ is given in terms of background covariantly (anti)chiral FP and NK ghost
superfields as
\beq
S_{gh} =  \int d^4x d^4 \theta ~ \Big[ \overline{c}' c - c'\overline{c} + .....
+ \overline{b} b \Big]
\label{ghosts}
\eeq

We now concentrate on the pure gauge part of the action and derive the Feynman
rules suitable for one--loop calculations, starting from
 \beq
  S_{inv} + S_{GF} =
-\frac{1}{2g^2} \int d^4x d^4 \theta~ {\rm Tr}[ (e_\ast^{V} \ast
{\overline{\boldnabla}}^\ad \ast e_\ast^{-V})\ast {D}^2 (e_\ast^{V} \ast
\overline{\boldnabla}_\ad \ast e_\ast^{-V}) ~+~\frac{1}{\a} V \ast
(\overline{\boldnabla}^2 {D}^2 + {D}^2 \overline{\boldnabla}^2) \ast V ]
 \eeq
Having chosen antichiral representation, we were able to replace one factor of
$\boldnabla^2 \equiv D^2$  by $\int d^2\theta$. As explained in the previous
Section, in the chiral representation the corresponding replacement of
$\boldnabla$ would have been fraught with difficulties in the NAC case.

We extract the quadratic part in $V$ and read the
$V$-$V$ propagator and the $V$-$V$-background vertices which are the only
vertices entering one--loop diagrams. We  proceed as in the ordinary
case, replacing products with star products.
The propagator is the ordinary one since the star product does not
affect quadratic actions. Working in Feynman gauge ($\a=1$) it is
\beq
\langle V^A(z) V^B(z') \rangle = \frac{\d^{AB}}{\Box_0} ~\d^{(8)}(z-z')
\label{VVprop}
\eeq
where $\Box_0 \equiv \frac12 \pa^{\a \adot}\pa_{\a \adot}$.
For the interaction terms, after a bit of algebra we find
\bea
&& -\frac{1}{2g^2} \int d^4x d^4 \theta~ {\rm Tr} \Bigg\{ V \ast \Big[
  -i [ \bold{\G}^a, \pa_a V]_\ast - i \{ \Tilde{\bold{W}}^{\a} , D_{\a} V \}_\ast
- i \{ \overline{\bold{W}}^{\adot} , \overline{D}_{\adot}V \}_\ast
\nonumber \\
&&~~~~~~~~~ - \frac{i}{2} [ \pa_a \bold{\G}^a, V ]_{\ast} - \frac12 [ \bold{\G}^a , [ \bold{\G}_a , V]]_\ast
- \{ \overline{\bold{W}}^{\adot} , [ \overline{\bold{\G}}_{\adot} , V]_{\ast} \}_\ast \Big]
\Bigg\}
\label{VVvertex}
\eea
where $\bold{\G} , \bold{W}$ are background quantities.

We now turn to the action for  full (not background) covariantly chiral matter
and review first the usual, anticommutative case, with
\bea
S =&&\int d^4x d^4 \theta ~ \overline{\Phi} \Phi =
\int d^4x d^4 \theta ~ \overline{\bold{\Phi}} e^V \bold{\Phi} e^{-V}
 \nonumber\\
&& =\int d^4x d^4 \theta ~ \overline{\bold{\Phi}} \bold{\Phi}
+\overline{\bold{\Phi}} [V,\bold{\Phi}] +\cdots
\label{actionchiral}
\eea
Here
$\Phi$ and $\overline{\Phi}$ are
related by complex conjugation. The first term in the expansion is the
kinetic term for background covariantly (anti)chiral fields.
In particular, ghosts fall in this category so the
following procedure can be applied to the action (\ref{ghosts}). The
remaining terms give rise to ordinary interactions with the quantum
field $V$ and can be treated in standard perturbative fashion. Here,
for one-loop calculations, we concentrate on the kinetic term.
The corresponding equations of motion
\beq
{\cal O} \left( \begin{matrix}{\bold{ \Phi }\cr \bold{\overline{\Phi}}
} \end{matrix} \right) = 0
\quad \qquad
{\cal O} = \left( \begin{matrix} {0 & \overline{\boldnabla}^2 \cr
                                 D^2 & 0}
                  \end{matrix} \right)
\eeq
can be formally derived from the functional determinant
\beq
\Delta ~=~  \int {\cal D}\Psi
e^{ \overline{\Psi} {\cal O} \Psi }\sim  ({\rm det}{\cal O})^{-\frac{1}{2}}
\eeq
where $\Psi$ is the column vector $\left( \begin{matrix}{\bold{\Phi}
    \cr \bold{\overline{\Phi}}} \end{matrix}\right)$.
If we perform the change of variables $\Psi = \sqrt{{\cal O}} \Psi'$,
with jacobian  ${\rm det}\sqrt{{\cal O}} = \Delta^{-\frac{1}{2}}$,
we can write
\beq
\Delta = \int {\cal D} \Psi'
~\Delta^{-1}~  e^{\overline{\Psi}' {\cal O}^2 \Psi'}
\label{Delta}
\eeq
or equivalently
\beq
\Delta^2 = \int {\cal D} \Psi
e^{\overline{\Psi} {\cal O}^2 \Psi}
\eeq
where
\beq
{\cal O}^2 = \left( \begin{matrix} {\overline{\boldnabla}^2 D^2 & 0 \cr
                                     0 & D^2 \overline{\boldnabla}^2}
                  \end{matrix} \right)
\label{calO}
\eeq
is a diagonal matrix. The corresponding equations of motion
 can be derived from  the (anti)chiral actions
\bea
&& \overline{S}' = \frac12 \int d^4x d^4 \theta ~ \overline{\bold{\Phi}}~
\overline{\boldnabla}^2 ~\overline{\bold{\Phi}} ~=~ \frac12 ~\int d^4x d^2
 \overline{\theta} ~
\overline{\bold{\Phi}} ~\boldbox_- ~\overline{\bold{\Phi}}
\nonumber\\
&& S' = \frac12 \int d^4x d^4 \theta ~ \bold{\Phi} ~D^2 ~\bold{\Phi} ~=~
\frac12  ~\int d^4x d^2 \theta ~ \bold{\Phi} ~\boldbox_+ ~\bold{\Phi}
\label{actions}
\eea
with $\boldbox_\pm$ defined in refs. (\cite{GSZ, superspace}).
Note that in the NAC case the second expression would not hold in chiral
 representation.

In the ordinary case the following chain of identities holds
\cite{superspace}
\beq
\D^2= \int {\cal D}
\bold{\Phi} {\cal D} \overline{\bold{\Phi}}~ e^{S' + \overline{S}'}
~=~ \Big| \int {\cal D} \bold{\Phi} e^{S'} \Big|^2
~=~ \Big( \int {\cal D} \bold{\Phi} e^{S'} \Big)^2
\label{chain}
\eeq
where we have used $\overline{S}' = (S')^\dag$ and the fact that they
both contribute in the same way to $\Delta$ \cite{superspace}.
Therefore, when $\D$ is real,  we can identify the original
$\Delta$ with $\int {\cal D} \bold{\Phi} e^{S'}$ and derive from here
the Feynman rules \cite{superspace}.

We now extend the previous derivation to the case of NAC euclidean superspace
where all the h.c. relations are relaxed and $\bold{\Phi}$,
$\overline{\bold{\Phi}}$ are two independent {\em but real} superfields.
 The matter action is still given by
(\ref{actionchiral}) and we can still define $\Delta$ as in (\ref{Delta}).
Therefore, we write
\beq
\Delta_\ast =
\int {\cal D} \Psi  ~
e^{\Psi^T \ast {\bf O}
\ast \Psi} \sim ({\rm det}({\bf O}))^{-\frac12}
\label{Delta2}
\eeq
 We can then proceed as before and square
the functional integral to obtain
\beq
\Delta_\ast^2 =  \int {\cal D} \Psi^T
e^{\Psi \ast {\bf O}^2 \ast \Psi}
\eeq
where ${\bf O}^2$ is given in (\ref{calO}) with the products promoted to
star products.
Now, if we introduce
\beq
\D_1 = \int {\cal D} \bold{\Phi}   e^{S'}
\qquad , \qquad
\D_2 = \int {\cal D}\bold{\Phib} e^{\overline{S}'}
\eeq
with $S'$, $\overline{S}'$ still given in (\ref{actions}) we can finally write
\beq
\D^2 = \D_1  \ast\D_2
\label{r1}
\eeq
In contradistinction to the ordinary case, now $\overline{S}' \neq (S')^\dag$.
Moreover, the star product, when expanded,
could in principle generate different terms in the two actions. Therefore,
the chain of identities (\ref{chain}) is not immediately generalizable to the
NAC case and we cannot identify  $\D = \D_1$.
However, working in antichiral representation, from the
Feynman rules it follows that  the one-loop contributions to the effective
action are actually equal, as we are now going to show. Therefore, in order
to compute the effective action $\D$ we can indifferently concentrate  on
$\D_1$ or $\D_2$.

Following closely the ordinary case \cite{superspace} we derive the Feynman
rules from $S'$ and $\overline{S}'$ by first extracting the quadratic part of
the actions and then reading the vertices from the rest. Since the identities
involving covariant derivatives are formally the same except that the products
are now star products, the procedure to obtain the analytic
expressions
associated to the vertices is formally the same. We then refer the reader
to Ref. \cite{superspace} for details while reporting here only the final
rules:

\noindent

\begin{itemize}
\item
Propagators
\bea \label{prop}
&& \langle \bold{\Phi} (z) \bold{\Phi}(z') \rangle = -\frac{1}{\Box_0}
\d^{(8)} (z - z')
\nonumber \\
&& \langle \overline{\bold{\Phi}} (z) \overline{\bold{\Phi}}(z') \rangle =
-\frac{1}{\Box_0} \d^{(8)} (z - z')
\eea
\item
Chiral vertices: At one loop the prescription for chiral superfields
requires one to associate with one vertex
\beq
\frac12 (\overline{\boldnabla}^2 -\overline{D}^2 )D^2
\label{v1}
\eeq
and with the other vertices
\beq
\frac{1}{2} (\boldbox_+ - \Box_0)
\label{v2}
\eeq
where, with the definition $\overline{\boldnabla}^2 D^2 \overline{\boldnabla}^2 =
 \boldbox_+ \overline{\boldnabla}^2$
\beq
\boldbox_+ = \boldbox_{cov} - i \bold{\widetilde{W}}^\a \ast \boldnabla_\a
-\frac{i}{2}(\boldnabla^\a \ast \bold{\widetilde{W}}_\a ) \qquad , \qquad \boldbox_{cov}=
\frac12 \boldnabla^{\a \ad} \ast \boldnabla_{\a \ad}
\eeq
\item
Antichiral vertices: The prescription is similar to the chiral case except that
we associate $\frac12 D^2 (\overline{\boldnabla}^2 - \overline{D^2})$  at one  vertex and
$\frac12(\boldbox_- - \Box_0)$ at the other vertices with $D^2 \overline{\boldnabla}^2 D^2=
\boldbox_- D^2$ or, explicitly,
\beq
 \boldbox_-
 = \boldbox_{cov} - i \overline{\bold{W}}^\ad \ast \overline{\boldnabla}_\ad
-\frac{i}{2}(\overline{\boldnabla}^\ad \ast \overline{\bold{W}}_\ad )
\eeq
These vertices are then
 expanded in powers of the background fields.
\end{itemize}

Now, consider the one-loop chiral contribution to the amplitude, obtained from
(\ref{v1}, \ref{v2}). Omitting irrelevant factors it can be written (recalling the
definition of $\boldbox_+$) as
\beq
{\rm Tr}\left\{
\cdots \cdots (\overline{\boldnabla}^2 - \overline{D}^2)D^2 \frac{1}{\Box_0}
(\overline{\boldnabla}^2 - \overline{D}^2)D^2 \frac{1}{\Box_0}
(\overline{\boldnabla}^2 - \overline{D}^2)D^2 \frac{1}{\Box_0} \cdots \cdots \right\}
\eeq
where the trace includes integrations and the propagators include the $\d$-functions.
As in the usual supergraph rules the $D^2$ factors can be moved to the right across
the propagators. Subsequently, at all but one of the vertices we can rewrite
factors $D^2(\overline{\boldnabla}-\overline{D}^2)$ as $(\boldbox_--\Box_0)$
and we immediately obtain the expression for the antichiral one-loop. Thus, we have
shown that
$\D=\D_1=\D_2$ and it is sufficient to calculate one of the contributions. Because
it is simpler in antichiral representation, we shall compute the contribution
from the antichiral loop.

The procedure can be easily extended to the case of massive chirals by
simply promoting the propagators (\ref{prop}) to massive propagators
$-1/(\Box_0 - m^2)$. We also note that these rules are strictly one-loop rules.
At higher orders there are no difficulties and ordinary rules apply, as described in
\cite{superspace} with obvious modifications required by noncommutativity.

We can write down a formal effective interaction
lagrangian that corresponds to the one-loop rules above. In the  case of massive
matter (chirals with mass $m$ and antichirals with $\overline{m}$) in the adjoint
representation of the gauge group, it is given by (from now on we avoid indicating
star products and drop the boldface notation when no confusion arises)
\beq \label{effective}
\overline{S}_0 + \overline{S}_1 + \overline{S}_2 \equiv \int d^4x d^4 \theta ~ {\rm Tr}
\left\{ \overline{\xi}(\Box_0 - m\overline{m})\xi + \frac{1}{2} \left[ \overline{\xi}
D^2 (\overline{\nabla}^2 - \overline{D}^2)
\xi ~+~ \overline{\xi} (\overline{\Box}_+ - \Box_0 ) \xi \right] \right\}
\label{effective-b}
\eeq
where $\xi ~,~ \overline{\xi}$ are {\em unconstrained} quantum fields and the first
vertex must appear once, and only once, in a one-loop diagram.

In terms of connections and field strengths and after some integration by parts
 we can  rewrite it as
\bea
&& \overline{S}_1 = \int d^4x d^4 \theta ~ {\rm Tr} \left\{ \left( \frac{i}{4}
\overline{\G}^{\adot}
[ \xi,
\overline{D}_{\adot} D^2 \overline{\xi} ]  - \frac{i}{4} \overline{\G}^{\adot}
[ \overline{D}_{\adot} \xi , D^2 \overline{\xi} ] \right) +  \left( - \frac{1}{4}
\overline{\xi} \{ \overline{\G}^{\adot} [ \overline{\G}_{\adot}, D^2
\xi ]\} \right) \right\}
\nonumber \\
&&~~~~\equiv \overline{S}_1 + \overline{S}'_1 \label{Lbar}
\\
&& \overline{S}_2 = \int d^4x d^4 \theta ~ {\rm Tr} \left\{ \left( \frac{i}{4}
\overline{\G}^{\a \adot} [\xi,\pa_{\a \adot} \overline{\xi} ]  - \frac{i}{4}
\overline{\G}^{\a \adot} [ \pa_{\a \adot} \xi , \overline{\xi} ] \right) +
\left( \frac{i}{4}
\overline{W}^{\adot} [\xi, \overline{D}_{\adot} \overline{\xi} ] - \frac{i}{4}
\overline{W}^{\adot} [ \overline{D}_{\adot} \xi , \overline{\xi} ] \right) \right.
\nonumber \\
&& \left.~~~~~~~~~~~~~~~~~~~~~~~~~~+~
\left( \frac14 [\overline{\G}^{\adot} , \xi][\overline{W}_{\adot} , \overline{\xi}]
+ \frac14 [\overline{W}^{\adot} , \xi][\overline{\G}_{\adot} , \overline{\xi}] \right)
+ \left( \frac14 [\overline{\G}^{\a\adot} ,\xi ][\overline{\G}_{\a \adot}
 ,\overline{\xi}] \right)
\right\}
\nonumber \\
&&~~~~\equiv \overline{S}_2 + \overline{S}'_2 + \overline{S}_2'' + \overline{S}_2'''
\label{Rbar}
\eea


\section{Feynman rules in momentum space} \label{Feynmanrules}

In this Section we describe the general procedure we follow to
perform one--loop calculations. As in the ordinary supersymmetric theories,
in general it is much more convenient to perform quantization and
renormalization directly in superspace without going to components
in the WZ gauge. In particular, this becomes unavoidable for NAC theories
where the implications of  nonanticommutativity on gauge invariance
are nontrivial \cite{pr}.

In the NAC case,  as in the ordinary noncommutative case, it is also convenient
to keep the star product implicit by delaying its expansion as much as possible.
This can be accomplished by moving to a {\em momentum superspace} setup.
Given a generic superfunction of superspace coordinates we Fourier transform
both the bosonic and the fermionic coordinates according to the prescription
\beq
\widetilde{\Phi}(p, \pi, \overline{\pi}) = \int d^4x d^2 \theta d^2 \overline{\theta}~
e^{ipx + i\pi \theta + i\overline{\pi} \overline{\theta}}~ \Phi(x, \theta, \overline{\theta})
\label{completeFT}
\eeq
In so doing the star products get traded for exponential factors dependent on the
spinorial momentum variables
\beq
\Phi (x,\theta, \overline{\theta}) \ast \Psi(x,\theta, \overline{\theta})
\longrightarrow e^{\pi \wedge \pi'} ~\widetilde{\Phi}(p,\pi,\overline{\pi})
\widetilde{\Psi}(p',\pi',\overline{\pi}')
\eeq
where we have defined $\pi \wedge \pi' \equiv \pi_\a {\cal F}^{\a \b} \pi'_\b$.

We then develop perturbative techniques in momentum superspace. As for the ordinary
anticommutative case this amounts to translating Feynman rules for propagators
and vertices to momentum language. In particular, spinorial $D$ derivatives
become $\widetilde{D}$ derivatives according to the relations (\ref{Dtilde}).

In the NAC case relevant changes occur due to the appearance of exponential
factors at the vertices. In fact, given a local $n$--point vertex of the form
$\int A_1 \ast \cdots  \ast A_n$ the corresponding expression in momentum superspace
becomes
\beq
\prod_{i <j }^n ~e^{\p_i \wedge \pi_j} ~\widetilde{A}_1(\Pi_1) \cdots
\widetilde{A}_n(\Pi_n)~\d^{(8)}(\sum_i\Pi_i)
\eeq
where $\Pi_{k} \equiv (p_{k}, \pi_{k},
\overline{\pi}_{k})$  denotes collectively the bosonic and fermionic momenta.
When contracting quantum lines coming out from the vertices inside Feynman
diagrams different ways of performing contractions lead to different
configurations of exponential factors. Due to spinorial momentum conservation
at each vertex the diagrams can be classified into {\em planar} diagrams
characterized by exponentials depending only on the external momenta and
{\em nonplanar} ones which have a  nontrivial exponential dependence on the
loop momenta. This pattern resembles closely what happens in the case
of bosonic noncommutative theories \cite{filk, minwalla}. Therefore,
we use the same prescriptions \cite{filk, minwalla}
to determine the overall exponential factor associated with a given diagram.

The ordinary $D$--algebra which allows  reducing a supergraph to an ordinary
momentum diagram gets translated into a $\widetilde{D}$--algebra in a
straightforward way. In particular, while in configuration superspace the
general rule to get a nonzero contribution from a given supergraph is to
perform $D$--algebra until we are left with a factor $D^2\overline{D}^2$ for
each loop, in momentum superspace it gets translated into the requirement of
performing $\widetilde{D}$--algebra until one ends up with a factor
$\pi^2 \overline{\pi}^2$ for each loop, where $(\pi, \overline{\pi})$ are the loop
spinorial momenta.
Therefore, once the exponential factor and the structure of the $\widetilde{D}$ derivatives
associated to a given diagram have been established we proceed by performing
$\widetilde{D}$--algebra. This amounts to reducing the number of spinorial derivatives
by using identities (\ref{D}), expanding the exponential factors as
\beq
 e^{\pi_1 \wedge \pi_2} = 1 + \pi_1^\a {\cal F}_{\a\b} \pi_2^\b -
\frac12 \pi_1^2 {\cal F}^2 \pi_2^2
\label{exp}
\eeq
and selecting those configurations of spinorial momenta which have a factor
$\pi^2 \overline{\pi}^2$ for each loop. It is important to note that while
$\overline{\pi}^2$ factors
only come from $\widetilde{\bar D}$ derivatives associated to the vertices as in
the ordinary case, $\pi^2$ factors can also come from the expansion (\ref{exp}),
giving extra nonvanishing contributions to a given diagram proportional to
the nonanticommutation parameter ${\cal F}$. This is the way
nonanticommutativity enters the calculations in our approach.

Finally, once $\widetilde{D}$--algebra has been performed, we are left with
ordinary momentum loop integrals. We evaluate them in dimensional
regularization ($n = 4 - 2\e$) and in the G--scheme in order
to avoid dealing with irrelevant constants coming from the expansion of gamma
functions.

We now list the propagators and the vertices in momentum superspace as used
in our calculations.
In order to simplify the notation, we omit the tildes
over the Fourier transformed superfields and for a generic superfield $\Psi$ we write
$\Psi(k) \equiv \Psi\left( \Pi_{k} \right)$.
When integration over superspace momenta is present,
$\int$ stands for
\beq
\int \prod_{k=1}^r ~d^8 \Pi_k ~\d^{(8)}\left( \sum_{k=1}^r
\Pi_r \right) \quad ,
\eeq
where $ r$ is the number of superfields appearing in the integral.
Finally, we find it convenient to introduce the structures
\beq
{\cal P}^{ABC}(\pi_1, \pi_2) \equiv i~ f^{ABC}
\cosh{(\pi_1 \wedge \pi_2)} + d^{ABC} \sinh{(\pi_1 \wedge \pi_2)}
\label{Pcal}
\eeq
\bea
{\cal Q}^{ABCD} ( \pi_1, \pi_2, \pi_3, \pi_4) &\equiv& ~{\rm
Tr}([T^A,T^B][T^C,T^D])~\cosh{(\pi_1 \wedge \pi_2)}~ \cosh{(\pi_3
\wedge \pi_4)}
\nonumber \\
&+& {\rm Tr}(\{T^A,T^B\}[T^C,T^D])~
\sinh{(\pi_1 \wedge \pi_2)}~\cosh{(\pi_3 \wedge \pi_4)}
\nonumber \\
&+& {\rm Tr}([T^A,T^B]\{T^C,T^D\})~ \cosh{(\pi_1 \wedge \pi_2)}~
\sinh{(\pi_3 \wedge \pi_4)}
\nonumber \\
&+& {\rm
Tr}(\{T^A,T^B\}\{T^C,T^D\})~ \sinh{(\pi_1 \wedge \pi_2)}~ \sinh{(\pi_3
\wedge \pi_4)}
\nonumber \\
&&~~~~
\label{Qcal}
\eea
arising from the star products at the vertices.

\begin{itemize}

\item
\underline{Quantum gauge superfields}

Following the gauge--fixing procedure and the results discussed in
Section \ref{BFmethod} we can derive the Feynman rules for the pure
gauge action (\ref{invaction}). In Feynman gauge the $V$ propagator reads
\beq \langle V^A(1)
~V^B(2) \rangle = -g^2 \d^{AB} ~\frac{1}{p_1^2} ~\d^{(8)}(\Pi_1 + \Pi_2)
\eeq
whereas the vertices useful for one-loop calculations are
\bea
&& S_{inv} + S_{GF} \rightarrow ~~~\frac{1}{2~g^2} \int {\cal P}^{ABC}
(\pi_2, \pi_3) ~\Big\{~V^A(1)~\overline{\Gamma}^{\a \adot B}(2)~ p_{3 \a \adot} V^C(3)
\nonumber \\
&&~~~~+\frac{1}{2}~V^A(1)~p_{2 \a \adot}~\overline{\Gamma}^{\a \adot
B}(2)~V^C(3) +i~V^A(1)~\overline{W}^{\adot
B}(2)~\left(\widetilde{\overline{D}}_{\adot}~V^C(3) \right)
\nonumber \\
&&~~~~+i~V^A(1)~\widetilde{W}^{\a B}(2)~\left(\widetilde{D}_{\a} V^C(3) \right)
\Big\}
\nonumber \\
&& + \frac{1}{2~g^2} \int {\cal Q}^{ABCD}(\pi_1, \pi_2, \pi_3, \pi_4)
~\Big\{ \frac{1}{2}~V^A(1) ~\overline{\Gamma}^{\a \adot
B}(2) ~\overline{\Gamma}_{\a \adot}^C (3)~V^D(4)
\nonumber \\
&&~~~~~~~~~~~~~\qquad \qquad+ ~V^A(1)
~\overline{W}^{\adot B}(2) ~\overline{\Gamma}_{\adot }^C (3)~V^D(4) \Big\}
\eea
where the structures ${\cal P}$ and ${\cal Q}$ have been defined in eqs.
 (\ref{Pcal}) and (\ref{Qcal}).

\item
\underline{Quantum (anti)chiral superfields}

We now consider the matter/ghost actions and concentrate only on the parts
which contribute to the pure gauge one--loop effective action.
As discussed in the previous Section its structure can be read from the
``effective'' action (\ref{effective}) and it is the same for both chiral and
antichiral sectors.

The propagator for the unconstrained superfields $\xi$, $\overline{\xi}$ is
(for ghosts we just set $m=\overline{m}=0$)
\beq \langle \overline{\xi}^A(1) \xi^B(2)
\rangle = \d^{AB} ~\frac{1}{p_1^2 + m \overline{m}} ~ \d^{(8)} \left( \Pi_1
+ \Pi_2 \right)
\eeq
whereas performing FT of the interaction terms we are interested in,
we find cubic vertices
\bea
&& \overline{S}_1 = \frac{i}{4} \int {\cal P}^{ABC}(\pi_1, \pi_2)
~\overline{\G}^{\adot A}(1) ~\big[~ \xi^{B}(2)
~\widetilde{\overline{D}}_{\adot}\widetilde{D}^2
\overline{\xi}^{C}(3) - \widetilde{\overline{D}}_{\adot}\xi^{B}(2)
~\widetilde{D}^2 \overline{\xi}^C(3) ~\big]
\nonumber \\
&&~~~~
\label{barS1} \\
&& \nonumber \\
&&\overline{S}_2 = \frac{1}{4} \int {\cal P}^{ABC}(\pi_1, \pi_2) ~\big[ - p_{2 \a
\adot} + p_{3 \a \adot}\big]~\overline{\G}^{\a \adot A}(1) ~
\xi^{B}(2) ~\overline{\xi}^{C}(3)
\label{barS2}\\
&& \nonumber \\
&& \overline{S}'_2 = \frac{i}{4} \int {\cal P}^{ABC}(\pi_1, \pi_2)
~\overline{W}^{\adot A}(1) ~\big[~ \xi^{B}(2)
~\widetilde{\overline{D}}_{\adot} \overline{\xi}^{C}(3) -
\widetilde{\overline{D}}_{\adot}\xi^{B}(2) ~ \overline{\xi}^C(3) ~\big]
\label{barS3}
\eea
and quartic vertices
\bea
&& \overline{S}'_1
= - \frac{1}{4}\int {\cal Q}^{ABCD} (\pi_1, \pi_2, \pi_3, \pi_4)~
\overline{\xi}^A(1)~\overline{\G}^{\adot B }(2)~\overline{\G}_{\adot}^C(3)~
\widetilde{D}^2 \xi^D(4)
\label{barS1'}
\\
&& \overline{S}''_2 = - \frac{1}{4}\int {\cal Q}^{ABCD} ( \pi_1,\pi_2, \pi_3, \pi_4)
~\times
\nonumber \\
&&~~~~~~~~~~~~~~~~~~~~~~~~\overline{\xi}^A(1)
 \left(\overline{\Gamma}^{\adot B }(2)~\overline{W}_{\adot}^C(3) +
\overline{W}^{\adot B }(2)~\overline{\G}_{\adot}^C(3)~\right)~\xi^D(4)
\nonumber \\
&&~~~~~
\label{barS2''}
\\
&& \overline{S}'''_2 =  -\frac{1}{4}\int {\cal Q}^{ABCD} ( \pi_1, \pi_2, \pi_3, \pi_4)
~\overline{\xi}^A(1)~
\overline{\G}^{\a \adot B }(2)~\overline{\G}_{\a \adot}^C(3)~\xi^D(4)
\label{barS2'''}
\eea
where again ${\cal P}$ and ${\cal Q}$ are given in (\ref{Pcal}) and (\ref{Qcal}).
\end{itemize}


\section{General structure of divergent terms} \label{structure}

We begin by considering a SYM theory with or without matter described by the
actions (\ref{action1}) or (\ref{invaction}) respectively, and concentrate on the
 perturbative gauge effective action and its divergence structure.
It is convenient to first investigate the
general background dependence of possible divergent contributions.
This can be accomplished by following the
procedure of \cite{BF}, \cite{R}, \cite{LR} based on dimensional analysis
and invariance under global (pseudo)symmetries.
This procedure has been used to constrain the structure
of the NAC Wess-Zumino effective action, first in components
\cite{BF} and then directly in superspace \cite{R} by imposing the invariance
under two $U(1)$ global (pseudo)symmetries.
A similar analysis has been used in \cite{LR} for $N=\frac{1}{2}$ SYM
theory in components, in the WZ gauge, by requiring invariance under
a global $R$--symmetry.
Here we apply the same analysis directly in superspace without fixing any
supergauge. We assign suitable $R$-charges to the various quantities
 in the classical action and determine the most general dependence of a divergent
term on the background superfields by requiring it to have null
R-charge and a non-negative power dependence
on the momentum UV cut--off $\Lambda$ \footnote{For this purpose we find
more convenient to work with a cut--off regularization rather than dimensional
regularization. In any case the selection of possible divergent structures
is independent of the regularization used.}.

The crucial observation is the following: since we work in antichiral
representation any superfield appearing in the final structure of a
divergent term can be written in terms of $\overline{\Gamma}^{\adot}$
(see discussion in Section 3). Moreover, since the general counterterm
has to be $N=1/2$ supersymmetric its structure in superspace might
contain explicit dependence on powers of $\overline{\theta}$.
For simplicity, we forget about spinorial and color indices.
With the following assignment of $R$--charges
\begin{center}
\setlength{\extrarowheight}{6pt}
\begin{tabular}{|c|c|c|}
\hline
  &  dim & R-charge \\
\hline
$\overline{\Gamma}^{\adot}$ & $1/2$    &  $-1$ \\
\hline
$D_{\a}$ &   $1/2$   & $1$ \\
\hline
$\overline{D}_{\adot}$ &   $1/2$   &  $-1$  \\
\hline
$\overline{\theta}$ & $- 1/2$ & $1$ \\
\hline
$\pa_{\a \adot}$ &  $1$   &  $0$  \\
\hline
${\cal F}^{\rho \gamma}$ &   $-1$  &  $-2$ \\
\hline
$\Lambda$ &  $1$   &  $0$ \\
\hline
\end{tabular}
\end{center}
the most general local term will have the form
\beq
\int d^4 x~ d^4 \theta~ \overline{\theta}^{\,\overline{\tau}}~ 
\Lambda^{\beta}~{\cal F}^{\alpha}~D^{\gamma}~
\overline{D}^{\overline{\gamma}}~
\pa^{\delta}~\overline{\Gamma}^{\overline{\sigma}}
\label{C}
\eeq
where $\overline{\tau}, \beta, \alpha, \gamma, \overline{\gamma}, \delta,
\overline{\sigma}$ are all non-negative integers satisfying
\beq \label{dim}
\beta = 2 + \alpha
-\frac{1}{2} (\gamma + \overline{\gamma}) -\delta -\frac{1}{2}\overline{\sigma}
 + \frac{1}{2} \overline{\tau} \eeq
from the condition that the integrand  have mass dimension four and
\beq
\overline{\sigma} = \gamma - \overline{\gamma} -2 \a + \overline{\tau}
\label{R}
\eeq
from requiring the vanishing of the total R-charge. In (\ref{C}) $\pa^\delta$ stands
for spacetime derivatives and gives the maximal power of momenta present in the
counterterm.

Since we are looking for divergent terms we impose $\beta \ge 0$.
Replacing $\a$ in (\ref{dim}) with its expression as obtained from eq. (\ref{R})
we have
\beq
\overline{\sigma} \le 2 + \overline{\tau} - \overline{\gamma} - \delta
\label{sigmabar}
\eeq
Note that $\overline{\sigma}$ counts the number of background superfields in
 (\ref{C}). It then gives the maximal number of gauge vertices present in
the corresponding diagram. Therefore, from the condition
(\ref{sigmabar}) and from the observation that $\overline{\tau} \le 2$ it
follows that $n$-point functions with $n \ge 5$ always give convergent contributions
and we can concentrate on two, three and four point functions when computing the
divergent part of the effective action.

Pushing the analysis a bit further we find an extra constraint on $\overline{\sigma}$.
Since $D^3=0$ and, having already exhibited the
whole $\pa$--dependence, $\{D, \overline{D} \} \sim 0$,
 the maximal number of superspace covariant derivatives
in (\ref{C}) must satisfy
\beq
\gamma \le 2 \overline{\sigma} - 2
\eeq
It follows that
$\gamma - \overline{\gamma} + \overline{\tau} \le 2 \overline{\sigma}$ and from
(\ref{R}) we have
\beq
\a \le \frac{1}{2} \overline{\sigma} \le 2
\eeq
This allows  a complete classification of all possible divergent contributions.
For $\a = 0$, only 2,3,4--point diagrams would be divergent. However, since
they correspond to the divergences of the ordinary ${\cal N} = 1$
SYM theory \cite{superspace, Improved}, 3,4--point divergent diagrams are ruled
out by the restored ${\cal N}=1$ supersymmetry ($\overline{\tau} = 0$) and we
are left only with nonvanishing 2--point divergent contributions. In this case we know that
all the terms are logarithmically divergent since supersymmetry prevents higher
divergences to be generated.

In the NAC case new divergent terms proportional to the NAC parameter appear for
\bea
&& \bullet \qquad \a = 1 \qquad
\rightarrow \qquad \overline{\sigma} = 2,3,4 \nonumber \\ && \bullet
\qquad \a = 2 \qquad \rightarrow \qquad \overline{\sigma} = 4
\eea
We now list all possible divergences proportional to ${\cal F}$ ($\a =1$) and
${\cal F}^2$ ($\a =2$).

\vskip 20pt
\begin{itemize}
\item[$\bullet$] {$\a =1 ~~\overline{\sigma} = 2 $

In this case we have
\beq
\gamma - \overline{\gamma} + \overline{\tau} = 4, \qquad \gamma \le 2\qquad
\rightarrow \qquad \gamma =2 , ~~\overline{\gamma} = 0, ~~\overline{\tau}=2 ~~~~~~ \delta
\le 2 \nonumber
\eeq
Therefore possible divergent terms are proportional to
\bea
&&{\cal F^{\a}_{~\a}} ~\int d^4 x~d^4 \theta~ \overline{\theta}^2~\overline{\Gamma}^{\adot}~D^2
\overline{\Gamma}_{\adot}  \\
&&{\cal F^{\a \b}} ~\int d^4 x~d^4 \theta~\overline{\theta}^2~(\pa_{\a}^{~\adot}~
\overline{\Gamma}_{\adot})~(\pa_{\b}^{~\bdot}~D^2
\overline{\Gamma}_{\bdot})
\eea
However, both these expressions vanish for symmetry reasons independently of the
color structure,
as a consequence of the fact that ${\cal F}$ is symmetric in its spinorial indices
(${\cal F}^{\a}_{~\a} \equiv {\cal F}^{\a\b} C_{\b\a}=0$ and ${\cal F}^{\a\b}$ contracted
with two identical spinors vanishes).}

\vskip 20pt
\item[$\bullet$] {$\a =1 ~~\overline{\sigma} = 3$

It follows that
\beq
\gamma - \overline{\gamma} + \overline{\tau} = 5, \qquad \gamma \le 4 \nonumber
\eeq
Therefore we have three cases:
\bea
&&\gamma =4, ~~\overline{\tau}= 2, ~~\overline{\gamma} = 1,
~~\delta = 0 ~~~\rightarrow~~~ {\cal F}^{\a}_{~\a} ~\int d^4 x~d^4 \theta~ \overline{\theta}^2~
~\overline{D}^{\adot} \overline{\Gamma}_{\adot} ~ D^2
\overline{\Gamma}^{\bdot}~ D^2 \overline{\Gamma}_{\bdot} \nonumber \\
&& \\
&&\gamma =4, ~~\overline{\tau}= 1,~~\overline{\gamma} = 0,
~~\delta = 0 ~~~\rightarrow~~~ {\cal F}^{\a}_{~\a} ~\int d^4 x~d^4 \theta~ \overline{\theta}^{\adot}~
~\overline{\Gamma}_{\adot} ~ D^2 \overline{\Gamma}^{\bdot}~ D^2
\overline{\Gamma}_{\bdot} \nonumber \\
&& \\
&&\gamma =3, ~~\overline{\tau}= 2,~~\overline{\gamma} = 0,
~~\delta = 1 ~~~\rightarrow~~~ {\cal F}^{\a \b}~\int d^4 x~d^4 \theta~ \overline{\theta}^2~
~\pa_{\a \adot}D_{\b} \overline{\Gamma}^{\adot} ~ \overline{\Gamma}^{\bdot}~ D^2
\overline{\Gamma}_{\bdot} \nonumber \\
\label{3pF}
\eea
Again, the first two structures vanish for symmetry reasons, while
the third one
can survive
together with similar expressions which differ in the contractions of
spinorial indices and position of the spacetime derivative.
We note that any nontrivial trace structure is allowed. }
\vskip 20pt
\item[$\bullet$] {$\a =1 ~~\overline{\sigma} = 4$

In this case we have
\beq
\gamma - \overline{\gamma} + \overline{\tau} = 6, ~~~ \gamma \le 6,
~~~ \overline{\tau} - \delta - \overline{\gamma}
\ge 2 ~~\rightarrow ~~ \delta = \overline{\gamma} = 0, ~~ \gamma
=4~~\overline{\tau} = 2
\eeq
and all possible divergent terms are proportional to
\beq \label{4pF}
{\cal F}^{\a \b} ~\int d^4 x~d^4 \theta~ \overline{\theta}^2~\overline{\Gamma}^{\adot}~D_{\a}
\overline{\Gamma}_{\adot} ~D_{\b} \overline{\Gamma}^{\bdot}~D^2
\overline{\Gamma}_{\bdot}
\eeq
and similar expressions obtained by changing the contraction of the indices.
Again, all possible nontrivial trace structures are allowed.}
\vskip 20pt
\item[$\bullet$] {$\a =2 ~~\overline{\sigma} = 4$

We have
\beq
\gamma - \overline{\gamma} + \overline{\tau}= 8, ~~~\gamma \le 6 ~~~ \overline{\tau} - 
\delta - \overline{\gamma}
\ge 2 ~~\rightarrow ~~\delta =
\overline{\gamma} = 0, ~~ \gamma =6 ~~\overline{\tau}= 2
\eeq
Therefore all possible divergent terms are proportional to
\beq \label{4pF^2}
{\cal F}^2~\int d^4 x~d^4 \theta~ \overline{\theta}^2~\overline{\Gamma}^{\adot}~D^2
\overline{\Gamma}_{\adot}~D^2 \overline{\Gamma}^{\bdot}~D^2
\overline{\Gamma}_{\bdot}
\eeq
and similar expressions.}
\end{itemize}

We note that in selecting possible divergent structures we did not
take into account the requirement of supergauge invariance. For
example, structures like the ones appearing in (\ref{3pF}) are not
gauge invariant on their own. As already discussed in \cite{pr} and in
Section 2, these terms are required to appear in suitable linear
combinations in order to guarantee supergauge invariance.  Therefore,
the previous analysis together with the requirement of supergauge
invariance allow us to conclude that in performing loop calculations we
can focus only on the usual divergent diagrams for the undeformed
theory and on diagrams which give rise to background structures of the
form
\begin{eqnarray}
&& {\cal
F}^{\rho \g} \int d^4x~d^4 \theta~ \overline{\theta}^2~{\rm Tr} \left(
\pa_{\rho \dot{\rho}} \overline{\Gamma}^{\adot} \right)  {\rm Tr}\left(
\overline{W}_{\adot} \overline{\Gamma}_{\g}^{~\dot{\rho}} \right)
\label{2+3} \\
&& {\cal F}^2 \int d^4 x ~d^4
\theta~\overline{\theta}^2~ {\rm Tr} \left( \overline{\Gamma}^{\adot}
\overline{W}_{\adot} \overline{W}^{\bdot}
\overline{W}_{\bdot}\right) \label{4a} \\
&& {\cal F}^2
\int d^4x~d^4 \theta~\overline{\theta}^2~{\rm Tr}
\left(\overline{\Gamma}^{\adot}  \overline{W}_{\adot}\right) {\rm
Tr}\left(\overline{W}^{\bdot} \overline{W}_{\bdot}\right) \label{4b}\\
&& {\cal F}^2 \int d^4 x~d^4
\theta~\overline{\theta}^2 ~{\rm Tr} \left( \overline{\Gamma}^{\adot}
\right)  {\rm Tr} \left(
\overline{W}_{\adot} \right) {\rm Tr} \left( \overline{W}^{\bdot}
\overline{W}_{\bdot}\right) \label{4c}
\end{eqnarray}
The 4--point structures (\ref{4a}-\ref{4c}) are supergauge invariant, while
the 3-point function (\ref{2+3}) is not. However, we include it in the list
of ``good'' terms since, as already noted, it is the completion of the
gauge--variant 2-point function $\int d^4x~d^4 \theta~{\rm
Tr}( \overline{\Gamma}^{\adot} )  {\rm Tr} (
\overline{W}_{\adot})$. 

Other 4-point terms proportional to ${\cal F}^2$ but with different color structures 
have not been included since they are identically zero for spinorial reasons. 
Finally, there are not 4-point
functions of order ${\cal F}$ which satisfy the requirement of supergauge invariance.

Finally, a very important conclusion can be drawn from the previous
analysis concerning the nature of the divergences. For values of the
constants $\a, \overline{\tau}, \g, \overline{\g}, \d, \overline{\s}$ corresponding to
divergent nonvanishing structures the power $\b$ of the cutoff
$\Lambda$ as computed from (\ref{dim}) is always zero.  Terms which
would allow for values of $\b$ strictly positive actually vanish for
symmetry reasons. Therefore, we have a direct proof in superspace that in $N=1/2$
SYM theories the divergences continue to be only {\em logarithmic} as
in the ordinary anticommutative case (A similar conclusion
can be reached by working in components \cite{LR,Beren}). This means that, in spite of the
supersymmetry breaking induced by the ${\cal F}^{\a\b}$ tensor the
deformed theories maintain the nice quantum properties of the ordinary
ones. The NAC mechanism of supersymmetry breaking can be considered a
{\em soft} mechanism.


\section{One--loop diagrams in  the background field method} \label{oldresults}

We begin by considering a SYM theory with matter in the adjoint representation of
the $U({\cal N})$ gauge group. As discussed in Sections 2,3 we find it
convenient to work in gauge antichiral representation where the classical action
is (\ref{action1}) \footnote{In Ref. \cite{pr} the divergent part of the one--loop
gauge effective
action was computed by working in a mixed chiral/antichiral setting
for supergauge covariant derivatives. Here we reproduce that calculation
in completely antichiral representation.}.
Following the prescription described in Section 3 we perform quantum--background
splitting, we add gauge--fixing terms and we end up with a quantized action
whose Feynman rules suitable for one--loop calculations are summarized at
the end of Section 3. The diagrams are then computed using momentum superspace
techniques as described in Section 4. Besides the divergent contributions already
present in the ordinary case, i.e. a gauge two--point function with a chiral loop
\cite{GSZ, superspace}, we find new divergent contributions up to 4--point functions
proportional to ${\cal F}$ and ${\cal F}^2$ due to nontrivial modifications to
the $D$--algebra induced by the star product.

Calculations are performed in dimensional regularization ($n = 4
-2\epsilon$) and the integrals useful for our goal are listed in
Appendix A. In particular, all the divergences are expressed in terms
of a self--energy integral ${\cal S}$ defined in (\ref{selfenergy}).

We now list the divergent nontrivial contributions at one loop. Up to
an overall divergent factor ${\cal S}$, they are:
\vskip 5pt
\begin{itemize}
\item
\underline{Ordinary terms: two--point function}

Following standard $D$-algebra arguments, ordinary one--loop divergent
contributions with gauge external fields come only from the
diagram in Fig.1 with a chiral matter/ghost quantum loop.
\vskip 10pt
\begin{minipage}{\textwidth}
\begin{center}
\includegraphics[width=0.25\textwidth]{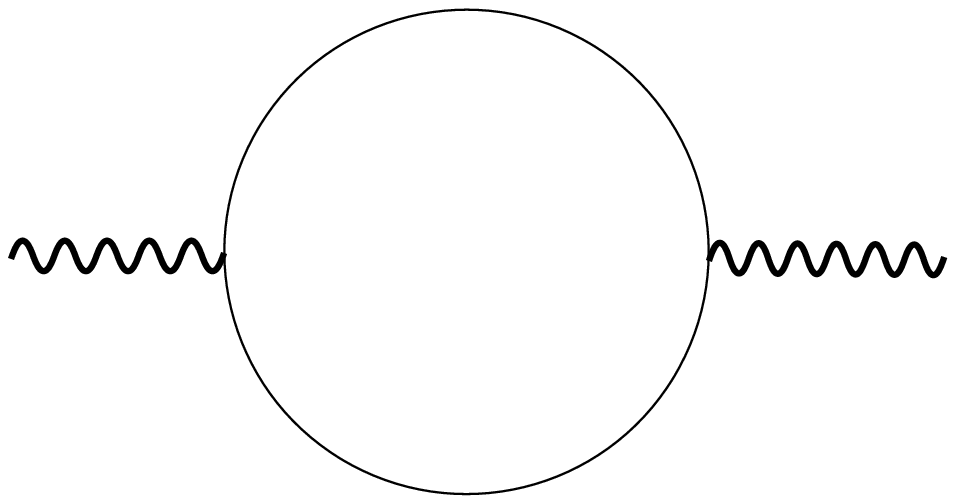}
\end{center}
\begin{center}
{\small{Figure 1: Gauge one--loop two--point functions from chiral loop. }}
\end{center}
\end{minipage}
\vskip 10pt
In antichiral representation it reads
\begin{equation} \label{Gamma2}
\Gamma_{2}^{(1)} = \frac{1}{2}(-3 + N_f) \int d^4 x~d^4 \theta~ \Big[~{\cal
N}~{\rm Tr}\left(\overline{\Gamma}^{\adot}
~\overline{W}_{\adot}\right) - {\rm Tr}\left(
\overline{\Gamma}^{\adot} \right){\rm Tr}\left(
\overline{W}_{\adot}\right)\Big]
\end{equation}
Here $N_f$ counts the number of matter flavors while $(-3)$ is the contribution from
the three ghosts in the action (\ref{ghosts}).

\item
\underline{NAC terms: three-- and four--point functions}

Three-- and four--point diagrams with vector loops are all finite,
whereas divergent contributions arise from the chiral loop diagrams in Figs. 2
and 3 for the three--point and four--point functions, respectively.
\vskip 10pt
\begin{minipage}{\textwidth}
\begin{center}
\includegraphics[width=0.50\textwidth]{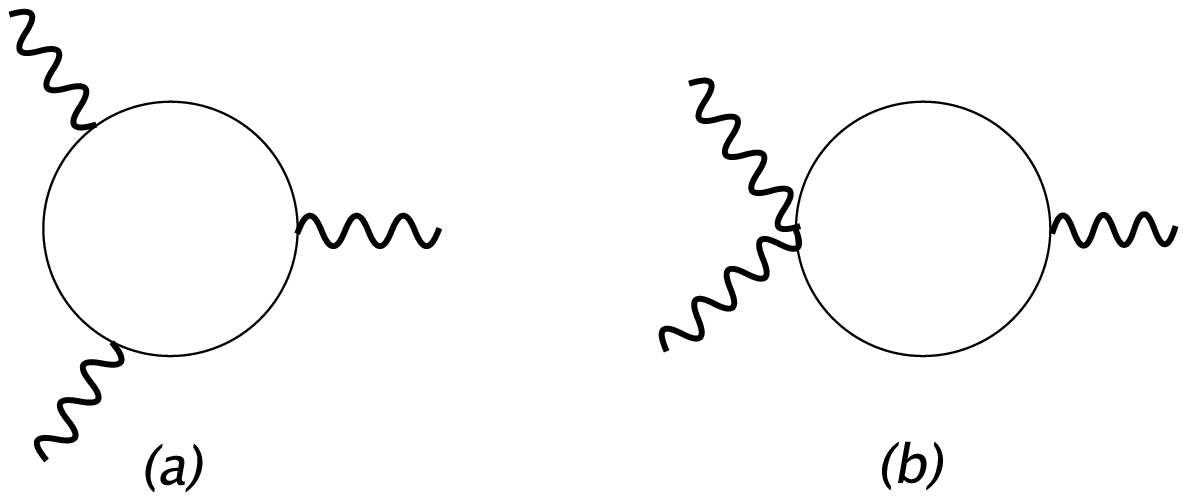}
\end{center}
\begin{center}
{\small{Figure 2: One--loop three--point functions with chiral loop. }}
\end{center}
\end{minipage}
\vskip 10pt
\vskip 5pt
\begin{minipage}{\textwidth}
\begin{center}
\includegraphics[width=0.50\textwidth]{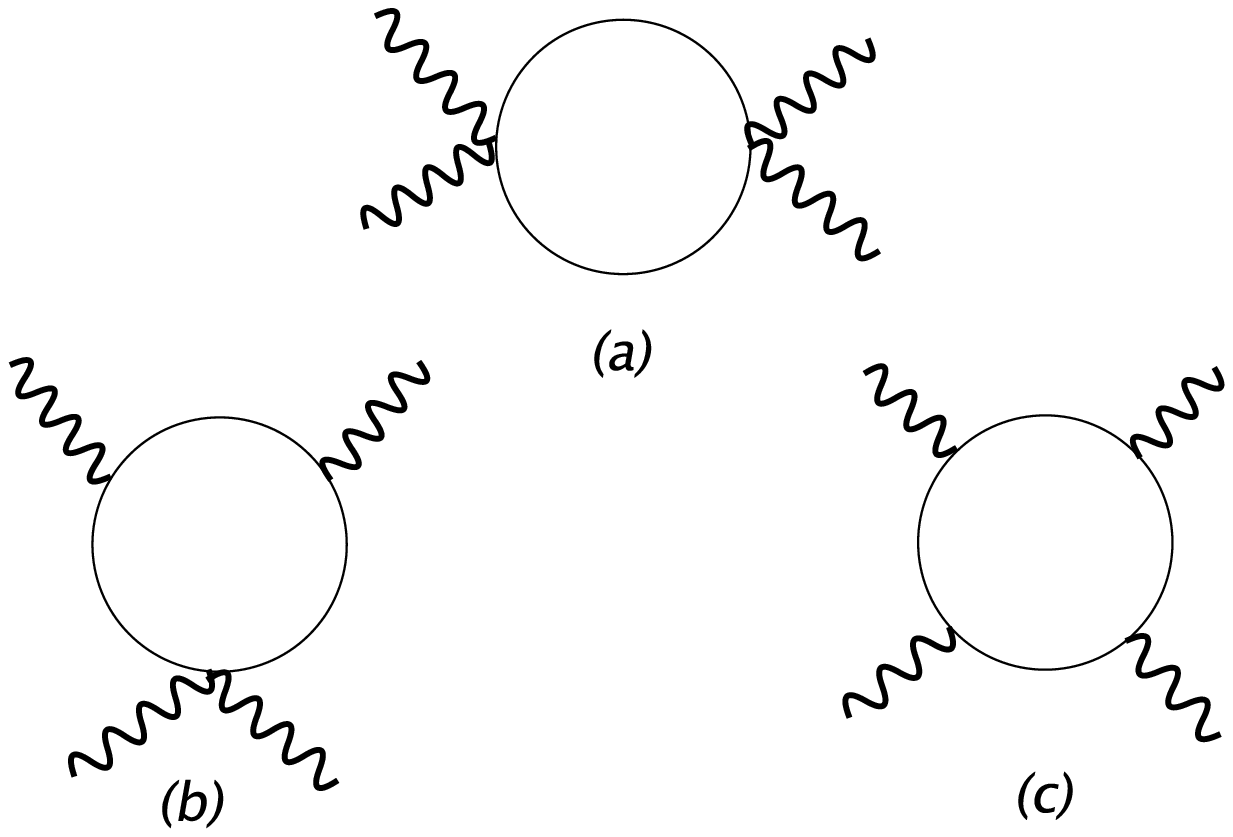}
\end{center}
\begin{center}
{\small{Figure 3: One--loop four--point functions with chiral loop. }}
\end{center}
\end{minipage}
\vskip 10pt
In antichiral representation they read
\begin{eqnarray}
&& \Gamma^{(1)}_3 = - 2 i (-3 + N_f) {\cal F}^{\rho \g} \int d^4x~d^4
\theta~~ \overline{\theta}^2~{\rm Tr} \left( \pa_{\rho \dot{\rho}}
\overline{\Gamma}^{\adot} \right) {\rm Tr}\left( \overline{W}_{\adot}
\overline{\Gamma}_{\g}^{~\dot{\rho}} \right) \label {Gamma3} \\
&& \Gamma^{(1)}_4 = \frac{1}{2}(-3 + N_f) {\cal F}^2 \int d^4x~d^4 \theta~~
\overline{\theta}^2~{\rm Tr}
\left(\overline{\Gamma}^{\adot}\overline{W}_{\adot}\right){\rm
Tr}\left(\overline{W}^{\bdot}\overline{W}_{\bdot}\right) \label{Gamma4}
\end{eqnarray}
We note in both  contributions the explicit dependence on
$\overline{\theta}^2$ which indicates the partial breaking of supersymmetry.
\end{itemize}
\vskip 7pt
To summarize, starting from the classical action
(\ref{action1}) the one-loop divergent contributions to the
gauge effective action have the form
\bea \label{eff-action1}
\G^{(1)}_{gauge} &\rightarrow&  \frac{1}{2}(-3 + N_f)  \times \Big\{
\int d^4x~d^4 \theta~ \Big[ {\cal N}{\rm Tr}\left(
\overline{\Gamma}^{\adot} ~ \overline{W}_{\adot}\right) - {\rm
Tr}\left( \overline{\Gamma}^{\adot}  \right){\rm Tr}\left(
\overline{W}_{\adot}\right) \Big] \nonumber \\
&& \qquad \qquad \qquad \qquad - 4 i {\cal F}^{\rho \g} \int d^4x~d^4 \theta~~
\overline{\theta}^2~{\rm
Tr} \left( \pa_{\rho \dot{\rho}}  \overline{\Gamma}^{\adot} \right)
{\rm Tr}\left( \overline{W}_{\adot}
\overline{\Gamma}_{\g}^{~\dot{\rho}} \right) \nonumber \\
&& \qquad \qquad \qquad \qquad + ~{\cal F}^2 \int d^4x~d^4 \theta~~ \overline{\theta}^2~{\rm
Tr} \left(\overline{\Gamma}^{\adot}\overline{W}_{\adot}\right){\rm
Tr}\left(\overline{W}^{\bdot}\overline{W}_{\bdot}\right) \Big\}
\eea
As proven in \cite{pr} this result is  supergauge invariant; while the
first and the last terms are invariant on their own, the gauge variance of
the third term compensates
that  of the abelian noninvariant ${\rm Tr}( \overline{\Gamma}^{\adot}
){\rm Tr}(\overline{W}_{\adot})$ term as discussed in Section 2.

Given the result (\ref{eff-action1}) for the divergent part of the
effective action, we can immediately conclude that the action
(\ref{action1}) is {\em not} one--loop renormalizable in a superspace setup.
In fact, given the classical gauge action in (\ref{action1}) we cannot
cancel by multiplicative renormalization the new divergent structures
which arise at one loop.

It may be useful to compare our result with
the one obtained in components, in the Wess-Zumino gauge \cite{JJW}. 
This is made possible by the fact that (super)gauge invariance
of $N=\frac{1}{2}$ SYM theories is not spoiled by quantum corrections
\cite{pr}.  Therefore, quantum properties like renormalizability do
not depend on the particular (super)gauge and we can fix a particular
one in order to simplify the study.  In Appendix C we
choose the WZ gauge and reduce the one--loop effective action
(\ref{eff-action1}) to components. This reduction confirms at the
component level the result already evident in superspace, i.e. that
the $N=\frac{1}{2}$ SYM theory defined by the action (\ref{action1})
as obtained by the natural deformation of the ordinary $N=1$ SYM
theory, is not one--loop renormalizable. Our result agrees with what
has been found in \cite{JJW}.


\section{The modified action}

Given that the gauge action (\ref{invaction}) is not renormalizable, the only way out
is to modify the classical action from the very
beginning by adding  new terms which allow for the cancellation of all the
divergent terms at one loop. This approach has been already applied to the theory
in components in \cite{JJW}, where the  one--loop renormalizable deformation
in components has been found.

Our aim is to do the same in the more general superspace setup. We
focus on the pure gauge deformed SYM theory (a brief discussion
in the presence of  matter superfields will be given in section
\ref{matter}) and start with a modified classical action which
contains extra pieces up to quartic order in the superfields  which
are allowed by the symmetries of the theory
\begin{eqnarray} \label{final-action}
&& S_{final} =
~\frac{1}{2~g^2}\int d^4 x~d^4 \theta~ {\rm Tr}\left(
\overline{\Gamma}^{\adot} \ast \overline{W}_{\adot}\right) \nonumber \\
&& \qquad
+ \frac{1}{2~g_0^2~{\cal N}} \int d^4x~d^4 \theta~\Bigg[~{\rm
Tr}\left( \overline{\Gamma}^{\adot} \right) \ast {\rm Tr}\left(
\overline{W}_{\adot}\right) \nonumber \\
&& \qquad \qquad \qquad \qquad\qquad \qquad + 4 i {\cal
F}^{\rho \g} ~ \overline{\theta}^2~{\rm Tr} \left(
\pa_{\rho \dot{\rho}} \overline{\Gamma}^{\adot} \right)\ast  {\rm Tr}\left(
\overline{W}_{\adot} \ast \overline{\Gamma}_{\g}^{~\dot{\rho}} \right)
\Bigg] \nonumber \\
&& \qquad + \frac{1}{2~h^2~{\cal N}} {\cal F}^2
\int d^4x~d^4 \theta~~ \overline{\theta}^2~{\rm Tr}
\left(\overline{\Gamma}^{\adot} \ast \overline{W}_{\adot}\right)\ast {\rm
Tr}\left(\overline{W}^{\bdot}\ast \overline{W}_{\bdot}\right) \nonumber \\
&& \qquad + \frac{1}{l^2} ~{\cal F}^2 \int d^4 x~d^4
\theta~\overline{\theta}^2 {\rm Tr} \left( \overline{\Gamma}^{\adot}\ast
\overline{W}_{\adot}\ast \overline{W}^{\bdot} \ast \overline{W}_{\bdot}\right)
\nonumber \\
&& \qquad + \frac{1}{r^2} ~{\cal F}^2 \int d^4 x~d^4
\theta~\overline{\theta}^2 {\rm Tr} \left( \overline{\Gamma}^{\adot}
\right) \ast {\rm Tr} \left(
\overline{W}_{\adot} \right)\ast {\rm Tr} \left( \overline{W}^{\bdot} \ast
\overline{W}_{\bdot}\right)
\end{eqnarray}
Five coupling constants $g, g_0, h, l, r$ have been introduced.
The coefficients of the various terms have been chosen in such a way that if  $g^2 = - g_0^2 = h^2$
(and setting $1/l^2=1/r^2=0$),
we maintain the original ratio
among the divergent terms coming from chiral/ghost loops (see eq. (\ref{eff-action1})).

The number of independent coupling constants is dictated by supergauge invariance.
In fact, as already noted, while the $g, h, l,$ and $r$ terms are separately supergauge
invariant, the terms proportional to $g_0$ have gauge variations which cancel
only if the ratio between their coefficients is the one given in (\ref{ACgaugeinv}).

We note that the new action (\ref{final-action})
is written with $\ast$-product between the superfields. However we
can replace it with the ordinary product without producing new terms.


\section{The modified action: One--loop diagrams}

We consider the case of a NAC gauge theory with no matter,
now described by the classical action (\ref{final-action})
and address the issue of its quantization.

We perform quantum--background splitting and expand each term
in (\ref{final-action}) to second order in the quantum fields.
In this manner we generate the kinetic terms for the gauge superfields
and vertices of the form quantum--quantum--background,
the only vertices which can contribute at one loop.

The first question concerns the gauge--fixing procedure and, in
particular, the choice of a convenient gauge--fixing function
and the corresponding gaussian weight factor in the functional integral.
In Appendix D we treat this problem in all detail whereas here we
briefly summarize the main points. The introduction of a new
quadratic term proportional to the coupling constant $g_0$
assigns different weights to the quadratic parts of the abelian $U(1)$
and non--abelian vector fields. As discussed in Appendix D,
in the ordinary anticommutative case, given gauge--fixing functions
suitable for the background field method we have different
possibilities for the choice of the gaussian weight factor in the functional
integral (see eqs. (\ref{naif0}) and (\ref{naif})). In particular, choosing the
gaussian factor (\ref{naif}) would lead to the same expression for the propagators
of the abelian and non--abelian fields, so in the ordinary case it might be the most
convenient choice, whereas the choice (\ref{naif0}) would lead to  different
propagators for the $U(1)$ and $SU({\cal N})$ fields. However, in the presence of the
$\ast$-product the first choice becomes inappropriate when working in
background field method since it generates a
quadratic term in the action which is not background gauge invariant.
Therefore, for our purposes
it is preferable to choose a gauge--fixing procedure which preserves background
gauge invariance but leads to different
propagators for the $SU({\cal N})$ and $U(1)$ vector superfields
as given in (\ref{convenient}) and (\ref{U1prop}). In momentum superspace they read
\begin{eqnarray}
&& \langle V^{a}(1)~V^{b}(2) \rangle = - \d^{ab} ~\frac{g^2}{p_1^2}~
\d^{(8)} \left( \Pi_1 + \Pi_2 \right)\nonumber \\
&& \langle V^{0}(1)~V^{0}(2) \rangle = - \frac{g^2}{p_1^2}~\left[ 1 - \left(
\frac{g^2}{g_0^2 + g^2} \right) \frac{
\widetilde{\overline{D}}^{\adot} \widetilde{D}^2
\widetilde{\overline{D}}_{\adot}}{p_1^2} \right] \d^{(8)} \left( \Pi_1
+ \Pi_2 \right)
\end{eqnarray}

Feynman rules for the vertices are still the ones
given at the end of Section 3 supplemented by new vertices coming
from the extra pieces in (\ref{final-action}). These new vertices
are listed in Appendix E. It is easy to realize that at this order
the new vertices will enter only diagrams with
vector fields inside the loop. Therefore the contributions from
the chiral loops, which in this case come only from ghosts, are not
modified and can be read from Section \ref{oldresults}.  Thus,
the ghost contribution to the divergent part of the effective
action is (\ref{eff-action1}) with $N_f =0$.

We then concentrate on the evaluation of the new divergent diagrams
built up with the new vertices in Appendix E for the modified action.
In contradistinction to the vertices coming from (\ref{invaction}) which
never enter one--loop divergent diagrams (see discussion in Section
6 and \cite{pr}), the new vertices can give rise to many new divergent
contributions. Despite the possibility of having more than one hundred
divergent diagrams, we can use the general arguments in Section
5 as a way to eliminate
from the very beginning diagrams which would not lead to the correct
structure. Moreover, superspace techniques allow  a
straightforward
cancellation of most of the remaining ones and the final list of nontrivial diagrams
is quite contained.

In what follows we report the results for the new divergent diagrams.

\begin{itemize}
\item
\underline{Three--point functions} \\
The contributions to the
three--point function are described by the diagrams in Fig.4, but they all {\em cancel
out} in a nontrivial way.
\vskip 10pt
\begin{minipage}{\textwidth}
\begin{center}
\includegraphics[width=0.60\textwidth]{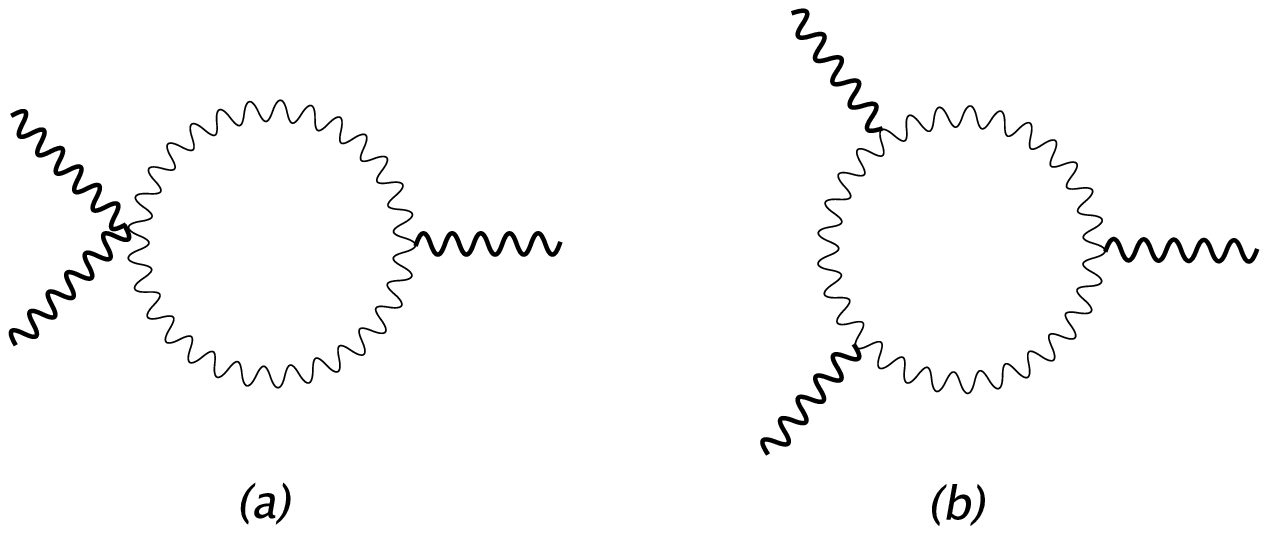}
\end{center}
\begin{center}
{\small{Figure 4: Gauge one--loop three--point functions from vector loop. }}
\end{center}
\end{minipage}
\vskip 10pt
We would have expected this result for the following reason: from gauge invariance
considerations \cite{pr}  any one--loop correction has to  be
gauge invariant. However, there is no way to build a gauge
invariant three-point function proportional to ${\cal F}$ unless it comes
together with a suitable two--point function as given in (\ref{ACgaugeinv}).
On the other hand, we know from Section \ref{structure} that divergent two--point
functions proportional to ${\cal F}$ cannot be produced:
In superspace, only
ordinary $\int d^4 \theta ~{\rm Tr} ( \overline{\Gamma}^{\adot}
\overline{W}_{\adot} )$ and $\int d^4 \theta ~{\rm Tr} (
\overline{\Gamma}^{\adot} ) {\rm Tr} ( \overline{W}_{\adot} )$ can
appear, and these corrections arise only from chiral loop diagrams
(not vectors ones!) for $D$--algebra reasons. Therefore, the absence of
a two--point divergent contribution from vector loops rules out the
possibility of having a nonvanishing three--point correction.

\item
\underline{Four--point functions} \\
The contributions to the four--point functions come from the diagrams in Fig.5
\vskip 10pt
\begin{minipage}{\textwidth}
\begin{center}
\includegraphics[width=0.60\textwidth]{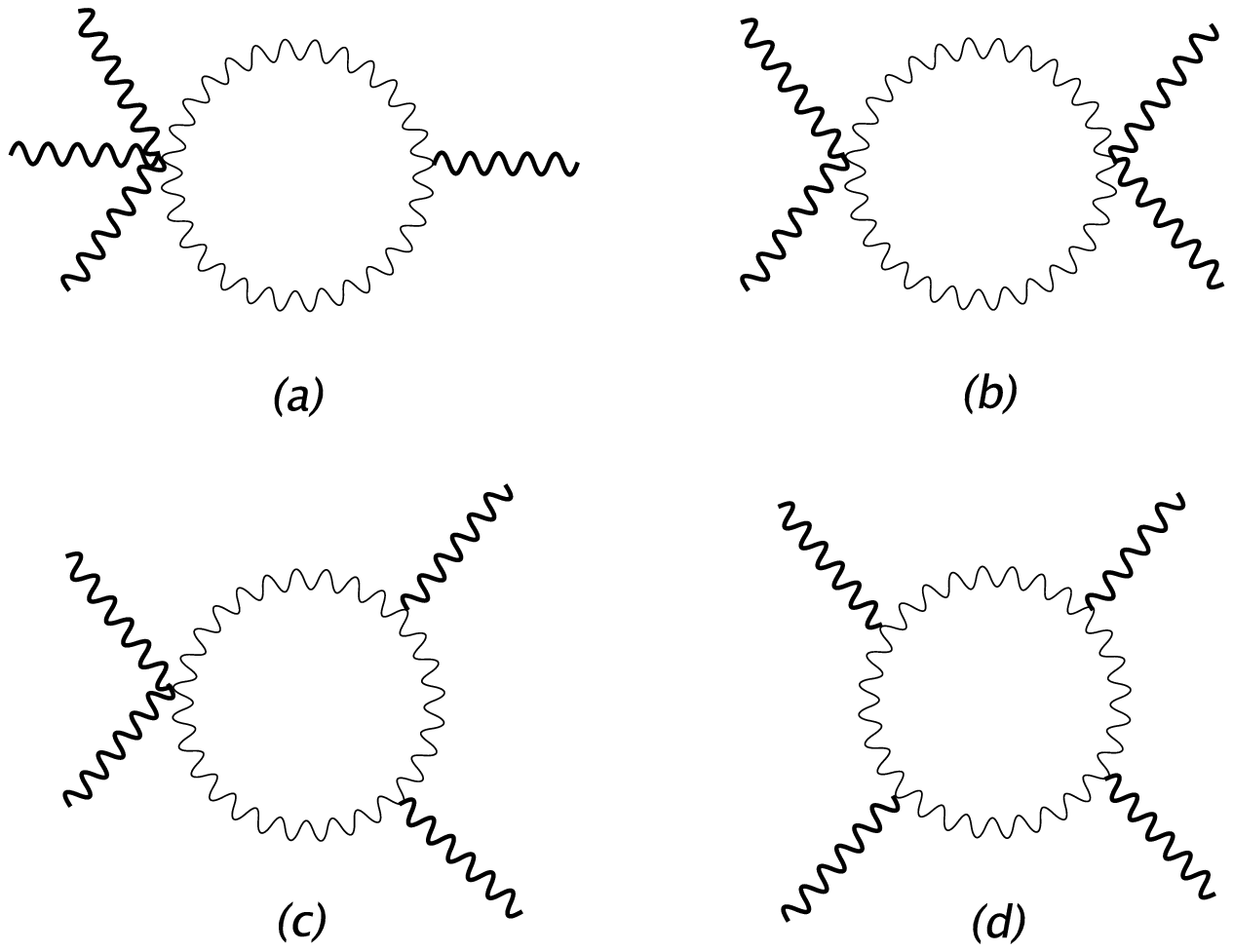}
\end{center}
\begin{center}
{\small{Figure 5: Gauge one--loop four--point functions from vector loop. }}
\end{center}
\end{minipage}
\vskip 10pt
Even if in this case there is no complete cancellation,
most of the possible divergent diagrams give null contributions due to
standard superspace $D$--algebra.
Moreover, it is easy to see that
diagrams with the structure in Fig. 5(b) cancel among themselves.
Therefore, summing all the remaining terms, we have four
kinds of contributions with the same trace structure but depending on four
different combinations of coupling constants
\begin{eqnarray}
&&{\rm Contribution} ~\sim~
\frac{\widetilde{g}^2}{g_0^2 ~{\cal N}} ~: \qquad  - 24
{\cal F}^2 \int d^4 x~d^4 \theta~\overline{\theta}^2  {\rm Tr}
\left( \overline{\Gamma}^{\adot}
~\overline{W}_{\adot}~\overline{W}^{\bdot}~\overline{W}_{\bdot} \right) \nonumber
\\
&&{\rm Contribution}~ \sim ~\frac{g^2}{{h^2~\cal N}} ~: \qquad
- 12 {\cal F}^2 \int d^4 x~d^4
\theta~\overline{\theta}^2 {\rm Tr} \left( \overline{\Gamma}^{\adot}
~\overline{W}_{\adot}~\overline{W}^{\bdot}~\overline{W}_{\bdot} \right) \nonumber \\
&&{\rm Contribution}~
\sim ~\frac{g^2 \widetilde{g}^2}{g_0^4 ~{\cal N}}~ : \qquad
- 12 {\cal F}^2 \int d^4 x ~d^4 \theta~\overline{\theta}^2 {\rm Tr}
\left( \overline{\Gamma}^{\adot}
~\overline{W}_{\adot}~\overline{W}^{\bdot}~\overline{W}_{\bdot} \right)
\nonumber
\\
&&{\rm Contribution} ~\sim~ \frac{g^2 {\cal N}}{l^2} ~: \qquad
\,\,-~6 {\cal F}^2 \int d^4 x ~d^4 \theta
~\overline{\theta}^2~{\rm Tr} \left( \overline{\Gamma}^{\adot}
~\overline{W}_{\adot}~\overline{W}^{\bdot}~\overline{W}_{\bdot}\right)
\nonumber
\end{eqnarray}
where $\widetilde{g}^2 $ is defined in equation (\ref{gtilde}).
\end{itemize}
We note that terms of the form $\int d^4
\theta~\overline{\theta}^2{\rm Tr} ( \overline{\Gamma}^{\adot}
\overline{W}_{\adot}) {\rm Tr} ( \overline{W}^{\bdot}~\overline{W}_{\bdot})$
and \\$\int d^4
\theta~\overline{\theta}^2{\rm Tr} ( \overline{\Gamma}^{\adot})
{\rm Tr} ( \overline{W}^{\bdot} )~{\rm Tr}
( \overline{W}_{\bdot} ~\overline{W}_{\adot})$ cancel along the calculation.

Finally, adding to the classical action  all the divergent contributions
 from ghost and vector loops to the one-loop gauge effective action we obtain
\begin{eqnarray} \label{total}
&& \Gamma^{(1)} = \frac{1}{2~g^2}\int d^4 x~ d^4 \theta~ {\rm
Tr}\left( \overline{\Gamma}^{\adot}\overline{W}_{\adot}\right) \times \Big[
1 -3  \frac{g^2~{\cal N}}{(4 \pi)^2~\epsilon}
\Big]\nonumber \\
&& ~ \qquad + \frac{1}{2~g_0^2~{\cal N}} \int
d^4x~d^4 \theta~\Bigg[{\rm Tr}\left( \overline{\Gamma}^{\adot} \right){\rm
Tr}\left( \overline{W}_{\adot}\right) \nonumber
\\
&& ~\qquad \qquad \qquad
\qquad \quad +4 i {\cal F}^{\rho \g} \overline{\theta}^2~{\rm Tr}
\left( \pa_{\rho \dot{\rho}} \overline{\Gamma}^{\adot} \right) {\rm
Tr}\left( \overline{W}_{\adot} \overline{\Gamma}_{\g}^{~\dot{\rho}}
\right) \Bigg] \times \Big[ 1 + 3 \frac{ g_0^2~{\cal N}}{(4
\pi)^2~\epsilon} \Big] \nonumber \\
&& ~\qquad + \frac{1}{2~h^2~{\cal
N}} {\cal F}^2 \int d^4x~d^4 \theta~~ \overline{\theta}^2~{\rm Tr}
\left(\overline{\Gamma}^{\adot}\overline{W}_{\adot}\right){\rm
Tr}\left(\overline{W}^{\bdot}\overline{W}_{\bdot}\right) \times \Big[
1 -3 \frac{ h^2~{\cal N}}{(4 \pi)^2~\epsilon}
\Big]\nonumber \\
&& ~\qquad + \frac{1}{l^2} ~{\cal F}^2 \int d^4 x ~d^4
\theta~\overline{\theta}^2 ~{\rm Tr} \left( \overline{\Gamma}^{\adot}~
\overline{W}_{\adot}~\overline{W}^{\bdot}~\overline{W}_{\bdot}\right)\times
\nonumber \\
&& \qquad \qquad \qquad \qquad\qquad \qquad \Big[
1 - 6 \frac{g^2}{(4 \pi)^2~\epsilon}\left( {\cal N} + \frac{2
l^2}{h^2~{\cal N}} + \left( 2 + \frac{g^2}{g_0^2}\right) \frac{2
l^2}{{\cal N} \left(g_0^2 + g^2\right)}\right) \Big] \nonumber \\
&& ~\qquad + \frac{1}{r^2} ~{\cal F}^2 \int d^4 x ~d^4
\theta~\overline{\theta}^2 ~{\rm Tr} \left( \overline{\Gamma}^{\adot}
\right) {\rm Tr} \left(
\overline{W}_{\adot} \right) {\rm Tr} \left( \overline{W}^{\bdot}~\overline{W}_{\bdot}\right)
\nonumber \\
\end{eqnarray}

Since the  divergent terms  have the same background structure as the
classical modified action we expect the theory described by
(\ref{final-action}) to be at least one--loop renormalizable. This will be the
subject of the next Section.


\section{Renormalization and $\b$-functions}

We now proceed to the renormalization of the theory.  We define
renormalized coupling constants as \footnote{in this setup there is no
need for wavefunction renormalization of the superfields.}
\begin{eqnarray}
&& g = \mu^{- \epsilon} Z_{g}^{-1} g_{(B)} \qquad \qquad g_0 =
\mu^{- \epsilon} Z_{g_{0}}^{-1} g_{0(B)} \nonumber \\ && h = \mu^{-
\epsilon} Z_{h}^{-1} h_{(B)} \qquad \qquad l = \mu^{- \epsilon}
Z_{l}^{-1} l_{(B)} \qquad \qquad r = \mu^{- \epsilon}
Z_{r}^{-1} r_{(B)}
\end{eqnarray}
where powers of the renormalization mass $\mu$ have been introduced in
order to deal with dimensionless renormalized couplings.
In order to cancel the divergences in (\ref{total}) we set
\begin{eqnarray} \label{zetas}
&& g Z_{g} = \Big[ g - \frac{3 ~{\cal N}}{2~(4
\pi)^2~\epsilon} ~g^3 \Big] \equiv g + \frac{g_{(1)}}{\epsilon} \nonumber
\\
&& g_0 Z_{g_{0}} = \Big[ g_0 +\frac{3~{\cal N}}{2~(4 \pi)^2~\epsilon}~ g_0^3 \Big] \equiv g_0 +
\frac{g_{0(1)}}{\epsilon} \nonumber \\
&& h Z_{h} = \Big[ h - \frac{3~{\cal N}}{2~(4 \pi)^2~\epsilon} ~
h^3\Big] \equiv h
+ \frac{h_{(1)}}{\epsilon} \nonumber \\
&& l Z_{l} = \Big[ l - 3
\frac{l~g^2}{(4 \pi)^2~\epsilon}\left( {\cal N} + \frac{2
l^2}{h^2~{\cal N}} +\left( 2 + \frac{g^2}{g_0^2} \right)~\frac{2
l^2}{ {\cal N} \left( g_0^2 + g^2 \right) }\right) \Big] \equiv l +
\frac{l_{(1)}}{\epsilon} \nonumber \\
&& r Z_{r} = r
\end{eqnarray}
We compute the $\beta$-functions by using the general prescription
\begin{equation}
\b_{\lambda_{j}} = - \epsilon~\lambda_{j} -\lambda_{j(1)} + \sum_{i}
\left( \lambda_{i} \frac{\pa \lambda_{j(1)}}{\lambda_{i}} \right)
\end{equation}
for any coupling $\lambda_{j}$. The $\b$-functions for this theory
turn out to be
\begin{eqnarray} \label{betas}
&& \b_{g} = - \epsilon ~g  - \frac{3~{\cal N}}{(4 \pi)^2}~g^3
\nonumber \\
&& \b_{g_0} = - \epsilon ~ g_0 +  \frac{3~{\cal N}}{(4 \pi)^2}~g_0^3 \nonumber \\
&& \b_{h} = -\epsilon~ h~ - \frac{3~{\cal
N}}{(4 \pi)^2}~h^3 \nonumber \\
&& \b_{l} = -\epsilon~l
- 6 \frac{l~g^2}{(4 \pi)^2}\left( {\cal N} + \frac{2 l^2}{h^2~{\cal
N}} + \left( 2 + \frac{g^2}{g_0^2} \right)~\frac{2 l^2}{{\cal N}
  \left( g_0^2
+ g^2 \right)}\right) \nonumber \\
&& \b_{r} = -\epsilon~ r
\end{eqnarray}
We note that, despite the fact that the modified action
(\ref{final-action}) has an explicit dependence on the NAC parameter,
there is no need to renormalize the constant two--form ${\cal
F}^{\a\b}$. Therefore, the star product does not get deformed by
quantum corrections, consistently with what has been found at the
component level \cite{JJW}.

It is manifest that the
renormalization of $r$ is trivial ($ Z_r =1$) since the term \\ $\int d^4
\theta~\overline{\theta}^2 {\rm Tr} ( \overline{\Gamma}^{\adot})
{\rm Tr} ( \overline{W}_{\adot} ) {\rm Tr} (
\overline{W}^{\bdot}~\overline{W}_{\bdot})$ does not receive any
divergent correction. Therefore, in order to find
the minimal renormalizable modification of the original deformation of
the SYM theories, we can avoid introducing that term.

Moreover,
from (\ref{zetas}) it is easy to see that the renormalization functions
of $g, g_0$ and $h$ are equal up to a sign.  This stems from the
fact that vector loops do not contribute to the terms proportional to
these coupling constants so that the ratio between the divergent
contributions is fixed as in (\ref{eff-action1}). In particular, this
allows us to set $h^2 = - g_0^2$, or $h^2 = g^2$ keeping  the renormalization
procedure consistent at one--loop order. Therefore, the minimal number of
coupling constants we need introduce to make the theory renormalizable at
one--loop is {\em three} and the
two minimal modified deformations of the $N=\frac{1}{2}$ SYM theory are
\bea \label{Smin1}
&& S_{min} =
~\frac{1}{2~g^2}\int d^4 x~d^4 \theta~ {\rm Tr}\left(
\overline{\Gamma}^{\adot}~\overline{W}_{\adot}\right) \nonumber \\
&& \qquad
+ \frac{1}{2~g_0^2~{\cal N}} \int d^4x~d^4 \theta~{\Bigg[~\rm
Tr}\left( \overline{\Gamma}^{\adot} \right){\rm Tr}\left(
\overline{W}_{\adot}\right) \nonumber \\
&& \qquad \qquad \qquad \qquad \qquad \qquad + 4 i {\cal
F}^{\rho \g} ~ \overline{\theta}^2~{\rm Tr} \left(
\pa_{\rho \dot{\rho}} \overline{\Gamma}^{\adot} \right) {\rm Tr}\left(
\overline{W}_{\adot} \overline{\Gamma}_{\g}^{~\dot{\rho}} \right)
 \nonumber \\
&& \qquad  \qquad \qquad \qquad \qquad \qquad - {\cal F}^2
~ \overline{\theta}^2~{\rm Tr}
\left(\overline{\Gamma}^{\adot}\overline{W}_{\adot}\right){\rm
Tr}\left(\overline{W}^{\bdot}\overline{W}_{\bdot}\right) ~~\Bigg] \nonumber \\
&& \qquad + \frac{1}{l^2} ~{\cal F}^2 \int d^4 x ~d^4
\theta~\overline{\theta}^2 ~{\rm Tr} \left( \overline{\Gamma}^{\adot}~
\overline{W}_{\adot}~\overline{W}^{\bdot}~\overline{W}_{\bdot}\right)
\nonumber \\
\end{eqnarray}
or
\bea \label{Smin2}
&& S'_{min} =
~\frac{1}{2~g^2} \int d^4 x~ d^4 \theta~ \Bigg[~~{\rm Tr}\left(
\overline{\Gamma}^{\adot}~\overline{W}_{\adot}\right) \nonumber \\
&& \qquad  \qquad \qquad  \qquad \qquad  ~~~+ {\cal F}^2
~ \overline{\theta}^2~{\rm Tr}
\left(\overline{\Gamma}^{\adot}\overline{W}_{\adot}\right){\rm
Tr}\left(\overline{W}^{\bdot}\overline{W}_{\bdot}\right) \Bigg] \nonumber \\
&& \qquad
+ \frac{1}{2~g_0^2~{\cal N}} \int d^4x~d^4 \theta~\Bigg[~~{\rm
Tr}\left( \overline{\Gamma}^{\adot} \right){\rm Tr}\left(
\overline{W}_{\adot}\right) \nonumber \\
&& \qquad \qquad \qquad \qquad \qquad ~~~+ 4 i {\cal
F}^{\rho \g} ~ \overline{\theta}^2~{\rm Tr} \left(
\pa_{\rho \dot{\rho}} \overline{\Gamma}^{\adot} \right) {\rm Tr}\left(
\overline{W}_{\adot} \overline{\Gamma}_{\g}^{~\dot{\rho}} \right) \Bigg]
 \nonumber \\
&& \qquad + \frac{1}{l^2} ~{\cal F}^2 \int d^4
\theta~\overline{\theta}^2 ~{\rm Tr} \left( \overline{\Gamma}^{\adot}~
\overline{W}_{\adot}~\overline{W}^{\bdot}~\overline{W}_{\bdot}\right)
\nonumber \\
\end{eqnarray}
One would be tempted to further reduce the number of independent couplings by
setting $g_0^2 = - g^2$. However, this would lead to a dangerous cancellation
between the quadratic terms of the $U(1)$ vector fields and, as it is
evident from (\ref{U1prop}), this would make our procedure
inconsistent.



\section{Adding matter and going beyond one--loop} \label{matter}

So far we have discussed the renormalization at one loop of a
$N=\frac{1}{2}$ pure gauge theory classically described by the
action (\ref{final-action}).
A natural extension of our analysis would follow two possible directions:

\noindent
1) The inclusion of (anti)chiral matter and the evaluation of the
corresponding effective action at one--loop;

\noindent
2) The evaluation of higher order corrections to the effective action
with or without matter.

We note that the evaluation
of higher loop corrections to the gauge effective action necessarily involves
lower order corrections to the ghost action. As explained in Section 3,
ghosts formally behave as matter fields. Therefore, computing quantum corrections to
the matter action gives direct informations about the effective action for
ghosts and allows for higher order calculations in the gauge theory. The two
lines of investigation are then strictly related.

In this Section we briefly address the main features of the calculation
of the matter effective action without entering into any detail.
We look for divergent corrections to the matter action
\beq
\int d^4 \theta ~\Phi \overline{\Phi}
\label{chiral}
\eeq
(we forget about the trace structure for the moment).
We recall that $\Phi$ and $\overline{\Phi}$ are covariantly chiral
and antichiral superfields; performing the quantum--background expansion of the
action we obtain interaction vertices of the form
chiral--antichiral--gauge. We can then look for one--loop corrections
to these terms.

Following exactly the same arguments of Section 5, we select the general
structure of matter--gauge corrections on the basis of dimensional arguments
and R-symmetries, but neglecting for the moment gauge invariance considerations.
For example we can look for terms of the form
\beq
\Lambda^{\beta} \int d^4 x~d^4 \theta~
\overline{\theta}^{\overline{\tau}}~{\cal
F}^{\alpha}~\Phi~\overline{\Phi}~D^{\gamma}~\overline{D}^{\overline{\gamma}}~
\pa^{\delta}~\overline{\Gamma}^{\overline{\sigma}}
\label{ct2}
\eeq
where $0 \leq \overline{\tau} \leq 2$ and
the dimensions and $R$-charges for the new objects are
\begin{center}
\setlength{\extrarowheight}{6pt}
\begin{tabular}{|c|c|c|}
\hline
  &  dim & R-charge \\
\hline
$\Phi$ & $1$ &  $-1$ \\
\hline
$\overline{\Phi}$ & $1$ &  $1$ \\
\hline
\end{tabular}
\end{center}
In particular, the R-charges of the covariantly (anti)chiral fields have
been fixed by requiring their masses to have vanishing R--charge.

Fixing the dimension and the R-charge of the generic counterterm (\ref{ct2})
leads to the following relations
\bea
&& \a +
\frac{\overline{\tau}}{2} - \frac{1}{2} \left( \g + \overline{\g} +
\overline{\sigma} \right) - \d \ge 0 \\
&& \overline{\sigma} + 2 \a =
\overline{\tau} + \g - \overline{\g}
\eea
from which it follows
\beq
\overline{\sigma} \le \overline{\tau} -\d - \overline{\g} \le
\overline{\tau} \le 2
\eeq
As discussed in Section 5 this is supplemented by the extra constraint
\beq
\g \le 2 \overline{\sigma}
\eeq
Other conditions come from the requirement of having all the spinorial indices
correctly contracted, together with a nontrivial dependence on ${\cal F}^{\rho \g}$
(avoiding  ${\cal F}^{\rho}_{~\rho} = 0$, of course)
\bea
&& 2 \a +
\d + \g = 2 n + 2 \qquad \qquad n \ge 1\\
&& \overline{\tau} + \d +
\overline{g} + \overline{\sigma} = 2 m \qquad \qquad m \ge 0 \qquad
\qquad
\eea
where $n,m$ are integers. Collecting all the constraints, it is easy to
see that the only possible structures are
\bea
&& {\cal F}^{\rho
\g} ~\int d^4x~d^4 \theta~\overline{\theta}^2~\Phi~\pa_{\rho}^{~\adot}
\overline{\Phi}~\overline{\Gamma}_{\g \adot} \label{3pch1} \\
&& {\cal F}^{\rho \g}
~\int d^4x~d^4
\theta~\overline{\theta}^2~\Phi~\overline{\Phi}~\pa_{\rho}^{~\adot}
\overline{\Gamma}_{\g \adot} \label{3pch2}\\
&& {\cal F}^{\rho \g} ~\int d^4x~d^4
\theta~\overline{\theta}^2~\Phi~\overline{\Phi}~\overline{\Gamma}_{\g}^{~\adot}~\overline{\Gamma}_{\rho
\adot} \label{4pch1} \\
&& {\cal F}^2~ \int d^4x~d^4
\theta~\overline{\theta}^2~\Phi~\overline{\Phi}~\overline{W}^{\adot}~\overline{W}_{\adot}
\label{4pch2}
\eea
Performing the reduction to components of each single term by using the
conventions introduced in Appendix D, one can easily see
the correspondence between these terms and the results obtained
directly in components \cite{JJW}.

In superspace these contributions can arise from the sets of diagrams in Figures
6--9.
\vskip 5pt
\begin{minipage}{\textwidth}
\begin{picture}(120,120)
\put(0,0){
\includegraphics[width=5.5 cm]{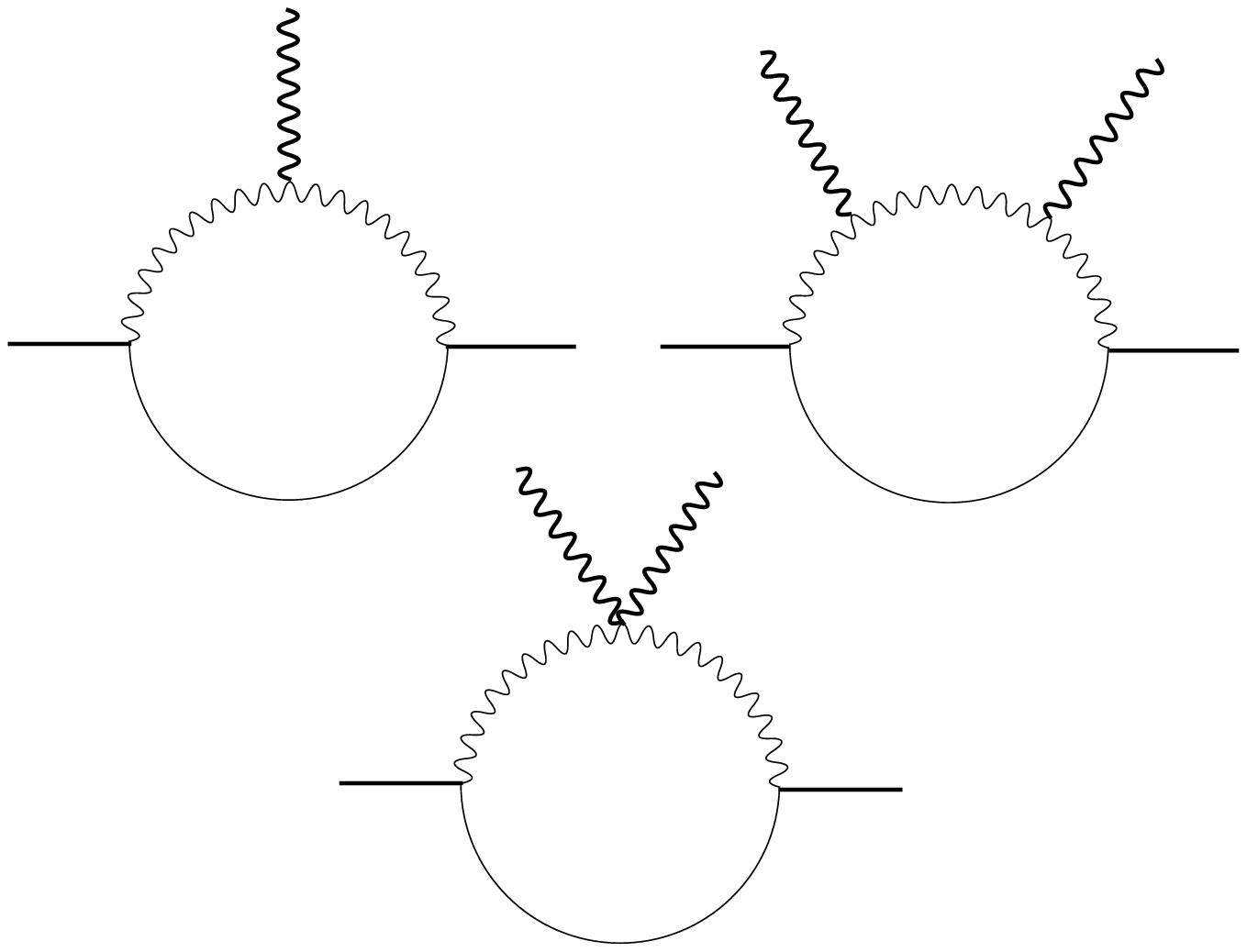}
}
\put(65,-40){{\small{Figure 6}}}
\put(220,-20){
\includegraphics[width=5.5 cm]{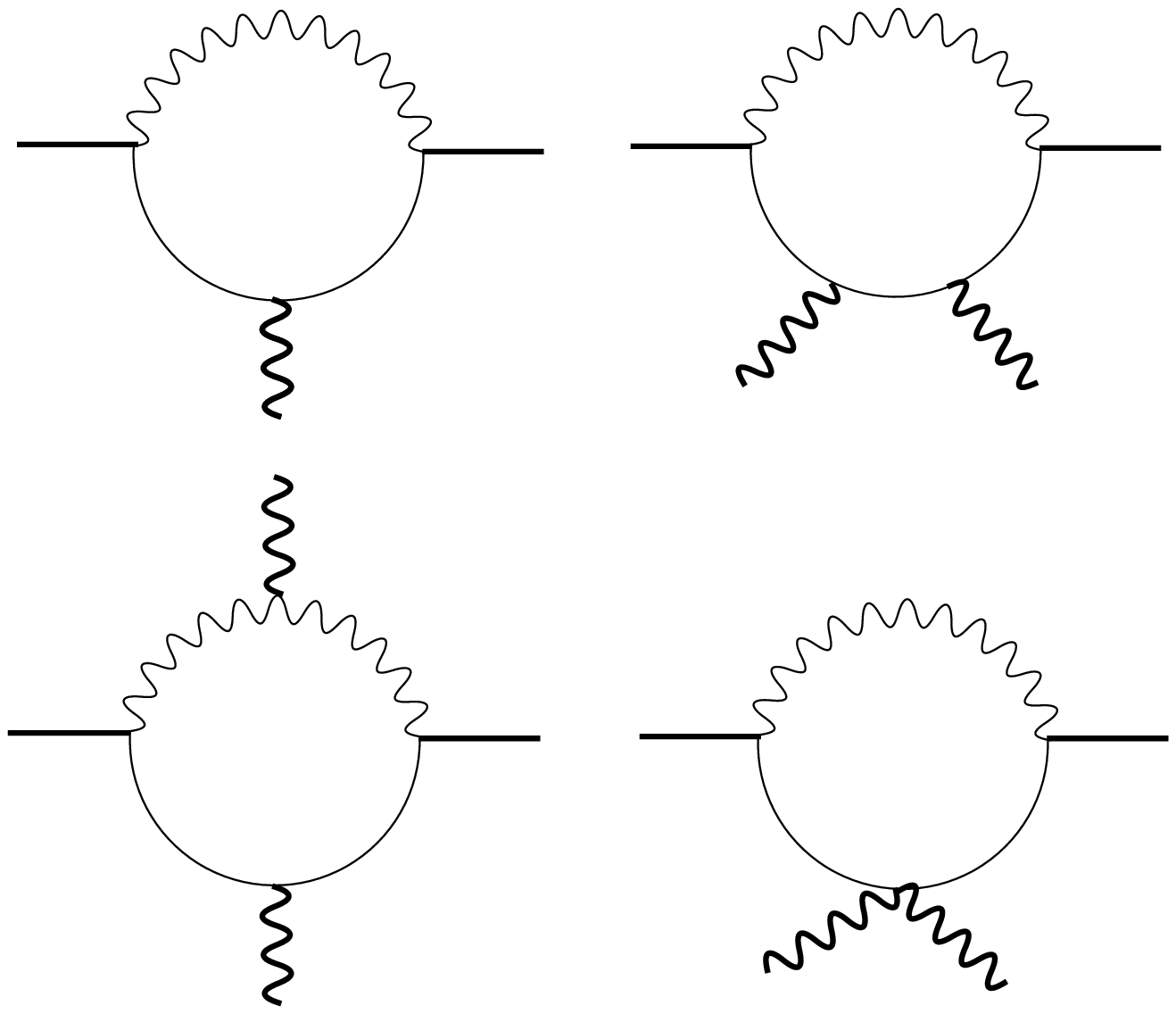}
}
\put(290,-40){{\small{Figure 7}}}
\end{picture}
\end{minipage}
\vskip 70pt

\begin{minipage}{\textwidth}
\begin{picture}
(120,120)
\put(0,0){
\includegraphics[width=5.5 cm]{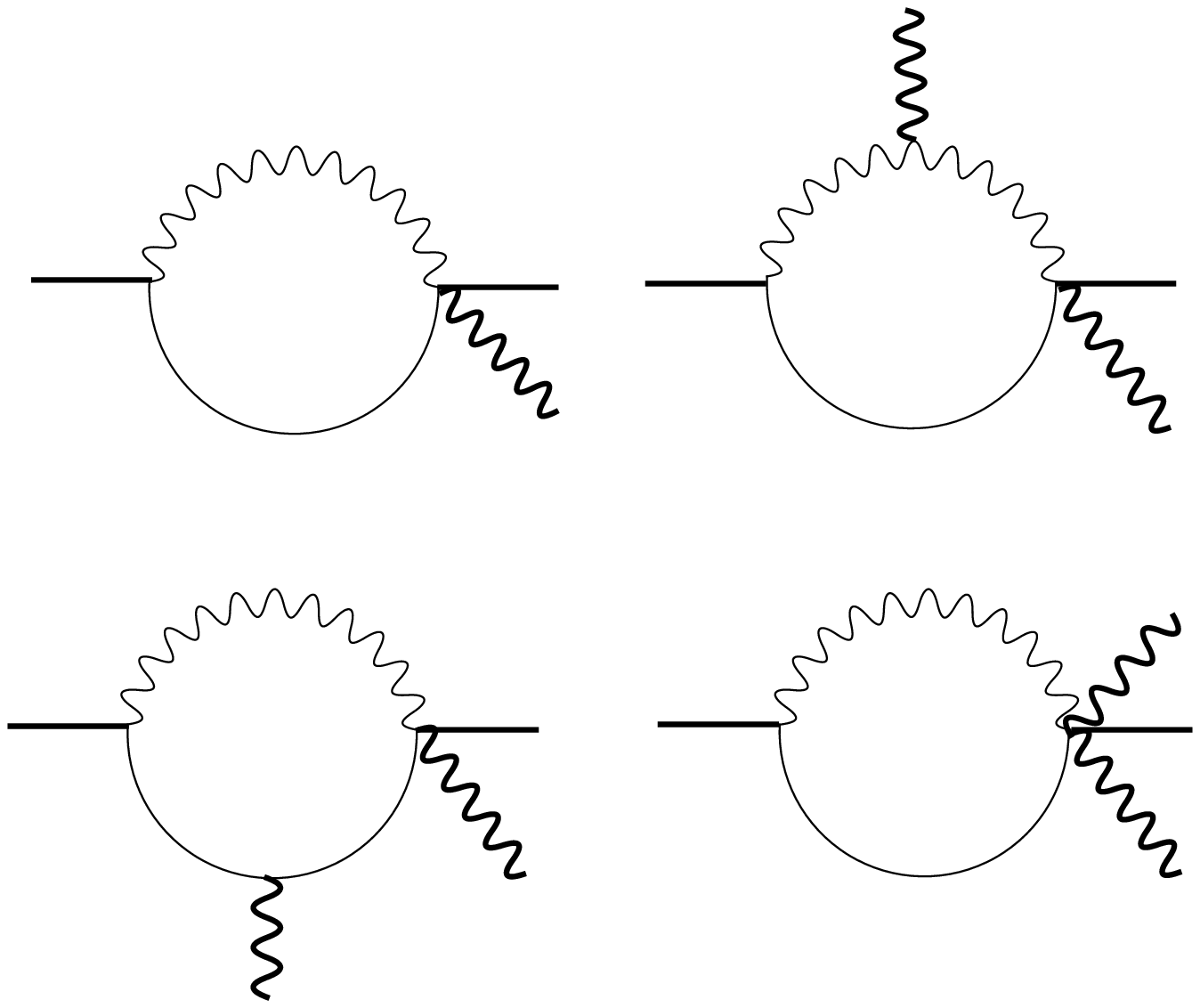}
}
\put(65,-30){{\small{Figure 8}}}
\put(200,10){
\includegraphics[width=7 cm]{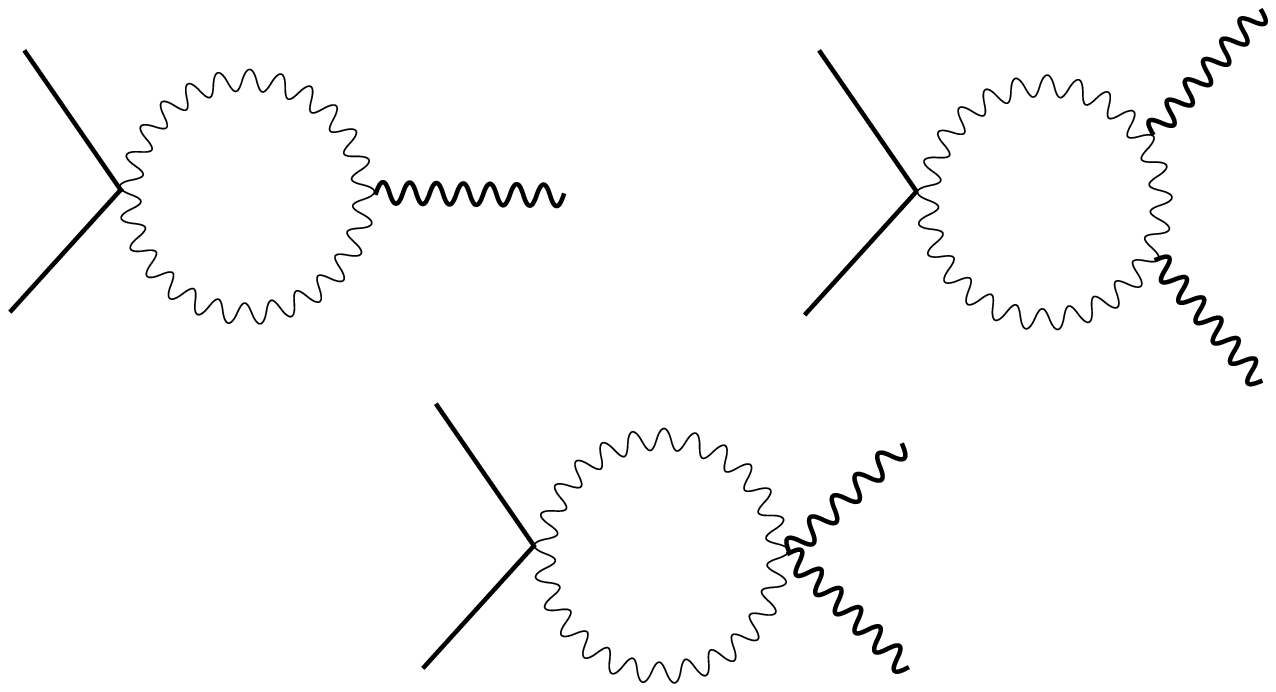}
}
\put(290,-30){{\small{Figure 9}}}
\end{picture}
\end{minipage}
\vskip 50pt

In Figs. 6 and 9 external gauge fields come from vertices arising from the
expansion of the pure gauge part of the action (\ref{final-action}).
In Fig. 7 and 8 external vector legs inserted on a matter propagator
and/or at a mixed vertex come from
\begin{equation} \label{var}
  \frac{\d \Phi (z)}{\d \Phi (z')} = \overline{\nabla}^2
  \d^{(8)} (z-z')
\end{equation}
where we expand the covariant derivatives in powers of background
gauge fields.

Without  performing the complete evaluation of these
contributions, we can give a straightforward argument to
argue that some of them will be certainly present.
The argument goes as follows: already in the ordinary anticommutative
case, the matter propagator receives corrections from the diagram in Fig. 10
\vskip 5pt
\begin{minipage}{\textwidth}
\begin{center}
\includegraphics[width=0.25\textwidth]{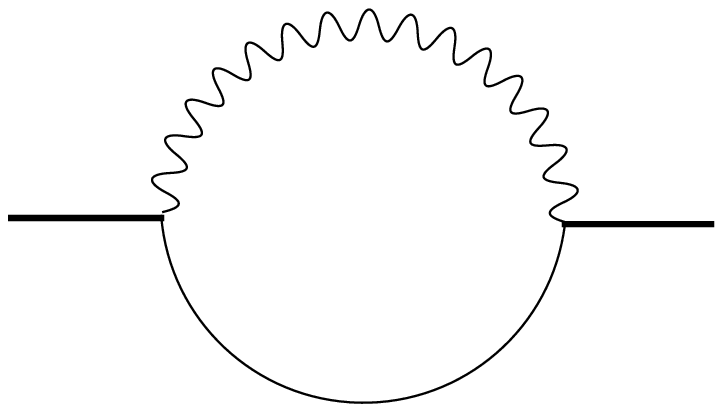}
\end{center}
\begin{center}
{\small{Figure 10}}
\end{center}
\end{minipage}
\vskip 5pt
\noindent
which gives rise to a divergent term of the form
\beq
\int d^4 x~d^4
\theta~ \left[ {\cal N} {\rm Tr} \left( \overline{\Phi} ~\Phi \right)
- {\rm Tr}\left( \overline{\Phi} \right){\rm Tr}\left( \Phi \right)
\right]
\label{ct3}
\eeq
This contribution is present also in the NAC case. In this case,
however, the double trace piece is not supergauge invariant \cite{pr}
and its supergauge variation is proportional to
\beq
{\cal F}^{\g \rho}~\int d^4 x~d^4 \theta~
\overline{\theta}^2~{\rm Tr} \left( \pa_{\rho}^{~\dot{\rho}}
\overline{\Lambda} ~ \pa_{\g \dot{\rho}} \overline{\Phi} \right)~{\rm
Tr}\left( \Phi \right)
\label{gaugevar2}
\eeq
Supergauge invariance of the effective action
then requires extra divergent terms to emerge from
diagrams 6--9 whose gauge variations compensate (\ref{gaugevar2}). In particular,
the variation (\ref{gaugevar2}) can be canceled by the variation of
\beq \label{3pchinv}
{\cal F}^{\rho \g}
~\int d^4x~d^4 \theta~\overline{\theta}^2~{\rm Tr}\left(
\overline{\Gamma}_{\rho}^{~ \dot{\rho}}~\pa_{\g \dot{\rho}}
\overline{\Phi} \right) ~{\rm Tr}\left( \Phi \right)
\eeq
Therefore, we expect such a a term to appear at one--loop order. This signals
the necessity of modifying the classical matter action to make the theory
renormalizable as we did for the pure gauge part. In components this
has been already done in \cite{JJW}.

\vskip 15pt

We now briefly discuss the evaluation of higher--loop contributions to
the pure gauge effective action. In higher--loop diagrams, lower order
corrections to the ghost propagator and vertices will be necessary.
Therefore,  in order to compute the gauge effective action we need go
through the evaluation of the ghost effective action as well.

As already mentioned, ghosts behave formally as massless (anti)chiral
matter, at least for what concerns Feynman rules. Therefore we can
partially exploit the previous discussion about matter contributions.
However, there is a crucial difference between matter
and ghosts which should be clarified: while matter undergoes
background--quantum splitting and one--loop divergent
contributions to the matter background effective action require renormalization,
ghosts are intrinsically quantum fields and do not appear in the
 background effective action.
Divergent loops with external ghost lines (which can be obtained from the diagrams
discussed above)  appear only
 as subdiagrams in higher--order diagrams with external  physical background lines.
 The corresponding
 subdivergences have to be BPHZ-subtracted out or compensated by introducing counterterms
 in the quantum action.
Their form, and possible modifications due to the NAC effects would be controlled, or in
fact determined by
BRST invariance;
but this is not the place for discussing these issues.

In the case of ghosts, the general argument given above to select
all possible
divergent structures which might appear in the ghost effective action should be modified
since the $R$-charge assignment for ghosts is different. However, in this case we prefer
to proceed by direct inspection and observe that the graphs contributing at one--loop are still the
ones in Figs. 6--9 where now we have ghosts on the (anti)chiral lines.
In particular, given the explicit form of the ghost action
\bea \label{ghostexp}
  S_{gh} &=&  \int d^4 x d^4 \theta \Big[ \overline{c}' c - c'
  \overline{c} + \frac{1}{2} (c' + \overline{c}') [ V, (c +
  \overline{c})] \nonumber \\
  && \qquad \qquad \qquad \qquad + \frac{1}{12} (c' + \overline{c}')
     [ V, [ V, (c - \overline{c})]] \Big] + {\cal O}(V^3)
\eea
it is quite immediate to realize that graphs in Figs. 6,7,8 cancel due to
an opposite sign in the $\langle \overline{c}' c \rangle$ and
$\langle c' \overline{c} \rangle$ propagators. Therefore in the ghost case
new structures, if any, will emerge only from diagrams in Fig. 9. These corrections
have to be taken into account when we compute two loop corrections
to the gauge effective action. We note that in the case of ghosts we cannot
apply the argument which follows from eq. (\ref{ct3}) to guarantee that
new structures will necessarily appear at the quantum level, as ghost  self--energy
corrections coming from diagrams as in Fig. 10 cancel \cite{Improved}.

\vskip 15pt

As a final comment we consider the case of the $N=4$ SYM theory in the presence of
NAC geometry. In $N=1$ superfield formalism, $N=4$ SYM theory is
described by a $N=1$ gauge theory plus three chiral scalars in the adjoint
representation, interacting with a cubic superpotential \cite{Improved} (the
deformed action in components is given in \cite{ABBP}). In this setup it is possible
to discuss the quantum properties of the theory when ordinary
products are replaced by $\ast$-products, so generically breaking $N=4$ to $N= 1/2$.
In particular we are interested in the renormalization of the gauge part.
Looking at the result (\ref{eff-action1}) it is evident that
the gauge action (\ref{invaction}) does not receive any divergent correction
at one--loop since in this case $N_f = 3$.
However, as already noticed, new matter--matter--gauge vertices of the form
(\ref{3pch1}--\ref{4pch2}) can arise at one loop when we look at the matter part of
the effective action.
If these new terms appear, renormalization requires one inserting them already in the
original action and, once quantized, they might spoil the cancellation of (\ref{eff-action1}).
Therefore, a complete analysis of one--loop contributions to the matter superfields
has to be performed in order to make clear statements about the renormalizability
of the gauge part of this theory. This issue is beyond the scope of the present
work.



\vskip 50pt
{\bf Acknowledgments }

The research described in this paper has been supported in part by
INFN, PRIN prot. 2003023852\_008, the European Commission RTN
program MRTN--CT--2004--005104 and by NSERC of Canada.

\newpage

\appendix


\section{Mathematical tools}

In this Appendix we list the main identities we used in the
calculations concerning color traces and momentum loop integrals.

First of all we fix the group conventions for the
$U({\cal N})$ group.  In the fundamental representation the
generators are ${\cal N} \times {\cal N}$ unitary matrices $T^A$, $A=
0, \cdots , {\cal N}^2 -1 $, where
$T^0=\frac{\mathbbmtt{1}}{\sqrt{{\cal N}}}$, whereas $T^a$ are the
$SU(\cal{N})$ generators. Their normalization is fixed by
\beq {\rm
Tr} (T^A T^B ) = \d^{AB}
\label{id1bis}
\eeq
The algebra of generators reads
\beq
[ T^A , T^B ]= i~ f^{ABC} T^C
\eeq
where $f^{ABC}$ are the structure constants given by
\beq
f^{abc} = -i{\rm Tr}(T^a [T^b,T^c]) \qquad , \qquad
f^{0AB} =0
\label{id2bis}
\eeq
We also introduce
\beq
d^{abc} = {\rm Tr}(T^a \{T^b,T^c\})  \qquad , \qquad
d^{0AB} = \frac{2}{\sqrt{N}} \d^{AB}
\label{id3}
\eeq
Useful relations are:
\begin{eqnarray}
&& {\rm Tr} \left( T^A T^B T^C \right) = \frac{1}{2} \left( i f^{ABC}
  + d^{ABC}\right) \\
&& {\rm Tr} \left( T^A T^B T^C T^D \right) = \frac{1}{4} \left( i f^{ABE}
  + d^{ABE}\right) \left( i f^{ECD} + d^{ECD}\right)
\end{eqnarray}

Given two scalar objects $M \equiv M^A T^A $ and $N \equiv N^A T^A$ in the adjoint
representation of the gauge group, we have the general identity
\bea
[ M , N]_\ast &=& \frac12 \{ T^A, T^B\} [M^A , N^B ]_\ast ~+~ \frac12 [T^A, T^B] \{ M^A , N^B \}_\ast
\nonumber \\
&=& \frac12 d^{ABC} [M^A , N^B ]_\ast T^C ~+~ \frac{i}{2} f^{ABC} \{ M^A , N^B \}_\ast T^C
\label{commut}
\eea
The generalization of the above identity in the presence of spinorial
quantities can be obtained straightforwardly.

We now consider momentum loop integrals.
As stated in the main text, all the divergent contributions are
expressed in terms of a self--energy
integral ${\cal S}$ which in dimensional regularization ($n = 4 - 2\e$) is
\begin{equation}
{\cal S} \equiv
\int d^4 q ~\frac{1}{((q-p)^2 + m \overline{m})(q^2 + m \overline{m})}
= \frac{1}{(4\pi)^2}~\frac{1}{\epsilon} + {\cal O}(1)
\label{selfenergy}
\end{equation}
Other one--loop divergent integrals are obtained in terms of ${\cal
S}$ through the following identities
\beq \int d^4 q ~\frac{q_{\a
\adot}}{((q-p)^2 + m \overline{m})(q^2 + m \overline{m})} = \frac12
p_{\a\adot} ~{\cal S}
\label{S1}
\eeq
\beq \label{S4}
\int d^4 q \frac{q_{\a\adot} q_{\b \bdot}}{(q^2 + m\overline{m})(
  (q+p)^2 + m\overline{m})
((q+r)^2 + m\overline{m})} \sim
\frac{1}{2} C_{\a\b} C_{\adot \bdot} ~ {\cal S} \\
\eeq
\bea \label{S5}
&& \int d^4 q \frac{q_{\a\adot} q_{\b \bdot}q_{\g\gdot} q_{\rho
    \dot{\rho}} }{(q^2 + m\overline{m})( (q+p)^2 + m\overline{m})
((q+r)^2 + m\overline{m})((q+s)^2 + m\overline{m})} \sim \nonumber \\
&& \qquad \qquad \frac{1}{6}  \big( C_{\a\b} C_{\adot \bdot}
C_{\g\rho} C_{\gdot \dot{\rho}} + C_{\a\g} C_{\adot \gdot} C_{\b\rho}
C_{\bdot \dot{\rho}} + C_{\a\rho} C_{\adot \dot{\rho}} C_{\b\g}
C_{\bdot \gdot} \big) ~{\cal S}\nonumber \\
\eea
Since the divergent part of these integrals is mass independent,
the same results hold also for $m = \overline{m} =0$ (ghost contributions).


\section{Momentum superspace}

In this Appendix we give a detailed description of all the mathematical
tools for Fourier transformed superspace.

We first consider the case of ordinary, anticommuting
superspace. We define a {\em momentum superspace} described by
coordinates $(p_{\a\adot}, \pi_\a , \overline{\pi}_{\adot})$ conjugate to
$(x^{\a\adot}, \theta^\a , \overline{\theta}^{\adot})$.
The derivatives with respect to the spinorial momenta
\begin{equation}
\frac{\partial}{\partial \pi^{\alpha}} \equiv  \eth_{\alpha} \qquad \qquad
\frac{\partial}{\partial \overline{\pi}^{\dot{\alpha}}} \equiv
\overline{\eth}_{\dot{\alpha}}
\end{equation}
satisfy $\eth_\a \pi^\b = \d_\a^{~\b}$, $\eth^\a \pi_\b = -\d_\b^{~\a}$.
In analogy with the case of ordinary superspace,
we define the integration as
\beq
\int d^2 \pi ~ \Phi(\pi) = \eth^2 \Phi| \qquad , \qquad \int d^2 \overline{\pi} ~
\Phi(\overline{\pi}) = \overline{\eth}^2 \Phi|
\label{id2}
\eeq
In particular, this implies
\beq
\int d^2 \pi ~\pi^2 \equiv \eth^2 \pi^2 = -1\qquad , \qquad
\int d^2 \overline{\pi} ~\overline{\pi}^2 \equiv \overline{\eth}^2 \overline{\pi}^2 = -1
\eeq
so that, consistently, the momentum delta functions are given by
$\d^{(2)}(\pi) \equiv -\pi^2$, $\d^{(2)}(\overline{\pi}) \equiv -
\overline{\pi}^2$.

We define the Fourier transform (FT) in the bosonic coordinates and its inverse as
\beq
\widetilde{\Phi}(p) = \int d^4 x ~ e^{i p x }~ \Phi(x) \qquad , \qquad
\Phi(x) = \int d^4 p ~ e^{-i p x}~ \widetilde{\Phi}(p)
\eeq
(omitting $(2\pi)^4 $ factors).

Analogously, the FT on the fermionic coordinates and its inverse are given by
\beq
\widetilde{\Phi}(\pi, \overline{\pi}) = \int d^2 \theta d^2 \overline{\theta}~
e^{i\pi \theta + i\overline{\pi} \overline{\theta}}~ \Phi(\theta, \overline{\theta})
\qquad , \qquad
\Phi(\theta, \thb) =
\int d^2 \pi d^2 \overline{\pi}~ e^{-i\pi \theta - i\overline{\pi} \overline{\theta}}~
\widetilde{\Phi}(\pi, \overline{\pi})
\label{FFT}
\eeq
The consistency of the two expressions in (\ref{FFT}) follows from
the identities
\bea
&&\int d^2 \pi d^2 \overline{\pi}~ e^{i\pi (\theta - \theta') + i\overline{\pi}
(\overline{\theta}
- \overline{\theta}')} = \delta^{(4)}(\theta - \theta')
\nonumber \\
&&\int d^2 \theta d^2 \overline{\theta}~ e^{i(\pi-\pi') \theta + i(\overline{\pi}
-\overline{\pi}')
\overline{\theta}'}=\delta^{(4)}(\pi - \pi')
\eea
where, as usual, $\delta^{(2)}(\theta - \theta') \equiv -(\theta - \theta')^2$.

From the definitions given above we have the following identifications
between conjugate variables
\bea
&& p_{\alpha \dot{\alpha}}= i
\partial_{\alpha \dot{\alpha}} \qquad , \qquad
x_{\alpha \dot{\alpha}}= -i \frac{\partial}{\partial p^{\alpha \dot{\alpha}}}
\nonumber \\
&& \pi_{\alpha} = i \partial_{\alpha} ~~\qquad ,\qquad
\theta_{\alpha} = - i  \eth_{\alpha}
\nonumber \\
&& \overline{\pi}_{\ad} = i \pab_{\ad} ~~\qquad ,\qquad
\thb_{\ad} = -i \overline{\eth}_{\adot}
\label{dermom}
\eea

\subsection{Covariant derivatives}

We use chiral representation for the superspace covariant derivatives $D_\a ,
\overline{D}_{\adot}$ \cite{superspace}.
Performing FT, in momentum superspace we obtain momentum operators
$\widetilde{D}_\a, \widetilde{\overline{D}}_{\adot}$
given by (see identities (\ref{dermom}))
\bea
&& D_{\alpha} = \partial_{\alpha} + i \overline{\theta}^{\dot{\alpha}}
\partial_{\alpha \dot{\alpha}}
\quad \rightarrow  \qquad\widetilde{D}_{\alpha} = - i \pi_{\alpha} -i
\overline{\eth}^{\dot{\alpha}} p_{\alpha \dot{\alpha}} \nonumber \\
&& \overline{D}_{\dot{\alpha}} = \overline{\partial}_{\dot{\alpha}} \qquad
\quad\qquad \rightarrow
\qquad \widetilde{\overline{D}}_{\dot{\alpha}} = - i \overline{\pi}_{\dot{\alpha}}
\label{Dtilde}
\eea
and
\bea \label{D^2}
&& \widetilde{D}^2 = - \pi^2 -  \pi^{\alpha}\overline{\eth}^{\dot{\alpha}}
p_{\alpha \dot{\alpha}} - \overline{\eth}^2 p^2 \nonumber \\
&& \widetilde{\overline{D}}^2 = - \overline{\pi}^2
\eea
(Anti)commutation rules for $\widetilde{D}$--derivatives \footnote{With
abuse of language we call them ``derivatives'' even if some are
actually multiplicative operators.} are
\beq
\{ \widetilde{D}_{\a} , \widetilde{\overline{D}}_{\adot} \} =
p_{\a \adot} \qquad \qquad
{\rm the~rest ~= 0}
\eeq
and the following identities hold
\bea
&& [ \widetilde{D}^{\alpha}, \widetilde{\overline{D}}^2 ] = p^{\alpha \dot{\alpha}}
\widetilde{\overline{D}}_{\dot{\alpha}} \qquad \qquad
[ \widetilde{\overline{D}}^{\adot}, \widetilde{D}^2 ] = p^{\alpha \dot{\alpha}}
\widetilde{D}_{\a}
\nonumber \\
&&\widetilde{D}^2  \widetilde{\overline{D}}^2 \widetilde{D}^2 =  -p^2
\widetilde{D}^2
\qquad ~~~~\widetilde{\overline{D}}^2 \widetilde{D}^2 \widetilde{\overline{D}}^2 =  -p^2
\widetilde{\overline{D}}^2
\nonumber \\
&&- p^2 = \widetilde{\overline{D}}^2 \widetilde{D}^2 +
\widetilde{D}^2 \widetilde{\overline{D}}^2 -
\widetilde{\overline{D}}^{\adot} \widetilde{D}^2 \widetilde{\overline{D}}_{\adot}
=  \widetilde{\overline{D}}^2 \widetilde{D}^2 +
\widetilde{D}^2 \widetilde{\overline{D}}^2 -
\widetilde{D}^{\a} \widetilde{\overline{D}}^2 \widetilde{D}_{\a}
\label{D}
\eea
Momentum covariant derivatives can be integrated by parts according to the
following rule
\beq
\int d^4p ~d^4 \pi  \widetilde{D}_\a (p, \pi) \widetilde{\Phi}(p, \pi)
\widetilde{\Psi}(-p, -\pi) ~=~
- \int d^4p ~d^4 \pi \widetilde{\Phi}(p, \pi)
\widetilde{D}_\a(-p, -\pi) \widetilde{\Psi}(-p, -\pi)
\eeq

\subsection{(Anti)Chirality conditions}

Given the superfield $V(x, \theta, \overline{\theta})$ expanded in powers
of spinorial coordinates, it is easy to determine its expansion in momentum
superspace by Fourier transforming term by term through the following
identities
\bea
&&\int d^2 \theta d^2 \overline{\theta}~ e^{i\pi \theta + i\overline{\pi} \overline{\theta}}
~ 1 =
\pi^2 \overline{\pi}^2 = \d^{(4)}(\pi)
\nonumber \\
&& \int d^2 \theta d^2 \overline{\theta}~ e^{i\pi \theta + i\overline{\pi} \overline{\theta}}
~
\theta^\a = - i\pi^\a \overline{\pi}^2 = i\pi^\a \d^{(2)}(\overline{\pi})
\nonumber \\
&& \int d^2 \theta d^2 \overline{\theta}~ e^{i\pi \theta + i\overline{\pi} \overline{\theta}}
~ \theta^2 =
\overline{\pi}^2 = -\d^{(2)}(\overline{\pi})
\label{id1}
\eea
and analogous ones for the antichiral sector. It is easy to see that
the expansion
\bea
&& V(x, \theta, \overline{\theta}) = C(x) +  \theta^{\a} \chi_{\a} (x)+
\overline{\theta}^{\adot} \overline{\chi}_{\adot}(x) - \theta^2 M (x)- \overline{\theta}^2
\overline{M}(x) \nonumber \\
&& ~~~\qquad \qquad \qquad  +  \theta^{\a} \overline{\theta}^{\adot} A_{\a \adot}(x) -
\overline{\theta}^2 \theta^{\a} \lambda_{\a}(x) - \theta^2
\overline{\theta}^{\adot} \overline{\lambda}_{\adot} (x)+ \theta^2 \overline{\theta}^2 D'(x)
\eea
in momentum superspace corresponds to
\bea
&& \widetilde{V}(p, \pi, \overline{\pi}) = \pi^2 \overline{\pi}^2 \widetilde{C}(p)
-i \overline{\pi}^2 \pi^{\a} \widetilde{\chi}_{\a}(p)
-i \pi^2 \overline{\pi}^{\adot} \widetilde{\overline{\chi}}_{\adot}(p)
- \overline{\pi}^2 \widetilde{M}(p)
- \pi^2 \widetilde{\overline{M}}(p) \nonumber \\
&& ~\qquad \qquad \qquad \qquad \qquad  - \pi^{\a} \overline{\pi}^{\adot} \widetilde{A}_{\a \adot}(p)
- i \pi^{\a} \widetilde{\lambda}_{\a}(p)
- i \overline{\pi}^{\adot} \widetilde{\overline{\lambda}}_{\adot}(p)
+ \widetilde{D'}(p)
\eea

We now discuss constrained superfields. Going to momentum superspace the
chiral constraint on a superfield becomes
\beq
\overline{D}_{\dot{\alpha}} \Phi (\theta, \overline{\theta})= 0 \qquad \rightarrow
\qquad
\widetilde{\overline{D}}_{\dot{\alpha}} \widetilde{\Phi} (\pi, \overline{\pi}) =
(-i \overline{\pi}_{\dot{\alpha}})
\widetilde{\Phi} (\pi, \overline{\pi}) = 0
\eeq
The solution to this constraint is necessarily of the form $\widetilde{\Phi}
(\pi, \overline{\pi}) = \overline{\pi}^2 \chi(\pi)$, where $\chi$ is an unconstrained
superfield independent of $\overline{\pi}$. The solution can be also written as
\beq
\widetilde{\Phi} (\pi, \overline{\pi}) =
\overline{\pi}^2 ~\left(\overline{\eth}^2 \widetilde{\Phi} (\pi, \overline{\pi})|_{\overline{\pi}
=0}\right) =
- \d^{(2)}(\overline{\pi}) ~\overline{\eth}^2 \widetilde{\Phi} (\pi, \overline{\pi})
\label{ch1}
\eeq
Going to components the most general expression for a chiral superfield is
\beq
\Phi (x,\theta, \overline{\theta}) = \phi(y) + \theta^\a \psi_{\a}(y) -  \theta^2
F(y)
\eeq
where $y^{\a \ad} =x^{\a \ad} - i \theta^\a \thb^{\ad}$.
If we perform the FT keeping $y$ as an independent
variable we obtain (see eqs. (\ref{id1}))
\beq \label{ch2}
\widetilde{\Phi}= -\overline{\pi}^2 (F + i\pi^\a \psi_\a - \pi^2 \phi )
= \d^{(2)}(\overline{\pi}) (F + i\pi^\a \psi_\a - \pi^2 \phi )
\eeq
A useful identity satisfied by a chiral superfield is
\beq
\widetilde{\overline{D}}^2 \overline{\eth}^2 \widetilde{\Phi} = \widetilde{\Phi}
\label{idchiral}
\eeq
For an antichiral superfield the constraint becomes
\beq
D_{\alpha} \overline{\Phi} (\theta, \overline{\theta})= 0 \qquad \rightarrow  \qquad
\widetilde{D}_{\alpha}
\widetilde{\overline{\Phi}} (\pi, \overline{\pi})= (-i\pi_{\alpha}-
ip_{\alpha \dot{\alpha}} \overline{\eth}^{\dot{\alpha}})
\widetilde{\overline{\Phi}} (\pi, \overline{\pi}) = 0
\eeq
This is equivalent to
$\overline{\eth}_{\dot{\alpha}} \widetilde{\overline{\Phi}} = - \frac{p_{\a \adot}}{p^2}
\pi^\a \widetilde{\overline{\Phi}}
$
which implies
\beq
\overline{\eth}_{\dot{\alpha}} \widetilde{\overline{\Phi}}|_{\pi =0} = 0
\label{antichiral}
\eeq
For the antichiral superfield the $(\pi, \overline{\pi})$-expansion is more
complicated. If we introduce $\overline{y}^{\a \adot} \equiv y^{\a \adot} -
i \theta^\a
\thb^{\adot}$, $D_\beta \overline{y}^{\a \adot}=0$, the expansion of an antichiral
superfield in coordinate superspace
\beq
\overline{\Phi} (\overline{y},\theta, \overline{\theta}) = \overline{\phi}(\overline{y}) +
\thb^{\adot} \overline{\psi}_{\adot}(\overline{y}) -  \thb^2 \overline{F}(\overline{y})
\eeq
when transformed to momentum superspace becomes
\beq
\widetilde{\overline{\Phi}} (p,\pi, \overline{\pi}) = \widetilde{\overline{\phi}}(p)
(\pi^2 \overline{\pi}^2 + \pi^\a \overline{\pi}^{\adot} p_{\a \adot} -p^2)
-i (\pi^2 \overline{\pi}^{\adot} + \pi_\a p^{\a \adot} )
\widetilde{\overline{\psi}}_{\adot}(p) -  \pi^2 \widetilde{\overline{F}}(p)
\eeq

The functional derivatives with respect to a (anti)chiral
superfield in momentum superspace are defined as
\bea
&& \frac{\d \widetilde{\Phi}(\pi_1)}{\d \widetilde{\Phi}(\pi_2)} ~=~
\widetilde{\overline{D}}^2
\d^{(4)}(\pi_1 - \pi_2)
\nonumber \\
&& \frac{\d \widetilde{\overline{\Phi}}(\pi_1)}{\d \widetilde{\overline{\Phi}}(\pi_2)}
~=~
\widetilde{D}^2 \d^{(4)}(\pi_1 - \pi_2)
\label{functder}
\eea
This can be easily understood by observing that the condition
\beq
\frac{\d }{\d \Phi_1(\theta)} \int d^2 \theta' \Phi_1(\theta')
\Phi_2(\theta') ~=~
\Phi_2(\theta)
\eeq
is translated in momentum superspace as
\beq
\frac{\d }{\d \widetilde{\Phi}_1(\pi)} \int d^4 \pi' \widetilde{\Phi}_1(\pi')
\frac{\widetilde{D}^2}{\Box} \widetilde{\Phi}_2(-\pi') ~=~
\widetilde{\Phi}_2(-\pi)
\eeq
This equation is satisfied by (\ref{functder}).

\subsection{The star product in momentum superspace}

We now extend the previous definitions to $N=\frac12$ superspace. This
requires rotating to euclidean space and turning on the nontrivial
anticommutators (\ref{nc}).
Since the derivatives $\frac{\pa}{\pa \theta^\a}$ do not get affected by
nonanticommutativity, the spinorial variables $(\pi, \overline{\pi})$ in
momentum superspace remain anticommuting.

As in the NC bosonic case, in momentum superspace the effects of
noncommutativity are visible in the product of fields through the appearance
of momentum phase factors.
We apply FT (\ref{FFT}) to the NC product of two superfields
as given in (\ref{star}). We begin by noting that
\beq
\left( e^{i \pi \theta}  \right)_{\ast} =\left( e^{i \pi \theta}  \right)
\eeq
as a consequence of the fact that the new terms generated by the
$\ast$-product are all
proportional to the structure
$\pi_{\alpha} {\cal F}^{\alpha \beta} \pi_{\beta}$
which vanishes, since ${\cal F}$ symmetric.
Moreover, the following identity holds
\beq
e^{-i \pi \theta} \ast e^{-i \pi' \theta} = e^{-i (\pi+\pi') \theta}
e^{\pi \wedge \pi'}
\eeq
Therefore, when computing the NC product of two superfields in momentum
superspace we find
\bea
\Phi (\theta, \overline{\theta}) \ast \Psi(\theta, \overline{\theta}) &=& \int d^4
\pi d^4 \pi'
\left( e^{-i \pi \theta - i \overline{\pi} \overline{\theta}}~
\widetilde{\Phi}(\pi,\overline{\pi}) \right)
\ast \left( e^{-i \pi' \theta - i \overline{\pi}' \overline{\theta}}~
\widetilde{\Psi}(\pi',\overline{\pi}')
\right) \nonumber \\
&=& \int d^4 \pi d^4 \pi'  e^{\pi \wedge \pi'}
e^{-i (\pi+\pi') \theta} e^{-i (\overline{\pi}
+ \overline{\pi}') \overline{\theta} }~\widetilde{\Phi}(\pi,\overline{\pi})
\widetilde{\Psi}(\pi',
\overline{\pi}')
\eea
Thus, in momentum superspace the star product manifests itself
through the phase factor $e^{\pi \wedge \pi'}$.


\section{One--loop non--renormalizability in components}

In this Appendix we perform the reduction to components of our one--loop
result in (\ref{eff-action1}) and prove at component level the non--renormalizability of the
original action (\ref{action1}).

We work in the Wess--Zumino gauge defined by the usual conditions
$V| = DV| = \overline{D} V| = D^2 V| = \overline{D}^2 V| =0$.
While we use the ordinary definitions for the following component fields
\bea
&& f_{\a\b} = -\frac12 \pa_\a^{~ \adot} A_{\b\adot} -\frac12 \pa_\b^{~ \adot} A_{\a\adot}
+ \frac{i}{2} [A_\a^{~ \adot} , A_{\b \adot} ] \nonumber \\
&& \overline{f}_{\adot \bdot} = -\frac12 \pa_{\adot}^{~ \a} A_{\a \adot} -\frac12
\pa_{~\bdot}^{\a} A_{\a\adot} + \frac{i}{2}
[A^\a_{~ \adot} ,
A_{\a \bdot} ] \nonumber \\
&& D^2 \overline{D}_{\adot}V \Big| = \overline{\lambda}_{\adot} \qquad
\overline{D}_{\bdot}   D^2 \overline{D}_{\adot}V \Big|  = -i C_{\adot \bdot} D'
+ \overline{f}_{\adot \bdot}
\eea
we perform the shift \cite{seiberg}
\beq \label{vartheta}
i \overline{D}^2 D_{\a} V \Big| = \lambda_{\a} + \vartheta_{\a} \qquad
\qquad \vartheta_{\a} = - \frac{1}{2} {\cal F}_{\a}^{~\rho} \big\{
\overline{\lambda}^{\bdot}, A_{\rho \bdot} \big\}
\eeq
in order to simplify the expansions. With these conventions we have
\bea
&& ~~~~\overline{W}^{\adot} \Big| = \overline{\lambda}^{\adot} \nonumber \\
&& \overline{D}^{\bdot} \overline{W}^{\adot} \Big| = -i C^{\adot \bdot} D' + \overline{f}^{\adot \bdot}
\nonumber \\
&& \overline{D}^2 \overline{W}^{\adot}\Big| = i \pa^{\a \adot} \left( \lambda_{\a} + \vartheta_{\a} \right)
+ [A^{\a \adot}, \lambda_{\a} + \vartheta_{\a}] \nonumber \\
&& \qquad ~~~~~~~ + {\cal F}^{\gamma \rho} \Big[ \frac12 \big\{ \overline{\lambda}^{\adot},
i \pa_{\rho}^{~\dot{\rho}}
A_{\gamma \dot{\rho}} \big\} + \frac{1}{2} \big\{\overline{\lambda}_{\dot{\rho}} , i \pa_{\rho}^{~\dot{\rho}}
A_{\gamma}^{~\adot}\big\} + \frac{1}{2} \big\{ i \pa_{\gamma}^{~\dot{\rho}}\overline{\lambda}^{\adot} ,
A_{\rho \dot{\rho}} \big\}+ \frac{1}{2} \big\{ i \pa_{\rho}^{~\dot{\rho}}\overline{\lambda}_{\dot{\rho}},
A_{\gamma}^{~\adot} \big\} \nonumber \\
&& \qquad \qquad \qquad~~~~~~~~~~ + \frac{1}{2}  \overline{\lambda}^{\bdot}A_{\gamma}^{~\adot}A_{\rho \bdot}
+ \frac{1}{2} A_{\gamma}^{~\bdot}\overline{\lambda}^{\adot}A_{\rho \bdot} + \frac{1}{2} A_{\gamma}^{~ \bdot}
A_{\rho}^{~\adot}\overline{\lambda}_{\bdot} \Big] \nonumber \\
&& \qquad ~~~~~~~ + {\cal F}^2 \Big[ \frac{1}{4}\overline{\lambda}^{\bdot}\overline{\lambda}^{\adot}
\overline{\lambda}_{\bdot} \Big]
\eea
Useful identities are
\bea
&& i \pa^{\a \adot} \vartheta_{\a} = \frac{1}{2} {\cal F}^{\gamma \rho} \pa_{\rho \dot{\rho}}
\Big[ \big\{ \overline{\lambda}^{\dot{\rho}}, A_{\gamma}^{~\adot}\big\} -  \big\{ \overline{\lambda}^{\adot},
A_{\gamma}^{~\dot{\rho}} \big\} \Big]
\nonumber \\
&& [A^{\a \adot}, \vartheta_{\a}]  - \frac{1}{2} {\cal F}^{\gamma \rho} \big[
\overline{\lambda}_{\bdot}A_{\gamma}^{~\adot}A_{\rho}^{~\bdot} + A_{\gamma \bdot}
\overline{\lambda}^{\adot}A_{\rho}^{~\bdot} + A_{\gamma \bdot} A_{\rho}^{~\adot}\overline{\lambda}^{\bdot} \big]
= \nonumber \\
&&  \qquad \qquad \qquad \qquad\frac{1}{2} {\cal F}^{\gamma \rho}\big[ 2
A_{\gamma}^{~\bdot}\overline{\lambda}^{\adot}A_{\rho \bdot} + \overline{\lambda}^{\adot}
A_{\gamma}^{~\bdot}A_{\rho \bdot} + A_{\gamma}^{~\bdot}A_{\rho \bdot}\overline{\lambda}^{\adot} \big]
\label{C1}
\eea
Using the property ${\rm Tr}~(\overline{\lambda}^{\bdot}\overline{\lambda}^{\adot}
\overline{\lambda}_{\bdot})=0$, the classical gauge action (\ref{invaction})
in components reads
\bea
S_{inv} &=& \frac{1}{g^2} ~\int d^4x ~\left\{ {\rm Tr} ( D'^{~2} -
\frac{1}{2} \overline{f}^{\adot \bdot} \overline{f}_{\adot \bdot} + i \partial^{\a \adot}
\overline{\lambda}_{\adot} \lambda_{\a} + \big[ A^{\a \adot} , \overline{\lambda}_{\adot} \Big]
\lambda_{\a} ) \right. \nonumber \\
&~&~~~~ \left. -i {\cal F}^{\gamma \rho}~ {\rm Tr} (
f_{\gamma \rho} \overline{\lambda}^{\adot} \overline{\lambda}_{\adot})
+ \frac12 ~ {\cal F}^2 ~ {\rm Tr}(\overline{\lambda}^{\adot}
\overline{\lambda}_{\adot} \overline{\lambda}^{\bdot} \overline{\lambda}_{\bdot})
\right\}
\label{invaction-comp}
\eea
Similarly, we perform the reduction of the one--loop divergent contributions (\ref{eff-action1})
obtaining
\bea
&& \Gamma^{(1)~comp}_{gauge}= \frac{(-3 + N_f)}{(4 \pi)^2
\epsilon} ~\int d^4x \Bigg\{ ~ {\cal N} ~{\rm Tr} \Bigg( D'^{~2} -
\frac{1}{2} \overline{f}^{\adot \bdot} \overline{f}_{\adot \bdot} + i \partial^{\a \adot}
\overline{\lambda}_{\adot} \lambda_{\a} + \big[ A^{\a \adot} , \overline{\lambda}_{\adot} \Big]
\lambda_{\a} \Bigg) \nonumber \\
&& \qquad \qquad \qquad \qquad \qquad ~~~ - \Bigg( {\rm Tr}\big( D'\big){\rm Tr}\big( D'\big) -
\frac{1}{2} {\rm Tr}(\overline{f}^{\adot \bdot}){\rm Tr}(\overline{f}_{\adot \bdot}) +
i \partial^{\a \adot}{\rm Tr}( \overline{\lambda}_{\adot}){\rm Tr}( \lambda_{\a})  \Bigg)\nonumber \\
&&~~~~~~\nonumber \\
&& \qquad \qquad \qquad \qquad \qquad ~~~-i {\cal N} {\cal F}^{\gamma \rho}~ {\rm Tr} (
f_{\gamma \rho} \overline{\lambda}^{\adot} \overline{\lambda}_{\adot}) + i~ {\cal F}^{\gamma \rho}~ {\rm Tr}
( f_{\gamma \rho})
{\rm Tr} (\overline{\lambda}^{\adot} \overline{\lambda}_{\adot}) \nonumber \\
&&~~~~\nonumber \\
&& \qquad \qquad \qquad \qquad \qquad ~~~ + \frac12 ~ {\cal F}^2 ~{\cal N} ~ {\rm Tr}(\overline{\lambda}^{\adot}
\overline{\lambda}_{\adot} \overline{\lambda}^{\bdot} \overline{\lambda}_{\bdot}) - \frac12 ~{\cal F}^2
~{\rm Tr} (\overline{\lambda}^{\adot} \overline{\lambda}_{\adot}) {\rm Tr} (\overline{\lambda}^{\bdot}
\overline{\lambda}_{\bdot}) \Bigg\} \nonumber \\
\label{eff-action1-comp}
\eea

In order to renormalize the theory it is convenient to perform the following rescaling on the fields
\beq \label{shift}
A, ~\lambda, ~\overline{\lambda},~D' \qquad \rightarrow \qquad g ~A, ~g~\lambda, ~g ~\overline{\lambda},~g~ D'
\eeq
so that cubic terms in the action are proportional to $g$, quartic terms to $g^2$, while kinetic terms are
independent of the coupling constant and will undergo wave function renormalization.

Since the one--loop divergent contributions independent of the NAC parameter are the same as for
the ordinary $N=1$ SYM, the renormalization functions are fixed by
\bea \label{wave-renorm}
&& Z_{A_{a}} = Z_{\lambda_{a}}^{\frac{1}{2}} \cdot  Z_{\overline{\lambda}_{a}}^{\frac{1}{2}}= Z_{D'_{a}} =
Z_{g}^{-2} = 1 - \frac{(N_{f} - 3)}{(4 \pi)^2 \epsilon} {\cal N}g^2 \nonumber \\
&& Z_{A_{0}} = Z_{\lambda_{0}}=  Z_{\overline{\lambda}_{0}} = Z_{D'_{0}} =1
\eea
where for any field we have defined $\Phi = Z^{-\frac12} \Phi_B$ and $g = Z_g^{-1} g_B$ for the coupling
constant.
Note that in the ordinary case one is forced to choose $Z_{\lambda_{a}} =
Z_{\overline{\lambda}_{a}}$, whereas in the present case they could be different, as we are working in
euclidean space where the two fermions are not related by h.c. conditions.

Armed with the renormalization functions (\ref{wave-renorm}) we then study the new contributions
proportional to the NAC parameter.
We concentrate on the term  $( {\cal F}^{\gamma \rho}~f_{\gamma \rho}\overline{\lambda}^2 )$ which
we rewrite as
\beq
{\rm Tr} (f_{\gamma \rho} \overline{\lambda}^{\adot} \overline{\lambda}_{\adot}) = C^{aBC}
f_{\gamma \rho}^{a} ~\overline{\lambda}^{\adot B} \overline{\lambda}_{\adot}^{C} + {\cal N}^{-\frac{1}{2}}
f_{\gamma \rho}^{0}~ \overline{\lambda}^{\adot B} \overline{\lambda}_{\adot}^{B}
\eeq
where $C^{aBC} = \frac{1}{2}(i~f^{aBC} + d^{aBC})$.

In the one--loop result (\ref{eff-action1-comp}) it is easy to see that the term proportional to
the U(1) field strength actually cancels, whereas the divergent contribution to
the term proportional to the nonabelian $f_{\gamma \rho}^{a}$ reads
\beq
-\frac{(N_{f} - 3)}{(4 \pi)^2 \epsilon} {\cal N}g^2~{\cal F}^{\gamma \rho}~ \int d^4 x~ C^{aBC}~i
f_{\gamma \rho}^{a}~\overline{\lambda}^{\adot B} ~\overline{\lambda}_{\adot}^{C}
\eeq
Therefore the terms in the classical action (\ref{invaction-comp})
\beq \label{3pts}
g~{\cal F}^{\gamma \rho}~f_{\gamma\rho}^{0}~\overline{\lambda}^{\adot~B}~\overline{\lambda}_{\adot}^{B}
= g~{\cal F}^{\gamma \rho}~f_{\gamma \rho}^{0}~\overline{\lambda}^{\adot~0}~\overline{\lambda}_{\adot}^{0}
+g~{\cal F}^{\gamma \rho}~f_{\gamma\rho}^{0}~\overline{\lambda}^{\adot~b}~\overline{\lambda}_{\adot}^{b}
\eeq
do not renormalize. This is made possible if separately the two terms in (\ref{3pts}) do not renormalize,
i.e. if
\bea
Z_{{\cal F}} Z_{g} &=& 1 \nonumber \\
Z_{{\cal F}} Z_{g} \left( Z_{\overline{\lambda}_{b}}^{\frac{1}{2}} \right)^{2} &=& 1
\eea
hold separately. Since in the NAC case  we can choose $\lambda$ and $\overline{\lambda}$
(and their renormalization functions) to be independent, the previous conditions have a
solution compatible
with (\ref{wave-renorm}). Precisely, we choose
\bea
&& Z_{\overline{\lambda}_{a}} = 1 \nonumber \\
&& Z_{\lambda_{a}}^{\frac{1}{2}} =  Z_{A_{a}} \nonumber \\
&& Z_{{\cal F}} = Z_{g}^{-1} = Z_{A_{a}}^{\frac{1}{2}}
\label{wave-renorm2}
\eea
At this point all the renormalization functions have been fixed and no freedom is left.
What we need check is whether the choice (\ref{wave-renorm2}) is good enough to cancel the
rest of the
divergences, i.e  the ones proportional
to $g~{\cal F}^{\gamma \rho}~C^{aBC}~f_{\gamma\rho}^{a}~\overline{\lambda}^{\adot~B}~
\overline{\lambda}_{\adot}^{C}$ and the ones proportional to the product of four $\overline{\lambda}$.
We analyze them separately.

\noindent
a) Summing all the contributions proportional to $g~{\cal F}^{\gamma \rho}~C^{aBC}~
f_{\gamma\rho}^{a}~
\overline{\lambda}^{\adot~B}~\overline{\lambda}_{\adot}^{C}$  up to one--loop we find
\beq
\left( 1 - \frac{(N_{f} - 3)}{(4 \pi)^2~\epsilon} {\cal N}~g^2 \right)
g~{\cal F}^{\gamma \rho}~C^{aBC}~f_{\gamma\rho}^{a}~\overline{\lambda}^{\adot~B}~
\overline{\lambda}_{\adot}^{C}
\eeq
Expressing the renormalized quantities in terms of bare ones we find the condition
\beq
Z_{g}^{-1}~ Z_{{\cal F}}^{-1} ~ Z_{A_{a}}^{-\frac12}
\left( 1 - \frac{(N_{f} - 3)}{(4 \pi)^2~\epsilon} {\cal N}~g^2 \right)  =1
\eeq
Inserting the explicit expressions for the renormalization functions as given in
(\ref{wave-renorm}) it is
easy to see that this condition is {\em not} satisfied.

\noindent
b) We now consider the terms proportional to $g^2 {\cal F}^2 (\overline{\lambda}\overline{\lambda})^2$.
From (\ref{eff-action1-comp}) we see that these terms receive nontrivial divergent
corrections at one loop.
However, given the conditions $Z_g Z_{{\cal F}}=1$ and $Z_{\overline{\lambda}}=1$ we do not
have any possibility to cancel these divergences.\\

Therefore we confirm at the component level the result already evident in superspace,
i.e. that the $N=\frac{1}{2}$ SYM theory defined by the
action (\ref{action1}) as obtained by the natural deformation of the
ordinary $N=1$ SYM theory, is {\em not} one--loop renormalizable and
 the classical action has to be extended by adding extra gauge invariant
terms.


\section{Gauge fixing of the modified action}

In this Appendix we discuss in detail the gauge--fixing procedure for
the modified action (\ref{final-action}).

In the ordinary anticommutative  $SU({\cal N})$ SYM theory described
by the action (\ref{invaction})
the kinetic operator ${\rm Tr} \left( V \overline{D}^{\adot} D^2
\overline{D}_{\adot} V \right)$ is not invertible and one
has to implement the gauge-fixing Faddeev--Popov prescription extended
to superspace (see for example \cite{superspace}). In the
background field method the standard procedure requires the introduction
in the functional integral of
quantum gauge--variant but background gauge covariant quantities (the standard
choice is ${\boldnabla}^2 V$ and $\overline{\boldnabla}^2 V$)
with a suitable weight factor
\begin{equation} \label{naif0}
Z= \int {\mathcal D} V~ {\mathcal D} f~ {\mathcal D} \overline{f}
\Delta^{-1}(V) \d[\overline{\boldnabla}^2 V - f] \d[{\boldnabla}^2 V - f] e^{S_{inv}}
\exp{\left[-\frac{1}{g^2 \a} \int d^4 x d^4 \theta{\rm Tr}
    (f\overline{f})\right]}
\end{equation}
where $f$ and $\overline{f}$ are background covariantly (anti)chiral functions
and the factor $\Delta^{-1}(V)$ is the ghost action.

Performing the $f$, $\overline{f}$ integration one obtains the
standard gauge-fixing action
\begin{equation}
S_{GF} = -\frac{1}{g^2 \a} \int d^4 x d^4 \theta{\rm Tr} \left(
\overline{\boldnabla}^2 V ~{\boldnabla}^2 V \right)
\end{equation}
which combined with the original kinetic term gives rise to the
invertible operator
\begin{equation}
\frac{1}{2~g^2} ~V^{a}~\left[
\overline{D}^{\adot} D^2 \overline{D}_{\adot} - \a^{-1} \left(
D^2 \overline{D}^2 + \overline{D}^2 D^2 ) \right) \right]~ V^{a}
\label{quadr}
\end{equation}
The propagator then reads
\begin{equation}
\langle V~V \rangle = \frac{g^2
~\a}{\Box}~\left[ 1 + (1 - \a^{-1}) \frac{ \overline{D}^{\adot} D^2
\overline{D}_{\adot}}{\Box} \right]
\label{generalprop}
\end{equation}
and in the Feynman gauge ($\a=1$) we recover the standard
vector propagator (\ref{VVprop}).

We now consider, still in the ordinary anticommutative case, the
$U({\cal N})$ SYM theory described by the action
\begin{equation}
\frac{1}{2 g^2} \int d^4 x d^2 \overline{\theta}
~{\rm Tr} ( \overline{W}^{\adot} \overline{W}_{\adot} )
~+~ \frac{1}{2 g_0^2 {\cal N}} \int d^4 x d^2 \overline{\theta}
~{\rm Tr} ( \overline{W}^{\adot}) {\rm Tr}(\overline{W}_{\adot})
\end{equation}
Including the second term amounts to assigning different coupling
constants to the $SU({\cal N})$ and $U(1)$ gauge fields.
The kinetic terms for the different components are in fact
\begin{equation}
\frac{1}{2~g^2} ~V^{a}~
\overline{D}^{\adot} D^2 \overline{D}_{\adot} V^{a} + \frac12\left(
\frac{1}{g^2} + \frac{1}{g_0^2} \right) V^0~\overline{D}^{\adot}
D^2 \overline{D}_{\adot} V^0
\end{equation}
where the label $a$ runs over the $SU({\cal N})$ indices.

In order to fix the gauge, a natural way to proceed is
to change the weight factor in (\ref{naif0}) by adapting it to the
different normalizations of the kinetic terms for the abelian
and nonabelian gauge fields. This amounts to choosing the weight factor
\begin{equation} \label{naif}
    \exp{\left[-\frac{1}{g^2 \a} \int d^4 x d^4 \theta{\rm Tr}
    (f\overline{f})\right]} \times \exp{\left[-\frac{1}{g_0^2 {\cal N}
    \a} \int d^4 x d^4 \theta{\rm Tr} (f) {\rm Tr} (\overline{f})\right]}
\end{equation}
which leads to the convenient propagators (in Feynman gauge)
\begin{eqnarray}
&&\langle V^a~V^b \rangle = \frac{g^2
~}{\Box}  \d^{ab} \label{convenient}\\
&& \langle V^0~V^0 \rangle =
\frac{\widetilde{g}^2}{\Box}
\label{U1convenient}
\end{eqnarray}
with
\begin{equation} \label{gtilde}
 \widetilde{g}^2 = \frac{g^2 g_0^2}{g_0^2 + g^2}
\end{equation}
An alternative possibility would be to keep the weight factor
as in (\ref{naif0}). In Feynman gauge this would lead to the
propagator (\ref{convenient}) for the $SU({\cal N})$ fields, whereas
the abelian propagator would necessarily have the more general structure
\begin{equation}
\langle V^0~V^0 \rangle = \frac{g^2}{\Box}~
\left[ 1 + \left( \frac{g^2}{g_0^2 + g^2} \right) \frac{ \overline{D}^{\adot} D^2
\overline{D}_{\adot}}{\Box} \right]
\label{U1prop}
\end{equation}
In the ordinary nonanticommutative case the two choices are equivalent
and selecting one or the other is simply a matter of convenience.
In the NAC case, instead, because of the presence of the $\ast$-product,
the choice of the weight factor (\ref{naif}) is somewhat inconvenient.
In fact, after functional
integration in $f$ and $\overline{f}$, we would end up with an extra
gauge-fixing term in the action
\begin{equation}
-\frac{1}{g_0^2 {\cal N}
    \a} \int d^4 x d^4 \theta{\rm Tr} (\overline{\boldnabla}^2 V) {\rm
    Tr} (\boldnabla^2 V)
\end{equation}
which is {\em not} background gauge invariant. Since
the power of the background field method lies on this invariance
we would be forced to add to (\ref{naif}) a suitable
completion dependent on background gauge fields in a way very similar
to what has been done in (\ref{ACgaugeinv}). This would also change the
Nielsen--Kallosh ghosts action.

Therefore in this situation it is more convenient to start with the original
weight factor (\ref{naif0}) which does not spoil the background gauge
invariance of the gauge--fixing action.  The invertible kinetic terms are then
\begin{eqnarray}
&& \frac{1}{2~g^2} ~V^{a}~\left[
\overline{D}^{\adot} D^2 \overline{D}_{\adot} - \a^{-1} \left(
D^2 \overline{D}^2 + \overline{D}^2 D^2 ) \right) \right]~ V^{a}
\label{quadr1}\\ && \frac{1}{2~\widetilde{g}^2} ~V^{0}~\left[
\overline{D}^{\adot} D^2 \overline{D}_{\adot} - \widetilde{\a}^{-1}
\left( D^2\overline{D}^2 + \overline{D}^2 D^2 ) \right) \right]~ V^{0}
\label{quadr2}
\end{eqnarray}
with $\widetilde{g}^2$ as in (\ref{gtilde}) and $\widetilde{\a} \equiv
\a~\frac{g^2}{\widetilde{g}^2} $. Choosing the Feynman gauge ($\a =1$) we
finally obtain the propagators (\ref{convenient}) for $SU({\cal N})$
and (\ref{U1prop}) for the $U(1)$ superfields.


\section{Background field expansion of the new vertices}

In this Appendix we perform the background field expansion of the
new vertices emerging from the modified action (\ref{final-action}).
In particular, we extract the terms quadratic in the quantum $V$ field
necessary for one--loop calculations.

In antichiral representation the quantum-background splitting reads
\begin{equation}
\overline{\nabla}^{\adot} = e^{V}_{\ast} \ast\overline{\boldnabla}^{\adot}
\ast e^{-V}_{\ast}
\equiv \overline{D}^{\adot} -i\overline{\Gamma}^{\adot} \qquad \qquad
\nabla^{\a} = \boldnabla^{\a}
\end{equation}
Therefore, expanding up to second order in the quantum V superfields, we find
\bea
&& \overline{\Gamma}^{\adot} \rightarrow \overline{\bold{\Gamma}}^{\adot} -
i \left[ \overline{\boldnabla}^{\adot} , V \right]_\ast + \frac{i}{2}
\left[  \left[\overline{\boldnabla}^{\adot} , V \right]_{\ast} , V \right]_{\ast}
\nonumber \\
&& ~~~~~~=  \overline{\bold{\Gamma}}^{\adot} - i \overline{D}^{\adot} V + \left[ V ,
\overline{\bold{\Gamma}}^{\adot}  \right]_{\ast} - \frac{i}{2}  \left[ V ,
\overline{D}^{\adot} V \right]_{\ast} + \frac{1}{2} \left[ V ,\left[ V ,
\overline{\bold{\Gamma}}^{\adot}  \right]_{\ast}\right]_{\ast}
\eea
The correct expressions for $\overline{W}^{\adot}$ and
$\overline{\Gamma}^{\a \adot}$ follow from the definitions
\begin{equation}
\overline{W}^{\adot} = D^2 \overline{\Gamma}^{\adot} \qquad \qquad
\overline{\Gamma}^{\a \adot} = - i D^{\a} \overline{\Gamma}^{\adot}
\end{equation}
We are now ready to list the contributions quadratic in $V$
from each new vertex and obtain the new Feynman rules suitable for one-loop
calculations. We use conventions introduced in Section \ref{Feynmanrules} and
Appendices A and B. In particular, the color tensors ${\cal P}$ and ${\cal Q}$
are defined in (\ref{Pcal}, \ref{Qcal}). Moreover we avoid ``bold''
notation for background superfields.

\vskip 10pt
\noindent
$\bullet$ `` 2-point function -- double trace term''\\
\bea
&&~~\frac{1}{2~g_0^2~{\cal N}}\int d^4x~d^4 \theta~{\rm Tr}\left(
\overline{\Gamma}^{\adot} \right){\rm Tr}\left(
\overline{W}_{\adot} \right) \rightarrow
\nonumber \\
&& ~- \frac{1}{g_0^2~{\cal N}}\int \Big\{ 2 i ~\sqrt{{\cal N}}~\sinh{(\pi_{1}
\wedge \pi_{2})}~ \delta^{A B} ~\delta^{C 0} ~V^{A}(1)  ~\overline{\Gamma}^{\adot B}(2)
~\left(  \widetilde{D}^2 \widetilde{\overline{D}}_{\adot} V^{C} (3)\right) \nonumber \\
&&  \qquad \qquad ~ +i~\sqrt{{\cal N}}~ \sinh{(\pi_{1} \wedge \pi_{2})}
~\delta^{A B} ~\delta^{C 0} ~V^{A} (1)~\left(
\widetilde{\overline{D}}^{\adot} V^{B} (2) \right) ~\overline{W}_{\adot}^{C}(3) \Big\}
\nonumber \\
&& ~+ \frac{1}{g_0^2~{\cal N}}\int \Big\{ 2 ~ \sinh{(\pi_{1} \wedge
  \pi_{2})} \sinh{(\pi_{3} \wedge \pi_{4})} ~\delta^{A B} ~\delta^{C
  D}~V^{A} (1)  ~\overline{\Gamma}^{\adot B}(2) ~ \widetilde{D}^2 \left( V^{C}(3)
~\overline{\Gamma}_{\adot}^{D}(4) \right) \nonumber \\
&& \qquad ~+ ~\sqrt{{\cal N}}~\sinh{(\pi_{1} \wedge \pi_{2} +
  \pi_{1} \wedge \pi_{3})} \delta^{D 0} ~{\cal
  P}^{ABC}(\pi_{2},\pi_{3})
V^A(1) ~V^{B}(2) ~\overline{\Gamma}^{\adot C}(3) ~ \overline{W}_{\adot}^{D}(4) \Big\}
\nonumber \\
\eea
\vskip 15pt
\noindent
$\bullet$ `` 3-point function''\\
\bea
&& ~~\frac{2 ~i~{\cal F}^{\rho \g} }{g_0^2~{\cal N}}~ \int d^4x~d^4 \theta~~
\overline{\theta}^2~{\rm Tr} \left( \pa_{\rho \dot{\rho}}  \overline{\Gamma}^{\adot} \right)
{\rm Tr}\left( \overline{W}_{\adot} \overline{\Gamma}_{\g}^{~\dot{\rho}} \right)
\rightarrow \nonumber \\
&&- ~\frac{2 ~{\cal F}^{\rho \g} }{g_0^2~{\cal N}}~\int
\Big\{ + i~\sqrt{{\cal N}}~\delta^{A 0 }~\delta^{B C}
~\left( p_{1 \rho
\dot{\rho}}~ \overline{\eth}^2~ \overline{\Gamma}^{\adot A} (1)\right) \left(
\widetilde{D}^2 \widetilde{\overline{D}}_{\adot} V^{B}(2) \right)
 \left( \widetilde{D}_{\gamma} \widetilde{\overline{D}}^{\dot{\rho}}
V^{C}(3) \right)  \nonumber \\
&& ~~~~- ~\sqrt{{\cal N}}~\delta^{A 0 }~\delta^{B C}~\left( p_{1
\rho \dot{\rho}}~ \overline{\eth}^2~\widetilde{\overline{D}}^{\adot}
V^{A}(1)\right)\left( \widetilde{D}^2 \widetilde{\overline{D}}_{\adot}
V^{B}(2) \right) ~\overline{\Gamma}_{\g}^{~\dot{\rho}C}(3) \nonumber \\
&& ~~~~ + i~\sqrt{{\cal N}}~\delta^{A 0 }~\delta^{B C}~ \left( p_{1 \rho
\dot{\rho}} ~\overline{\eth}^2~\widetilde{\overline{D}}^{\adot}
V^{A} (1) \right)~\overline{W}_{\adot}^{B} (2)~ \left( \widetilde{D}_{\gamma}
\widetilde{\overline{D}}^{\dot{\rho}} V^{C} (3)\right) \Big\} \nonumber \\
&& - ~\frac{2 ~{\cal F}^{\rho \g} }{g_0^2~{\cal N}}~\int
\Big\{ - \frac{1}{2}~\sqrt{{\cal N}}~\delta^{A 0
}~f^{BCD}~\left( p_{1 \rho \dot{\rho}}~\overline{\eth}^2~ \overline{\Gamma}^{\adot
A} (1)\right)~\widetilde{D}^2 \left( \widetilde{\overline{D}}_{\adot}V^{B} (2)~
V^{C}(3) \right)~ \overline{\Gamma}_{\g}^{~\dot{\rho} D} (4)  \nonumber \\
&& ~~~~~ + i~\sqrt{{\cal N}}~\delta^{A 0 }~f^{BCD}~\left( p_{1 \rho
\dot{\rho}}~\overline{\eth}^2~ \overline{\Gamma}^{\adot A} (1)\right)
~\widetilde{D}^2
\left( \overline{\Gamma}_{\adot}^{B} (2)~ V^{C} (3)\right) \left(
\widetilde{D}_{\gamma} \widetilde{\overline{D}}^{\dot{\rho}} V^{D} (4)
\right) \nonumber \\
&& ~~~~~ -~\sqrt{{\cal N}}~\delta^{A 0 }~f^{BCD}~\left( p_{1 \rho
\dot{\rho}}~\overline{\eth}^2~ \widetilde{\overline{D}}^{\adot} V^{A} (1)\right) ~
\widetilde{D}^2 \left( \overline{\Gamma}_{\adot}^{B}(2) V^{C} (3) \right)
~\overline{\Gamma}_{\g}^{~\dot{\rho} D} (4) \nonumber \\
&& ~~~~~ + \frac{1}{2}~\sqrt{{\cal N}}~\delta^{A 0 }~{\cal
  P}^{BCD}(\pi_3, \pi_4)
\left( p_{1 \rho \dot{\rho}}~
\overline{\eth}^2~\overline{\Gamma}^{\adot A}(1) \right)~\overline{W}_{\adot }^{B}(2)
~\widetilde{D}_{\gamma} \left( \widetilde{\overline{D}}^{\dot{\rho}}
V^{C}(3)~ V^{D}(4)\right)
\nonumber \\
&& ~~~~~ + ~\sqrt{{\cal N}}~\delta^{A 0 }~{\cal P}^{BCD}(\pi_3, \pi_4)~\left( p_{1
\rho \dot{\rho}}~\overline{\eth}^2~ \overline{\Gamma}^{\adot A} (1)
\right) \left( \widetilde{D}^2 \widetilde{\overline{D}}_{\adot} V^{B}(2)\right)
~\widetilde{D}_{\gamma}  \left( \overline{\Gamma}^{\dot{\rho} C}(3)~  V^{D}(4) \right)
 \nonumber \\
&& ~~~~~ + ~\sqrt{{\cal N}}~\delta^{A 0 }~{\cal P}^{BCD} (\pi_3, \pi_4)~\left( p_{1
\rho \dot{\rho}}~\overline{\eth}^2~ \widetilde{\overline{D}}^{\adot} V^{A}
(1)\right) ~\overline{W}_{\adot}^{B}(2) ~\widetilde{D}_{\gamma}
\left(\overline{\Gamma}^{\dot{\rho} C}(3)~ V^{D}(4) \right) \nonumber \\
&& ~~~~~ + 2~i~\delta^{A B }~ \delta^{C D}~\sinh{(\pi_1 \wedge
\pi_2)}~\left( p_{1 \rho \dot{\rho}} + p_{2 \rho \dot{\rho}} \right) ~\overline{\eth}^2
\overline{\Gamma}^{\adot A}(1) ~ V^B(2)
 ~\left( \widetilde{D}^2 \widetilde{\overline{D}}_{\adot} V^{C}(3) \right)
~\overline{\Gamma}_{\g }^{~\dot{\rho} D} (4)\nonumber \\
&& ~~~~~ + 2~\delta^{A B }~ \delta^{C D}~\sinh{(\pi_1 \wedge
\pi_2)}~\left( p_{1 \rho \dot{\rho}} + p_{2 \rho \dot{\rho}}
\right)~\overline{\eth}^2 \overline{\Gamma}^{\adot A}(1) ~ V^B(2)
~\overline{W}_{\adot}^{C}(3)~ \left( \widetilde{D}_{\gamma}
\widetilde{\overline{D}}^{\dot{\rho}} V^{D}(4)\right) \nonumber \\
&& ~~~~~ + i~ \delta^{A B }~ \delta^{C D}~\sinh{(\pi_1 \wedge \pi_2)}~
\left( p_{1 \rho \dot{\rho}}+ p_{2 \rho \dot{\rho}}\right) \left(
~\overline{\eth}^2~\widetilde{\overline{D}}^{\adot} V^{A} (1)\right)
~V^{B} (2)~\overline{W}_{\adot}^{C}(3)~ \overline{\Gamma}_{\g}^{~\dot{\rho} D}(4) \Big\}
 \nonumber \\
&& - ~\frac{2 ~{\cal F}^{\rho \g} }{g_0^2~{\cal N}}~\int
%
\Big\{ + \frac{i}{2}~\sqrt{{\cal N}}~\delta^{A
0}~f^{FDE}~{\cal P}^{BCF} (\pi_2, \pi_3) \times \nonumber \\
&& \qquad \qquad \qquad \qquad \qquad \qquad \times \left(p_{1 \rho \dot{\rho}}
~\overline{\eth}^2~ \overline{\Gamma}^{\adot A}(1)\right) ~ \widetilde{D}^2
\left( \overline{\Gamma}_{\adot}^{B}(2)~V^{C}(3) ~V^{D}(4) \right)~
\overline{\Gamma}_{\g}^{~\dot{\rho} E}(5) \nonumber \\
&& ~~~~~ + \sqrt{{\cal N}}~\delta^{A 0}~f^{BCF}~{\cal
  P}^{DEF}~(\pi_4, \pi_5)~\times \nonumber \\
&& \qquad \qquad \qquad \qquad \qquad \qquad \times
\left( p_{1 \rho \dot{\rho}}
~\overline{\eth}^2~\overline{\Gamma}^{\adot A}(1) \right) ~ \widetilde{D}^2
\left( \overline{\Gamma}_{\adot}^{B}(2)~ V^{C}(3) \right) ~\widetilde{D}_{\gamma}
\left( \overline{\Gamma}^{\dot{\rho} D}(4) ~V^{E}(5) \right) \nonumber \\
&& ~~~~~ + 2~i~\delta^{A B }~f^{CDE}~\sinh{(\pi_1 \wedge \pi_2)}~~\times \nonumber \\
&& \qquad \qquad \qquad \qquad \qquad \qquad \times
\left( p_{1 \rho \dot{\rho}} + p_{2 \rho \dot{\rho}} \right)
~\overline{\eth}^2 \overline{\Gamma}^{\adot A}(1)~ V^{B}(2) ~ \widetilde{D}^2
\left( \overline{\Gamma}_{\adot}^{C}(3)~ V^{D}(4) \right)~
\overline{\Gamma}_{\g}^{~\dot{\rho} E}(5) \nonumber \\
&& ~~~~ - 2~i~\delta^{A B }~\sinh{(\pi_{1} \wedge \pi_{2})}~{\cal
  P}^{CDE}(\pi_4, \pi_5)~~\times \nonumber \\
&& \qquad \qquad \qquad \qquad \qquad \qquad \times
\left( p_{1 \rho \dot{\rho}} + p_{2 \rho \dot{\rho}}
\right) ~\overline{\eth}^2 \overline{\Gamma}^{\adot A}(1) ~ V^{B}(2) ~
\overline{W}_{\adot}^{C}(3)~ \widetilde{D}_{\gamma}
\left( \overline{\Gamma}^{\dot{\rho} D}(4) ~ V^{E}(5) \right) \nonumber \\
&& ~~~~ + \delta^{D E }~ \sinh{(\pi_{1} \wedge \pi_{3} + \pi_{2}
\wedge \pi_{3})}~{\cal P}^{ABC}(\pi_{1}, \pi_{2}) ~\times \nonumber \\
&& \qquad \qquad \qquad \qquad \qquad \qquad \times
~\left( p_{1 \rho \dot{\rho}} +p_{2 \rho \dot{\rho}}+ p_{3 \rho \dot{\rho}}\right)
~\overline{\eth}^2 \overline{\Gamma}^{\adot A}(1)~ V^{B}(2)~V^{C}(3)
~ \overline{W}_{\adot}^{D}(4) ~\overline{\Gamma}_{\g}^{~\dot{\rho} E}(5) \nonumber \\
&& ~~~~ - \frac{i}{2} ~\sqrt{{\cal N}}~ \delta^{A 0}~ {\cal P}^{BFE}
(\pi_{3}+ \pi_{4},\pi_{5}) ~
{\cal P}^{CDF}(\pi_{3}, \pi_{4}) ~\times \nonumber \\
&& \qquad \qquad \qquad \qquad \qquad \qquad \times
~\left( p_{1 \rho \dot{\rho}}~\overline{\eth}^2~ \overline{\Gamma}^{\adot A}(1)\right)
\overline{W}_{\adot}^{B}(2)~ \widetilde{D}_{\gamma}
\left( \overline{\Gamma}^{\dot{\rho} C}(3) ~ V^{D}(4) ~ V^{E}(5) \right) \Big\} \nonumber \\
\eea
\vskip 15pt
\noindent
$\bullet$ `` 4-point function -- double trace term''\\
\bea
&& ~~ \frac{{\cal F}^2 }{2~h^2~{\cal N}}~ \int d^4x~d^4 \theta~~
\overline{\theta}^2~{\rm Tr} \left(\overline{\Gamma}^{\adot}
\overline{W}_{\adot}\right){\rm Tr}\left(\overline{W}^{\bdot}
\overline{W}_{\bdot}\right) \rightarrow \nonumber \\
&&~~~  \frac{{\cal F}^2}{h^2~{\cal N}}~ \int
\Big\{ ~  \delta^{AB}~ \delta^{C D}~\left(
~\overline{\eth}^2~\widetilde{\overline{D}}^{\adot} ~V^{A}(1) \right)
~\left( \widetilde{D}^2 \widetilde{\overline{D}}_{\adot} V^{B}(2)
\right) ~\overline{W}^{\bdot C}(3) ~\overline{W}_{\bdot}^{D}(4) \nonumber \\
&& \qquad \qquad ~+2~ \delta^{AB}~\delta^{C D}~\left(
~\overline{\eth}^2~\widetilde{\overline{D}}^{\adot} ~V^{A}(1)
\right)~\overline{W}_{\adot}^{B}(2)~\left( \widetilde{D}^2
\widetilde{\overline{D}}^{\bdot} V^{C}(3) \right)~\overline{W}_{
\bdot}^{D}(4) \Big\} \nonumber \\
&&+ \frac{{\cal F}^2}{h^2~{\cal N}} \int
\Big\{ - f^{ABC}~\delta^{DE}~\left(
\overline{\eth}^2~V^{A}(1) \right) \left( \widetilde{\overline{D}}^{\adot}
V^{B}(2) \right) ~ \overline{W}_{\adot}^{C}(3) ~\overline{W}^{\bdot D}(4)
~\overline{W}_{\bdot}^{E}(5) \nonumber \\
&& \qquad \qquad ~-2 ~ f^{ABC}~\delta^{DE}~\left(
~\overline{\eth}^2~\widetilde{\overline{D}}^{\adot}~ V^{A}(1) \right) ~
\widetilde{D}^2 \left( V^{B}(2) \overline{\Gamma}_{\adot}^{C}(3) \right)
~\overline{W}^{\bdot D}(4) ~\overline{W}_{\bdot}^{E}(5) \nonumber \\
&& \qquad \qquad ~ -4 ~ f^{ABC}~\delta^{DE}~\left(
~\overline{\eth}^2~V^{A}(1) \right) ~\overline{\Gamma}^{\adot
B}(2)~\overline{W}_{\adot}^{C}(3)~\left( \widetilde{D}^2
\widetilde{\overline{D}}^{\bdot} V^{D}(4) \right)~\overline{W}_{
\bdot}^{E}(5) \Big\} \nonumber \\
&&+ \frac{{\cal F}^2}{h^2~{\cal N}} \int
\Big\{  ~ f^{ABG}~ f^{GCD}~\delta^{EF}~\left(
\overline{\eth}^2~V^{A}(1)\right)~\overline{\Gamma}^{\adot B}(2) ~
\widetilde{D}^2 \left(V^{C}(3) \overline{\Gamma}_{\adot}^{D}(4)
\right)~\overline{W}^{\bdot E}(5) ~\overline{W}_{\bdot}^{F}(6) \nonumber \\
&& \qquad \qquad ~ +2
~f^{ABC}~f^{DEF}~\left(\overline{\eth}^2~V^{A}(1)\right)~\overline{\Gamma}^{\adot
B}(2)~\overline{W}_{\adot}^{C}(3)~\widetilde{D}^2 \left(V^{D}(4)
\overline{\Gamma}^{\bdot E}(5) \right)~\overline{W}_{\bdot}^{F}(6)\nonumber \\
&& \qquad \qquad ~-i~ f^{DAG}~{\cal P}^{GBC} (\pi_{2}, \pi_{3})~
\delta^{E F}
\left(\overline{\eth}^2~V^{A}(1)\right) ~V^{B}(2)~\overline{\Gamma}^{\adot C}(3)
~\overline{W}_{\adot}^{D}(4) ~\overline{W}^{\bdot E}(5)\overline{W}_{\bdot}^{F}(6)
\Big\} \nonumber \\
\eea
\vskip 15pt
\noindent
$\bullet$ `` 4-point function -- single trace term''\\
\bea
&& ~~~\frac{{\cal F}^2}{l^2} ~\int d^4 \theta~\overline{\theta}^2
    {\rm Tr} \left( \overline{\Gamma}^{\adot}~
\overline{W}_{\adot}~\overline{W}^{\bdot}~\overline{W}_{\bdot}\right)
\rightarrow \nonumber \\
&&~~~\frac{{\cal F}^2}{l^2} ~\int
\Big\{ ~ \frac{1}{2} ~
d^{ABE} d^{ECD}
~ \left( \overline{\eth}^2
~\widetilde{\overline{D}}^{\adot} V^A(1) \right) ~\left(
\widetilde{D}^2 \widetilde{\overline{D}}_{\adot} V^B(2) \right)
~\overline{W}^{\bdot C}(3)~\overline{W}_{\bdot}^D(4) \nonumber \\
&& \qquad \qquad + ~d^{ABE}~d^{ECD}~\left( \overline{\eth}^2
~\widetilde{\overline{D}}^{\adot} V^A(1) \right) ~\overline{W}_{
\adot}^B(2)~\left( \widetilde{D}^2 \widetilde{\overline{D}}^{\bdot} V^C(3)
\right)~\overline{W}_{\bdot}^D(4) \Big\} \nonumber \\
&&~+ \frac{{\cal F}^2}{l^2}~ \int
\Big\{ ~- \frac{1}{2}~f^{ABF}d^{FCG}~\left( i f^{GDE} + d^{GDE} \right)~\times
\nonumber \\
&& \qquad \qquad \qquad \qquad \qquad\qquad \qquad
\left( \overline{\eth}^2~ V^A(1) \right)~ \left(
\widetilde{\overline{D}}^{\adot} V^B(2) \right)~\overline{W}_{
\adot}^C(3)~\overline{W}^{\bdot D}(4)~\overline{W}_{\bdot}^E (5) \nonumber
\\
&& ~~\qquad \qquad -  ~d^{AFG}f^{BCF}~\left( i f^{GDE} + d^{GDE}
\right) ~\times
\nonumber \\
&& \qquad \qquad \qquad \qquad \qquad \qquad \qquad\left( ~\overline{\eth}^2~\widetilde{D}^2
\widetilde{\overline{D}}^{\adot} V^A(1) \right)~
V^B(2)~ \overline{\Gamma}_{\adot}^C(3)~\overline{W}^{D
\bdot}(4)~\overline{W}_{\bdot}^E(5) \nonumber \\
&& ~~\qquad \qquad - 2 ~d^{ABF}f^{CDG}d^{FGE}~\left(~\overline{\eth}^2 ~\widetilde{D}^2
\widetilde{\overline{D}}^{\adot} V^A(1) \right)~\overline{W}_{
\adot}^B(2)~ V^C(3) ~\overline{\Gamma}^{\bdot D}(4) ~\overline{W}_{
\bdot}^E(5) \Big\}\nonumber \\
&&~+ \frac{{\cal F}^2}{l^2} \int
\Big\{  \frac{1}{2} ~ f^{ABG}~ f^{CDH}~\left( i f^{GHI}
+d^{GHI}\right)\left( i f^{IEF} + d^{IEF}\right)~\times
\nonumber \\
&& \qquad \qquad \qquad \qquad \qquad\qquad \qquad \left(
\overline{\eth}^2~V^{A}(1)\right)~\overline{\Gamma}^{\adot B}(2) ~
\widetilde{D}^2 \left(V^{C}(3) \overline{\Gamma}_{\adot}^{D}(4)
\right)~\overline{W}^{\bdot E}(5) ~\overline{W}_{\bdot}^{F}(6) \nonumber \\
&& \qquad \qquad ~ +
~f^{ABG}d^{GCH}f^{DEI}d^{IFH}~\times
\nonumber \\
&& \qquad \qquad \qquad \qquad \qquad\qquad \qquad
\left(\overline{\eth}^2~V^{A}(1)\right)~\overline{\Gamma}^{\adot
  B}(2)~\overline{W}_{\adot}^{C}(3)~\widetilde{D}^2 \left(V^{D}(4)
\overline{\Gamma}^{\bdot E}(5) \right)~\overline{W}_{
  \bdot}^{F}(6)\nonumber \\
&& \qquad \qquad ~-~\frac{i}{2}~ f^{AGH} {\cal P}^{BCG}
(\pi_{2}, \pi_{3})
~d^{HDI} ~\left( i f^{IEF} + d^{IEF} \right)~\times
\nonumber \\
&& \qquad \qquad \qquad \qquad \qquad\qquad \qquad \left(\overline{\eth}^2~V^{A}(1)\right)
~V^{B}(2)~\overline{\Gamma}^{\adot C}(3) ~\overline{W}_{\adot}^{D}(4)
~\overline{W}^{\bdot E}(5)\overline{W}_{\bdot}^{F}(6) \Big\} \nonumber \\
\eea
\vskip 10pt
\noindent
$\bullet$ `` 4-point function -- triple trace term''\\
\bea
&& ~\frac{{\cal F}^2}{r^2} ~\int d^4 \theta~\overline{\theta}^2
    {\rm Tr} \left( \overline{\Gamma}^{\adot} \right)~{\rm Tr} \left(
\overline{W}_{\adot} \right)~{\rm Tr} \left( \overline{W}^{\bdot}~\overline{W}_{\bdot}\right)
\rightarrow \nonumber \\
&&~~\frac{{\cal F}^2~{\cal N}}{r^2} \int
\Big\{
\left( \d^{A0} \d^{B0} \d^{CD} + \d^{AB} \d^{C0} \d^{D0}\right)
\left( \overline{\eth}^2
~\widetilde{\overline{D}}^{\adot} V^A(1) \right) \left(
\widetilde{D}^2 \widetilde{\overline{D}}_{\adot} V^B(2) \right)
\overline{W}^{\bdot C}(3)~\overline{W}_{\bdot}^D(4) \nonumber \\
&& \qquad \qquad + 4 ~\d^{A0} \d^{B0} \d^{CD}~\left( \overline{\eth}^2
~\widetilde{\overline{D}}^{\adot} V^A(1) \right) ~\overline{W}_{
\adot}^B(2)~\left( \widetilde{D}^2 \widetilde{\overline{D}}^{\bdot} V^C(3)
\right)~\overline{W}_{\bdot}^D(4) \qquad \Big\} \nonumber \\
&&+ \frac{{\cal F}^2~{\cal N}}{r^2} \int
\Big\{ ~-~f^{ABC}~\d^{D0}~\d^{E0}~
\left( \overline{\eth}^2~ V^A(1) \right) \left(
\widetilde{\overline{D}}^{\adot} V^B(2) \right)~\overline{W}_{
\adot}^C(3)~\overline{W}^{\bdot D}(4)~\overline{W}_{\bdot}^E (5) \nonumber
\\
&& ~~\qquad \qquad -  2 ~ f^{ABC}~\d^{D0}~\d^{E0}~
\left( ~\overline{\eth}^2~\widetilde{D}^2
\widetilde{\overline{D}}^{\adot} V^A(1) \right)~
V^B(2)~ \overline{\Gamma}_{\adot}^C(3)~\overline{W}^{D
\bdot}(4)~\overline{W}_{\bdot}^E(5) \nonumber \\
&& ~~\qquad \qquad - 4 ~\d^{A0}~\d^{B0}~f^{CDE}~\left(~\overline{\eth}^2 ~\widetilde{D}^2
\widetilde{\overline{D}}^{\adot} V^A(1) \right)~\overline{W}_{
\adot}^B(2)~ V^C(3) ~\overline{\Gamma}^{\bdot D}(4) ~\overline{W}_{
\bdot}^E(5) ~~~\Big\}\nonumber \\
&&+ \frac{{\cal F}^2~{\cal N}}{r^2} \int
\Big\{  f^{ABG}f^{GCD}\d^{E0}\d^{F0}
\left(
\overline{\eth}^2~V^{A}(1)\right)\overline{\Gamma}^{\adot B}(2) ~
\widetilde{D}^2 \left(V^{C}(3) \overline{\Gamma}_{\adot}^{D}(4)
\right)\overline{W}^{\bdot E}(5) ~\overline{W}_{\bdot}^{F}(6) \nonumber \\
&& \qquad \qquad ~-~i~ f^{AHD} {\cal P}^{BCH}
(\pi_{2}, \pi_{3})\d^{E0} \d^{F0}
\left(\overline{\eth}^2~V^{A}(1)\right)
V^{B}(2)~\overline{\Gamma}^{\adot C}(3) ~\overline{W}_{\adot}^{D}(4)
~\overline{W}^{\bdot E}(5)\overline{W}_{\bdot}^{F}(6) \Big\} \nonumber \\
\eea
These vertices together with the ones reported at the end of Section 4
allow us to compute all the divergent terms which arise at one loop for
the theory described by the modified action (\ref{final-action}).

\newpage



\begin{thebibliography}{100}

\bibitem{OV} H.~Ooguri and C.~Vafa,
  ``The C-deformation of gluino and non-planar diagrams,''
  Adv.\ Theor.\ Math.\ Phys.\  {\bf 7}, 53 (2003), hep-th/0302109;\\
  ``Gravity induced C-deformation,''
  Adv.\ Theor.\ Math.\ Phys.\  {\bf 7} (2004) 405, hep-th/0303063.

\bibitem{seiberg} N.~Seiberg,
  ``Noncommutative superspace, N = 1/2 supersymmetry, field theory and  string
  theory,''
  JHEP {\bf 0306} (2003) 010, hep-th/0305248.

\bibitem{seiberg2} N.~Berkovits and N.~Seiberg,
  ``Superstrings in graviphoton background and N = 1/2 + 3/2 supersymmetry,''
  JHEP {\bf 0307} (2003) 010, hep-th/0306226.

\bibitem{peter} J.~de Boer, P.~A.~Grassi and P.~van Nieuwenhuizen,
  ``Non-commutative superspace from string theory,''
  Phys.\ Lett.\ B {\bf 574} (2003) 98, hep-th/0302078.

\bibitem{lerda}
  M.~Billo, M.~Frau, I.~Pesando and A.~Lerda,
  ``N = 1/2 gauge theory and its instanton moduli space from open strings  in
  R-R background,''
  JHEP {\bf 0405} (2004) 023, hep-th/0402160; \\
  M.~Billo, M.~Frau, F.~Lonegro and A.~Lerda,
  ``N = 1/2 quiver gauge theories from open strings with R-R fluxes,''
  JHEP {\bf 0505} (2005) 047, hep-th/0502084.

\bibitem{ferrara}S.~Ferrara and M.~A.~Lledo,
  ``Some aspects of deformations of supersymmetric field theories,''
  JHEP {\bf 0005} (2000) 008, hep-th/0002084.

\bibitem{KPT} D.~Klemm, S.~Penati and L.~Tamassia,
  ``Non(anti)commutative superspace,''
  Class.\ Quant.\ Grav.\  {\bf 20} (2003) 2905, hep-th/0104190.

\bibitem{ferrara2} S.~Ferrara, M.~A.~Lledo and O.~Macia,
  ``Supersymmetry in noncommutative superspaces,''
  JHEP {\bf 0309} (2003) 068, hep-th/0307039.


\bibitem{AIO} T.~Araki, K.~Ito and A.~Ohtsuka,
  ``Supersymmetric gauge theories on noncommutative superspace,''
  Phys.\ Lett.\ B {\bf 573} (2003) 209, hep-th/0307076.

\bibitem{BF} R.~Britto and B.~Feng,
  ``N = 1/2 Wess-Zumino model is renormalizable,''
  Phys.\ Rev.\ Lett.\  {\bf 91} (2003) 201601, hep-th/0307165.

\bibitem{LR} O.~Lunin and S.~J.~Rey,
  ``Renormalizability of non(anti)commutative gauge theories with N = 1/2
  supersymmetry,''
  JHEP {\bf 0309} (2003) 045, hep-th/0307275.

\bibitem{JJW} I.~Jack, D.~R.~T.~Jones and L.~A.~Worthy,
  ``One-loop renormalisation of N = 1/2 supersymmetric gauge theory,''
  Phys.\ Lett.\ B {\bf 611} (2005) 199, hep-th/0412009;
  ``One-loop renormalisation of general N = 1/2 supersymmetric gauge theory,''
  Phys.\ Rev.\ D {\bf 72} (2005) 065002,
  hep-th/0505248.

\bibitem{Ryttov}
  T.~A.~Ryttov and F.~Sannino,
  ``Chiral models in noncommutative N = 1/2 four dimensional superspace,''
  Phys.\ Rev.\ D {\bf 71} (2005) 125004, hep-th/0504104.

\bibitem{Britto0}
  R.~Britto, B.~Feng and S.~J.~Rey,
  ``Deformed superspace, N = 1/2 supersymmetry and (non)renormalization
  theorems,''
  JHEP {\bf 0307} (2003) 067, hep-th/0306215.

\bibitem{Terashima0}
  S.~Terashima and J.~T.~Yee,
  ``Comments on noncommutative superspace,''
  JHEP {\bf 0312} (2003) 053, hep-th/0306237.

\bibitem{Britto2}
  R.~Britto, B.~Feng and S.~J.~Rey,
  ``Non(anti)commutative superspace, UV/IR mixing and open Wilson lines,''
  JHEP {\bf 0308} (2003) 001, hep-th/0307091.

\bibitem{us} M.~T.~Grisaru, S.~Penati and A.~Romagnoni,
  ``Two-loop renormalization for nonanticommutative N = 1/2 supersymmetric  WZ
  model,''
  JHEP {\bf 0308} (2003) 003, hep-th/0307099.
\bibitem{R} A.~Romagnoni,
  ``Renormalizability of N = 1/2 Wess-Zumino model in superspace,''
  JHEP {\bf 0310} (2003) 016, hep-th/0307209.

\bibitem{Beren}
  D.~Berenstein and S.~J.~Rey,
  ``Wilsonian proof for renormalizability of N = 1/2 supersymmetric field
  theories,''
  Phys.\ Rev.\ D {\bf 68} (2003) 121701, hep-th/0308049.

\bibitem{Banin0}
  A.~T.~Banin, I.~L.~Buchbinder and N.~G.~Pletnev,
  ``Chiral effective potential in N = 1/2 non-commutative Wess-Zumino  model,''
  JHEP {\bf 0407} (2004) 011, hep-th/0405063.

\bibitem{pr} S.~Penati and A.~Romagnoni,
  ``Covariant quantization of N = 1/2 SYM theories and supergauge invariance,''
  JHEP {\bf 0502} (2005) 064, hep-th/0412041.

\bibitem{Azorkina:2005mx}
  O.~D.~Azorkina, A.~T.~Banin, I.~L.~Buchbinder and N.~G.~Pletnev,
  ``Generic chiral superfield model on nonanticommutative N = 1/2 superspace,''
  Mod.\ Phys.\ Lett.\ A {\bf 20} (2005) 1423, hep-th/0502008.

\bibitem{Hatanaka}
  T.~Hatanaka, S.~V.~Ketov, Y.~Kobayashi and S.~Sasaki,
  ``Non-anti-commutative deformation of effective potentials in supersymmetric
  gauge theories,''
  Nucl.\ Phys.\ B {\bf 716} (2005) 88, hep-th/0502026.

\bibitem{Ihl}
  M.~Ihl and C.~Saemann,
  ``Drinfeld-twisted supersymmetry and non-anticommutative superspace,''
  hep-th/0506057.


\bibitem{ABBP2} O.D. Azorkina, A.T. Banin, I.L. Buchbinder, N.G. Pletnev,
``Construction of the effective action in nonanticommutative
  supersymmetric field theories'', hep-th/0509193.

\bibitem{superspace}{S.J. Gates, Jr., M.T. Grisaru, M. Ro\v cek and W. Siegel,
{\em Superspace}, Benjamin Cummings, (1983) Reading, MA.}

\bibitem{JJW3} I.~Jack, D.~R.~T.~Jones and L.~A.~Worthy,
  ``Renormalisation of supersymmetric gauge theory in the uneliminated
  component formalism,'' hep-th/0509089.

\bibitem{ILZ} E.~Ivanov, O.~Lechtenfeld and B.~Zupnik,
  ``Nilpotent deformations of N = 2 superspace,''
  JHEP {\bf 0402} (2004) 012, hep-th/0308012.

\bibitem{GSZ} M.~T.~Grisaru and W.~Siegel,
  ``Supergraphity. 2. Manifestly Covariant Rules And Higher Loop Finiteness,''
  Nucl.\ Phys.\ B {\bf 201} (1982) 292
  [Erratum-ibid.\ B {\bf 206} (1982) 496]; \\
  M.~T.~Grisaru and D.~Zanon,
  ``Covariant Supergraphs. 1. Yang-Mills Theory,''
  Nucl.\ Phys.\ B {\bf 252} (1985) 578.

\bibitem{filk} T.~Filk,
  ``Divergencies in a field theory on quantum space,''
  Phys.\ Lett.\ B {\bf 376} (1996) 53.

\bibitem{minwalla} S.~Minwalla, M.~Van Raamsdonk and N.~Seiberg,
  ``Noncommutative perturbative dynamics,''
  JHEP {\bf 0002} (2000) 020, hep-th/9912072.

\bibitem{Improved}
  M.~T.~Grisaru, W.~Siegel and M.~Rocek,
  ``Improved Methods For Supergraphs,''
  Nucl.\ Phys.\ B {\bf 159} (1979) 429.

\bibitem{ABBP} R. Abbaspur, A. Imaanpur,
``Nonanticommutative Deformation of N=4 SYM Theory: The Myers Effect and Vacuum States'',
hep-th/0509220.


\end{thebibliography}
\end{document}